\documentclass{ws-ijgmmp}
\newcommand{\ttbar}{\ensuremath{{t\bar{t}}}}
\newcommand{\mttbar}{\ensuremath{m^{t\bar{t}}}}
\newcommand{\mtt}{\ensuremath{m^{t\bar{t}}}}
\newcommand{\DeltaPhitt}{\ensuremath{\Delta\phi^{t\bar{t}}}}

\newcommand{\pt}{\ensuremath{{p_\mathrm{T}}}}
\newcommand{\pout}{\ensuremath{{p_\mathrm{out}}}}

\newcommand{\ptt}{\ensuremath{{p^{t}_\mathrm{T}}}}
\newcommand{\ptto}[1]{\ensuremath{{p^{t#1}_\mathrm{T}}}}
\newcommand{\pttt}{\ensuremath{{p^{\ttbar{}}_\mathrm{T}}}}

\newcommand{\HT}{\ensuremath{{H_\mathrm{T}}}}

\newcommand{\HTj}{\ensuremath{{H^{j}_\mathrm{T}}}}
\newcommand{\Etmiss}{\ensuremath{E^\mathrm{miss}_\mathrm{T}}}

\newcommand{\Yboost}{y_{\mathrm{boost}}^{\ttbar{}}}
\newcommand{\Rtt}{R^{t1,t2}}
\newcommand{\Chittbar}{\chi^{t\bar{t}}}

\newcommand{\Delphes}{\textsc{Delphes}\xspace}
\newcommand{\Pythia}{\textsc{Pythia}\xspace}
\newcommand{\GeV}[1]{{#1$\,\textrm{GeV}$}\xspace}
\newcommand{\MadGraph}{\textsc{MadGraph5}\xspace}
\newcommand{\BumpHunter}{\textsc{BumpHunter}\xspace}

\usepackage{graphicx}
\usepackage{graphics}
\usepackage{amsmath,amssymb}
\usepackage{xspace}
\usepackage{xcolor}
\usepackage{url}
\usepackage{hyperref}

\begin{document}

\markboth{Ji\v{r}\'{\i} Kvita}{Boosted and semi-boosted all-hadronic $t\bar{t}$ reconstruction performance on kinematic variables for selected BSM models using a 2D extension of the \BumpHunter{} algorithm}

%
\catchline{}{}{}{}{}
%

\title{Boosted and semi-boosted all-hadronic $t\bar{t}$ reconstruction performance on kinematic variables for selected BSM models using a 2D extension of the \BumpHunter{} algorithm}
\author{JI\v{R}\'{I} KVITA
}

\address{Joint Laboratory of Optics of Palack\'{y} University and Institute of Physics AS CR\\
  Faculty of Science, Palack\'{y} University, Olomouc, Czech Republic \\
\email{jiri.kvita@upol.cz}
}

\maketitle

\begin{history}
\received{(June 2023)}
\revised{}
\end{history}

\begin{abstract}
We explore the usage of boosted as well as semi-boosted topologies in all-hadronic $t\bar{t}$ final states in simulated $pp$ collisions at $\sqrt{s} = 14\,$TeV, with top quarks decaying into a boosted hadronic top-jet or a $W$-jet and an isolated $b$-jet.
Correlations between selected kinematic variables and their shapes are studied for scalar and vector resonances decaying to a pair of top quarks, and also for a models of $t\bar{t}$-associated production with an invisible dark matter particle pair. Stacked signal$+$background samples have been investigated in terms of the ability to resolve an excess of events over the Standard Model background in terms of the $t\bar{t}$ invariant mass, top quark transverse momentum and other 1D and 2D spectra using a parameterized detector simulation. A 2D extension of the {\textsc BumpHunter} algorithm is proposed, resulting in an improved signal sensitivity in specific 2D areas. We identify the most promising variables with the largest signal significance and smaller sensitivity to experimental uncertainties realated to the jet energy calibration. We compare to statistical tests computing the background-only hypothesis compatibility and a likelihood fit of the signal strength.

\end{abstract}


\section{Introduction}

The top quark has been studied intensively ever since its discovery at the Tevatron in 1995 by the CDF and D0 collaborations~\cite{Abe_1995,Abachi_1995} and its rediscovery at the LHC experiments in 2010~\cite{Aad_2011,Chatrchyan_2011}.
Due to its large mass, the top quark decays, before hadronizing, by the weak interaction into a~$W$ boson and a~$b$ quark. The classification of final states involving top quarks then depends on the way the $W$ boson decays.
Besides serving as a~test ground for precision tests of the Standard Model (SM) predictions computed at varied orders of the perturbation theory, its large mass suggests a~possible connection to the mechanism of the electroweak symmetry breaking. Top quark can play a~special r\^{o}le in Higgs physics and could couple more strongly to hypothetical new physics sectors, compared to other quarks.
Top quarks are produced singly or in pairs. At higher orders of the perturbation theory this strict distinction is however not valid.
Recently, evidence of four top quarks was reported by the ATLAS and CMS collaborations~\cite{ATLAS:2020hpj,CMS:2019rvj}.
Additionally, the search for and measurements of rare final states like the three top quarks production, and associated production of the $\ttbar$ pair with other SM particles, namely the vector $W/Z$ bosons and photon \cite{ATLAS:2021fzm,Knolle:2019hpk,ATLAS:2019fwo,CMS:2022tkv,ATLAS:2020yrp,CMS:2022lmh} or the Higgs boson~\cite{ATLAS:2018mme,CMS:2018uxb}, is continuing.

In this paper, we study the $\ttbar$ production with the focus on a~case where both $W$ bosons from top quarks decays lead to hadronic final states. \emph{i.e.} when each of the two $W$ bosons decay into a~pair of a~quark and an antiquark, the definition of the all-hadronic $\ttbar{}$ decay channel. The other two classes of the $\ttbar{}$ final state are 1) the case where one $W$ boson decays leptonically, leading to an isolated lepton, jets (two of which should be $b$-jets, \emph{i.e.} having properties as coming from a~$b$ quark fragmentation), and large missing transverse energy ($\Etmiss{}$) because of the undetected neutrino, denoted as $\ell$+jets channel; or 2) the case where both $W$ bosons decay leptonically, denoted as the dilepton channel, characterized by two isolated leptons, two $b$-jets, and large $\Etmiss{}$.

Searches for the $\ttbar$ resonances have been performed by both the ATLAS (\cite{ATLAS:2012dgv,Aaboud:2019roo,Aaboud:2018mjh,ATLAS:2020lks}) and CMS (\cite{Sirunyan:2018ryr} experiments, including searches for a~resonance decaying into a~$W$ boson and a~top quark~\cite{Sirunyan:2021fkj}; or for resonances decaying into a~$t$ and a~$b$ quark~\cite{CMS:2017zod}). Searches in both $\ell+$jets and all-hadronic channels have set limits for masses of possible beyond-the-standard model (BSM) $Z'$ particles of various scenarios, see \emph{e.g.}~\cite{Feldman:2006wb,Hayden:2013sra}, usually related to extensions of the Standard Model symmetry group.

In this paper we concentrate on using generic scalar and vector resonance models to study possible improvements in reconstructing the $\ttbar$ kinematics in various topologies of the final state, especially with regard to the degree of the Lorentz boost of the jets and with reference to variables sensitive to the presence of a~signal for the selected BSM models.

As the Large Hadron Collider (LHC) at CERN has provided a~large dataset and the current upgrade aim at increasing the center-of-mass energy as well as the luminosity, boosted topologies in hadronic final states are becoming tools heavily used in both measurements and searches at the LHC. They provide a~handle on processes occurring at large momentum transfer, with a~chance of observing possible BSM signals.

In this study we employ the usage of events where both boosted top-like and/or $W$-like hadronic jets are reconstructed at large or medium transverse momenta, respectively, in order to study the enhancement of events also in regions between the resolved and boosted topologies where the boosted jet mass and structure is consistent with a $W$-like or a top-like jet. This has been studied \emph{e.g.} by CMS~\cite{CMS:2014rsx,CMS:2012bti,CMS:2012jea,CMS:2020poo} and by ATLAS~\cite{ATLAS:2015lxh,ATLAS:2019kwg,ATLAS:2015irp,ATLAS:2018wis}.

We examine shapes of a~multiple of kinematic variables at the particle and detector levels using a~parameterized detector simulation for the presence of three broad BSM signal cases of a~scalar or vector particle decaying into a~pair of top quarks, and for the case of an associated production of top quark pairs with a~pair of invisible dark matter (DM) particles.

The signal significance is evaluated for various BSM processes of the \ttbar{} pair production at the central mass system energy of the $pp$ system of 14~TeV, using a state-of-the-art event generator and a parameterized detector simulation.
The presence of a~signal peak or other modification of spectra is  studied in boosted as well as semiboosted topologies, where, in the latter, one or two boosted jets consistent with the hadronic $W$ candidate are identified, and the top quark candidates are formed by adding a~four-momentum of the $b$-tagged jet.

In precision measurements, cross-sections of SM processes are often fitted by using simultaneous likelihood fits using binned variables with the power to separate signal from background, together with detector and modelling systematic uncertainties being allowed to vary and be fitted (profiling). In contrast, originally the searches for new physics used to set limits based on a~single one-dimensional (1D) sensitive variable like the invariant mass of the $\ttbar{}$ pair, searching for an excess of events over a~simulated, (semi) data-driven or fitted background, often validated in control regions. However, recent searches employ multivariate (MV) discriminants using multiple variables and thus broader information from the event, with MV approaches ranging from neural networks and boosted decision trees to more recent deep learning algorithms~\cite{Guest:2018yhq,10.21468/SciPostPhys.8.6.090}. 

We propose new pairs of sensitive variables with a~potential to separate a~BSM signal from the SM background, quantifying an excess of events over the expected background in a~two-dimensional (2D) phase space, as well as using ratios of variables. We thus suggest dimensional as well as dimensionless variables (which are ratios of other variables) and their pairs as 2D spectra for use in searches for BSM signals. We prove that their discriminating power is not a~straightforward function of a~variables correlation factor.

\section{Samples}
     
\begin{figure*}[!h]
\begin{center}
\begin{tabular}{ccc}
  \includegraphics[width=0.99\textwidth]{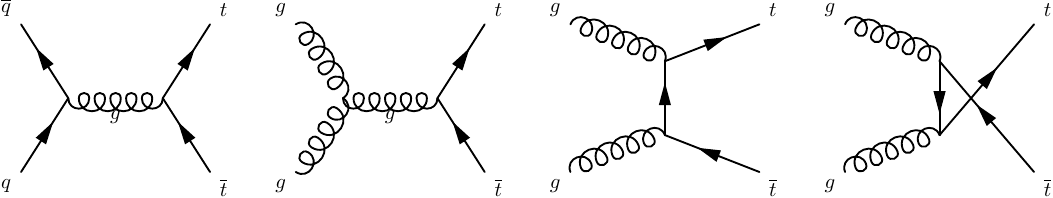} \\
    \includegraphics[width=0.99\textwidth]{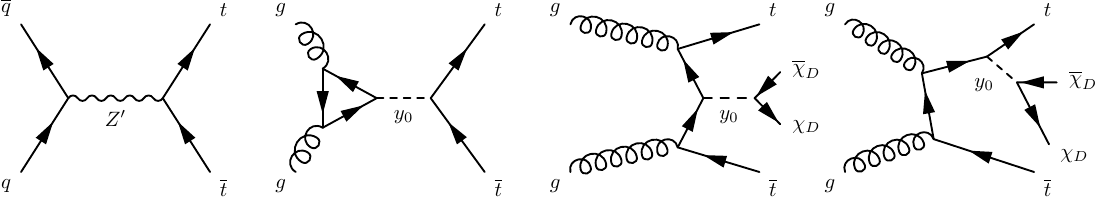} \\
\end{tabular}
\caption{Dominant leading-order Feynman diagrams for SM (top) and selected BSM (bottom) processes of the $\ttbar{}$ pair production. Created using the {\sl FeynMf} package~\protect\cite{Ohl:1995kr}.}
\label{fig_feynman}
\end{center}
\end{figure*}

Using the \MadGraph{} version {\tt 2.6.4} simulation toolkit~\cite{Alwall:2014hca}, proton-proton collision events at $\sqrt{s} = 14$ TeV were generated for the SM process $pp \rightarrow \ttbar{}$ and for the resonant  $s$-channel \ttbar{} production via an additional narrow-width (sub-GeV) vector boson $Z'$ as $pp \rightarrow Z' \rightarrow \ttbar{}$ (using the model \cite{FeynModelZprime,Christensen:2008py,Wells:2008xg}) with the {\tt NNPDF23\_nlo\_as\_0119} PDF set~\cite{Ball_2013} via the LHAP interface~\cite{Buckley_2015} v6.1.6. These samples were generated in the all-hadronic $\ttbar$ decay channel at next-to-leading order (NLO) in QCD in production, using the MLM matching technique~\cite{Hoche:2006ph}, \emph{i.e.} with additional processes with extra light-flavoured jets produced within the matrix element, matched and resolved for the phase-space overlap of jets generated by the parton shower using \MadGraph{} defaults settings. The parton shower and hadronization were simulated using \Pythia{}8~\cite{Sjostrand:2014zea}. Masses of the hypothetical $Z'$ particle, serving effectively as a~source of semi-boosted and boosted top quarks, were selected as \GeV{750, 800, 900, 1000, 1250 and 1500}. In addition, a~model with a~scalar particle decaying to a~pair of top quarks $y_0 \rightarrow \ttbar{}$ was also adopted~\cite{Christensen:2008py} at the leading-order (LO) in the $\ttbar$ production with the gluon-gluon fusion loop (more details in~\cite{Mattelaer:2015haa,Backovic:2015soa,Neubert:2015fka,Das:2016pbk,Kraml:2017atm,Albert:2017onk,Arina:2017sng,Afik:2018rxl}), with inclusive $\ttbar{}$ decays, selecting the all-hadronic channel later in the analysis. As the last BSM process considered, the possibility of generating a~pair of invisible dark matter (DM) fermions $\chi_D$ in association with a~$\ttbar{}$ pair using the scalar $y_0$ particle as a~mediator was also studied. We set the probed masses to $m_{y_0} = 1000\,$GeV and for its natural width we chose $\Gamma_{y_0} =\,10$, $100$ and $300\,$GeV. For the associated production of the DM particles masses we set $m_{y_0} = 1000\,$GeV; and $m_{\chi_D} = 10$, $100$ and $300\,$GeV, all with $\Gamma_{y_0} = 10\,$GeV. The mass of the top quark was set to \GeV{173} (\MadGraph{} default). A~selection of representative \MadGraph{} processes are depicted as Feynman diagrams in~Figure~\ref{fig_feynman} while the cross-sections and numbers of generated events for the considered processes are listed in~Table~\ref{tab:xsects}. As typical background processes, the production of hadronically decaying $W^\pm$ bosons in association with a~$b\bar{b}$ pair, accompanied by up to two additional light-flavoured jets produced at the matrix element level, was also simulated using the MLM matching with the \Pythia{}8 parton shower using \MadGraph{} default matching settings (referred to as $Wbbjj$), and also the production of two $W$ bosons decaying hadronically, accompanied by a~$b\bar{b}$ pair at LO, referred to as $WWbb$.
In order to include a~purely QCD process, the production of a~$b\bar{b}$ pair in association with up to two light-jets (denoted as $bbjj$, or $b\bar{b}$+jets) was simulated as a~MLM-matched sample for the same three jet \pt{} regions as the \ttbar{} sample. This background was designed to follow the selection, which requires at least two $b$-jets to be present in the event. Due to the high cross section, but yet very low efficiency to pass selection except for the 0B2S topology (see Section~\ref{sec:Selection}), the $bbjj$ sample with generator cuts of  $60\,\mathrm{GeV} < p_\mathrm{T}^{j1,j2} < 200\,\mathrm{GeV}$ was not used in the 2B0S and 1B1S topologies due to statistical fluctuations leading to distributions not smooth enough for a~stable background subtraction.

\begin{table*}[!]
  \centering
  {\footnotesize
  \begin{tabular}{lllrr}
    \hline
    Sample              &  Order & Parameters, comment & Events & Cross-section \\ \hline

    $pp \rightarrow \ttbar(j)\rightarrow$ hadrons  & NLO  &  $m_t = 173\,$GeV, matched & 5,283,343 & 162 pb \\ 
    $60\,\mathrm{GeV} < p_\mathrm{T}^{j1,j2}$         &      &  control sample & & \\ \hline
    $pp \rightarrow \ttbar(j)\rightarrow$ hadrons  & NLO  &  $m_t = 173\,$GeV, matched & 1,841,859 &  4.61 pb \\ 
    $200\,\mathrm{GeV} < p_\mathrm{T}^{j1,j2}$  &  &  &  & \\ 
    
    $pp \rightarrow \ttbar(j)\rightarrow$ hadrons  & NLO  &  $m_t = 173\,$GeV, matched & 1,896,629 &  138 pb \\ 
    $60\,\mathrm{GeV} < p_\mathrm{T}^{j1,j2} < 200\,\mathrm{GeV}$  &  &  &  & \\ 

    $pp \rightarrow \ttbar(j)\rightarrow$ hadrons  & NLO  &  $m_t = 173\,$GeV, matched & 1,544,855 & 19.5 pb \\ 
    $200\,\mathrm{GeV} < p_\mathrm{T}^{j1}$  &  &  &  & \\ 
    $60\,\mathrm{GeV} < p_\mathrm{T}^{j2} < 200\,\mathrm{GeV}$  &   &   &  & \\ \hline

    $pp \rightarrow Z' \rightarrow \ttbar(j)\rightarrow$ hadrons  & NLO   &  $m_{Z'}  =  750\,$GeV, matched & 293,402  & 0.244 fb \\ 
    $pp \rightarrow Z' \rightarrow \ttbar(j)\rightarrow$ hadrons  & NLO   &  $m_{Z'}  =  800\,$GeV, matched & 289,892 & 0.210 fb  \\ 
    $pp \rightarrow Z' \rightarrow \ttbar(j)\rightarrow$ hadrons  & NLO   &  $m_{Z'}  =  900\,$GeV, matched & 284,500 & 0.158 fb  \\ 
    $pp \rightarrow Z' \rightarrow \ttbar(j)\rightarrow$ hadrons  & NLO   &  $m_{Z'}  = 1000\,$GeV, matched & 279,442 & 0.121 fb  \\ 
    $pp \rightarrow Z' \rightarrow \ttbar(j)\rightarrow$ hadrons  & NLO   &  $m_{Z'}  = 1250\,$GeV, matched &  215,290 & 0.065 fb \\ 
    $pp \rightarrow Z' \rightarrow \ttbar(j)\rightarrow$ hadrons  & NLO   &  $m_{Z'}  = 1500\,$GeV, matched &  261,620 & 0.037 fb \\ 
    \hline
    
    $pp \rightarrow y_0 \rightarrow \ttbar$ incl. & loop SM  &  $m_{y_0}  = 1000\,$GeV, $\Gamma_{y_0} = 10\,$GeV   & 1,000,000  & 305 fb  \\    
    $pp \rightarrow y_0 \rightarrow \ttbar$ incl. & loop SM  &  $m_{y_0}  = 1000\,$GeV, $\Gamma_{y_0} = 100\,$GeV  &   500,000  & 31.1 fb \\    
    $pp \rightarrow y_0 \rightarrow \ttbar$ incl. & loop SM  &  $m_{y_0}  = 1000\,$GeV, $\Gamma_{y_0} = 300\,$GeV  &   500,000  & 10.0 fb \\    \hline
    $pp \rightarrow y_0 \,\ttbar \rightarrow \chi_D \bar{\chi}_D \ttbar\rightarrow$ hadr. & LO    & $m_{y_0} = 1\,$TeV, $\Gamma_{y_0} = 10\,$GeV  & 500,000 & 2.44 fb \\
    & & $m_{\chi_D} = 10\,$GeV & & \\
    $pp \rightarrow y_0 \,\ttbar \rightarrow \chi_D \bar{\chi}_D \ttbar\rightarrow$ hadr. & LO    & $m_{y_0} = 1\,$TeV, $\Gamma_{y_0} = 10\,$GeV & 1,000,000& 2.29 fb  \\
    & & $m_{\chi_D} = 100\,$GeV & & \\
    $pp \rightarrow y_0 \,\ttbar \rightarrow \chi_D \bar{\chi}_D \ttbar\rightarrow$ hadr. & LO    & $m_{y_0} = 1\,$TeV, $\Gamma_{y_0} = 10\,$GeV & 500,000 & 1.24 fb \\ 
    & & $m_{\chi_D} = 300\,$GeV & & \\ \hline
    $Wbbjj \rightarrow $ hadr., $p_\mathrm{T}^{j1,j2} > 60\,$GeV  & LO & matched 0--2 additional jets & 617,952 & 82.6 pb \\ \hline
    $WWbb \rightarrow $ hadr., $p_\mathrm{T}^{j1,j2} > 60\,$GeV  & LO &  & 2,000,000 & 126 pb \\ \hline

    $pp \rightarrow b\bar{b}+$jets  & LO  & matched 0--2 additional jets & 11,499,884 &   0.604 nb \\ 
    $200\,\mathrm{GeV} < p_\mathrm{T}^{j1,j2}$  &  &  &  & \\ 
    
    $pp \rightarrow b\bar{b}+$jets  & LO  & matched 0--2 additional jets & 2,275,190 &   21.8 nb \\ 
    $60\,\mathrm{GeV} < p_\mathrm{T}^{j1,j2} < 200\,\mathrm{GeV}$  &  &  &  & \\ 

    $pp \rightarrow b\bar{b}+$jets  & LO  & matched 0--2 additional jets &  2,482,646 &  2.66 nb \\ 
    $200\,\mathrm{GeV} < p_\mathrm{T}^{j1}$  &  &  &  & \\ 
    $60\,\mathrm{GeV} < p_\mathrm{T}^{j2} < 200\,\mathrm{GeV}$  &   &   &  & \\ 
    \hline
  \end{tabular}
  }
\caption{Cross sections and numbers of events for the samples used, generated with \MadGraph{}+\Pythia{}8, with cuts on the transverse momentum of the leading and sub-leading jets indicated as $j1$ and $j2$, respectively. For NLO processes, the merged cross-sections are quoted (and used in the analysis) as evaluated at the \MadGraph{} middle scale of 67.5~GeV.}
\label{tab:xsects}
\end{table*}

\section{Objects reconstruction}
\label{sec:Reconstruction}

By using the \Delphes{} (version {\tt 3.4.1}) detector simulation \cite{deFavereau:2013fsa} with a~modified ATLAS card, jets with two distance parameters of $R=0.4$ and 1.0 were reconstructed via the anti-$k_t$ algorithm~\cite{Cacciari:2008gp} to form corresponding small-$R$  and large-$R$ jets, using the FastJet algorithm~\cite{Cacciari:2011ma} at both particle and detector levels.
The \Delphes{} card was also modified to remove the DM particles from the both detector jet clustering and the particle-level jets.

Although no additional $pp$ interactions (pile-up) were added to the simulated hard-scatter processes, the trimming jet algorithm \cite{Krohn:2009th} as part of the \Delphes{} package was used to obtain jets at both the particle and detector levels with removed soft components, using the parameter of $R_\mathrm{trim} = 0.2$ and modified \pt{} fraction parameter $f^{\pt}_\mathrm{trim} = 0.03$ (originally 0.05). The trimming algorithm was chosen over the standard non-groomed jets, soft-dropped~\cite{Larkoski:2014wba} and pruned jets~\cite{Ellis:2009me}, with parameters varied according to the narrowness of the mass peaks, see Appendix~\ref{app:groom_cmp} for details.
The jet masses shapes of the other algorithms differed in the expected peak positions at $W$ and $t$ masses, although they performed similarly and could be also tuned for the purpose of this analysis. In detail, in the default settings the $W$ and top mass peaks were shifted to higher values by $10$--$20\,$GeV, with a~worse shape and shift for the $W$ case which is important for the semiboosted regimes. The new settings help establish both the peaks shape as well as position at the expected values. While the new $\ttbar{}$ invariant mass distribution for the $Z'$ sample with $m_{Z'} = 1\,$TeV shifts down by about $50\,$GeV, the relative peak resolution remains the same.

This procedure has lead to a~more pronounced reconstructed $W$ as well as top-quark mass peaks in terms of corresponding candidate large-$R$ jet masses, and also to a~better correlation between the detector and particle levels for derived quantities like the transverse momentum of the top quarks or the \ttbar{} invariant mass.

Custom jet energy scale (JES) correction was applied to large-$R$ jets at the detector level, resulting in an up-scaling correction of 5--15\% depending on $\eta$ and \pt{}\footnote{Pseudorapidity $\eta$ is defined using the angle $\theta$ measured from the $z$ axis (coinciding with one of the proton beams) as $\eta \equiv -\ln\tan\frac{\theta}{2}$; the transverse momentum $\pt \equiv \sqrt{p_x^2 + p_y^2}$.}. Additional custom JES was also applied to the default \Delphes{} JES for small-$R$ jets, resulting in an additional correction of 5\% in central rapidities and of 4\% at low \pt{}. JES closure tests in terms of the ratio of the particle and angularly matched detector-level jet result in agreement within~2\%. See \ref{app:JES} for details.

\begin{figure*}
\begin{center}
\begin{tabular}{ccc}
\includegraphics[width=0.32\textwidth]{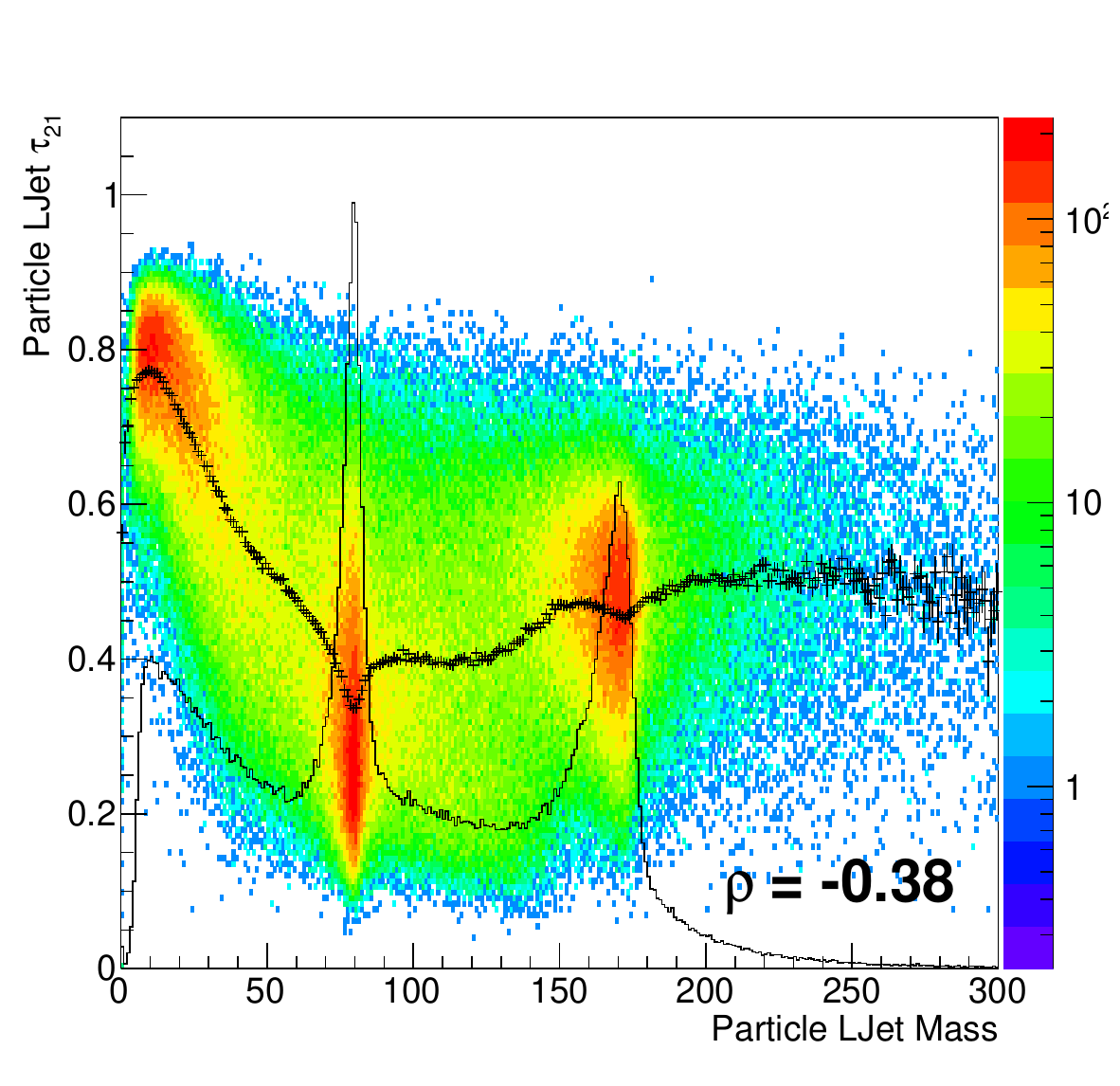} &
\includegraphics[width=0.32\textwidth]{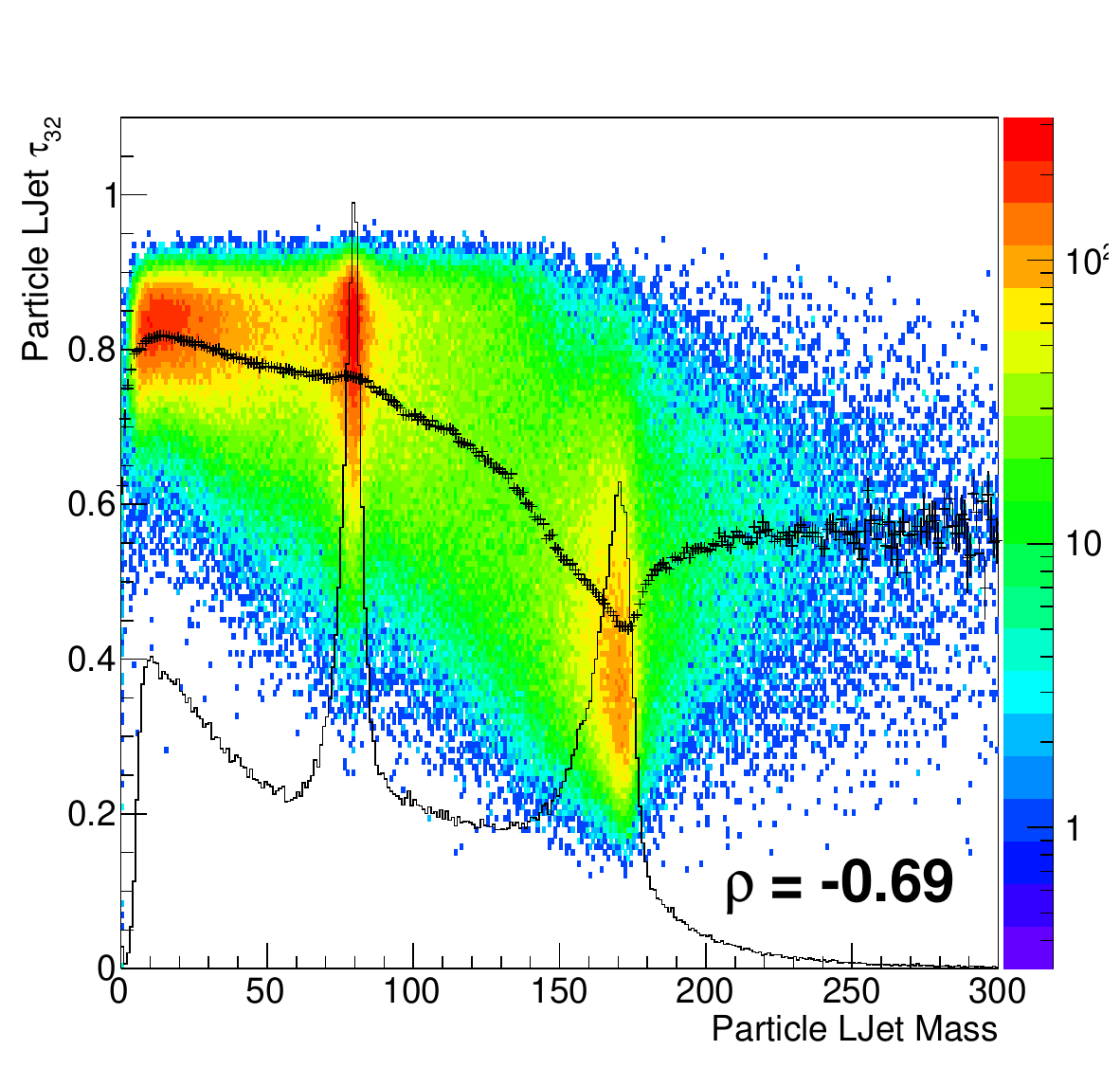} &
\includegraphics[width=0.32\textwidth]{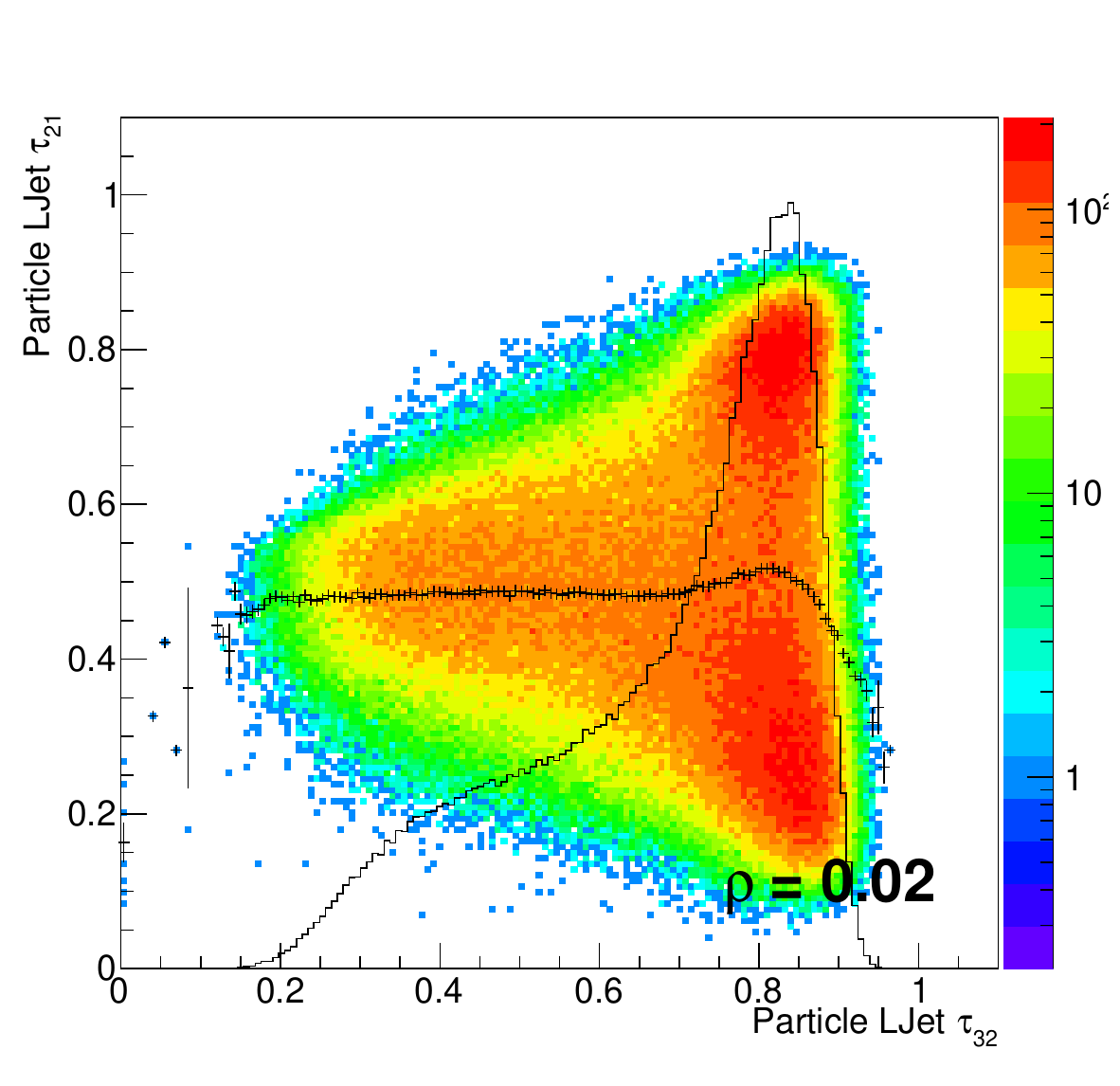} \\
\includegraphics[width=0.32\textwidth]{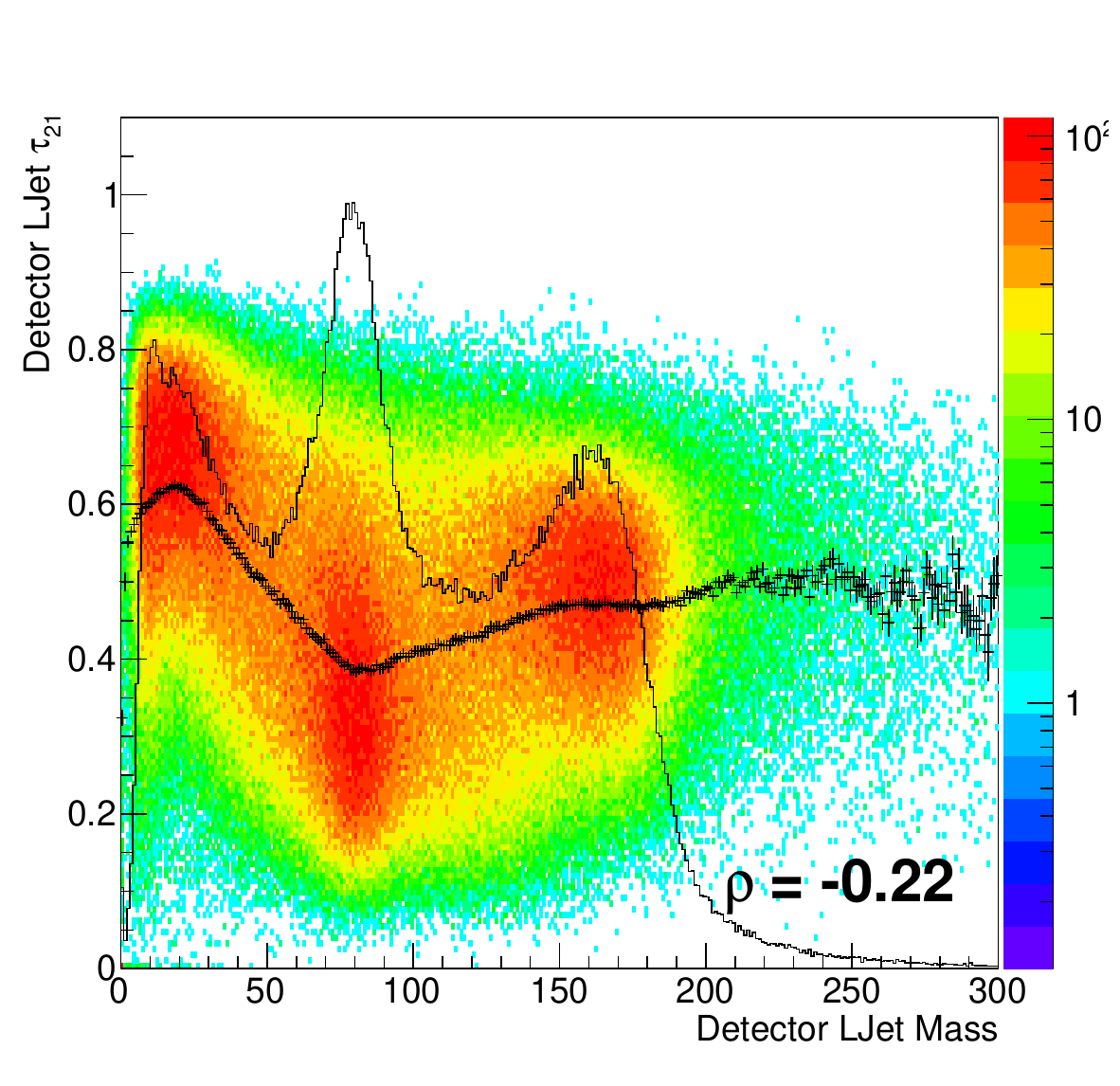} &
\includegraphics[width=0.32\textwidth]{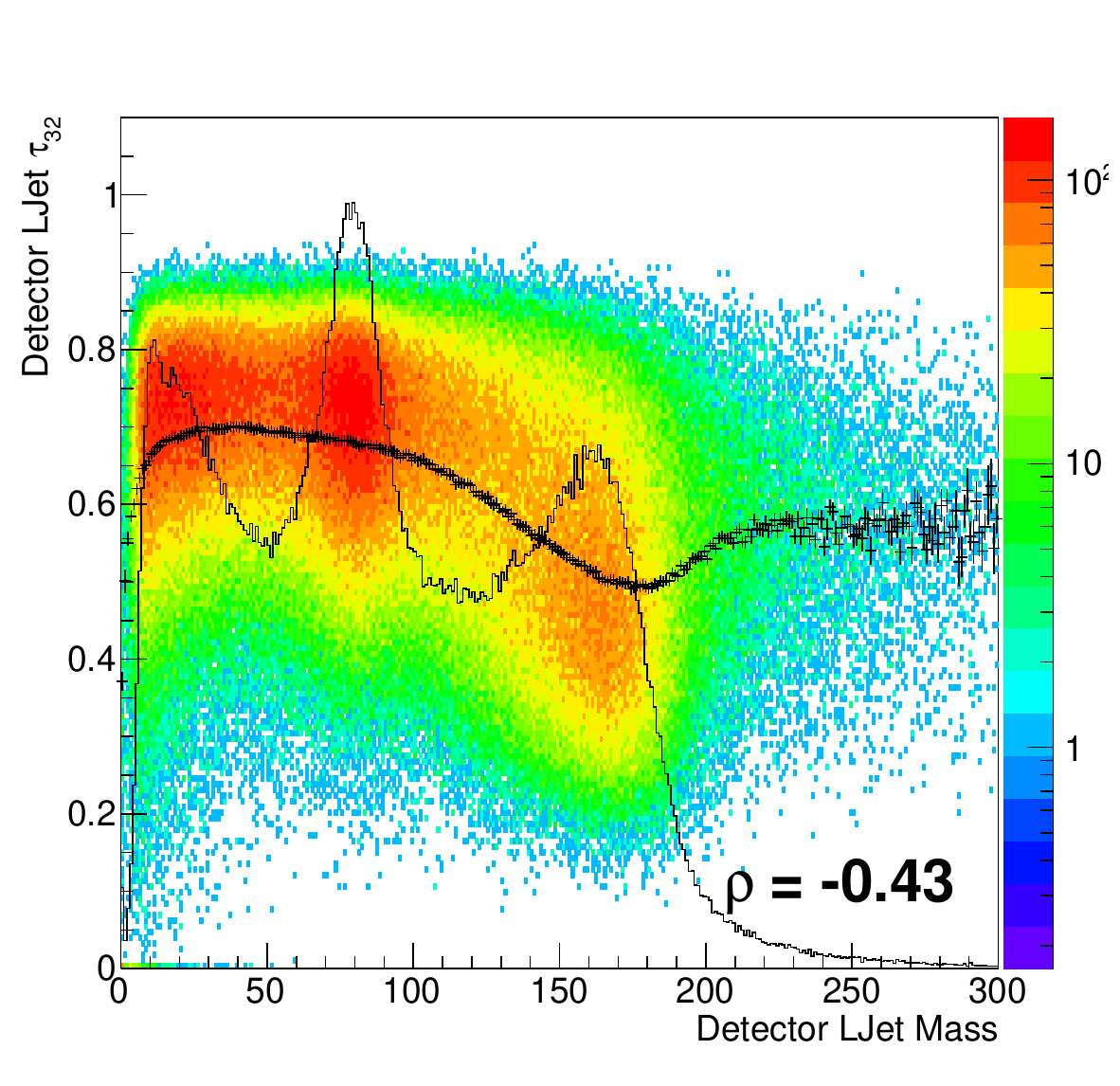} &
\includegraphics[width=0.32\textwidth]{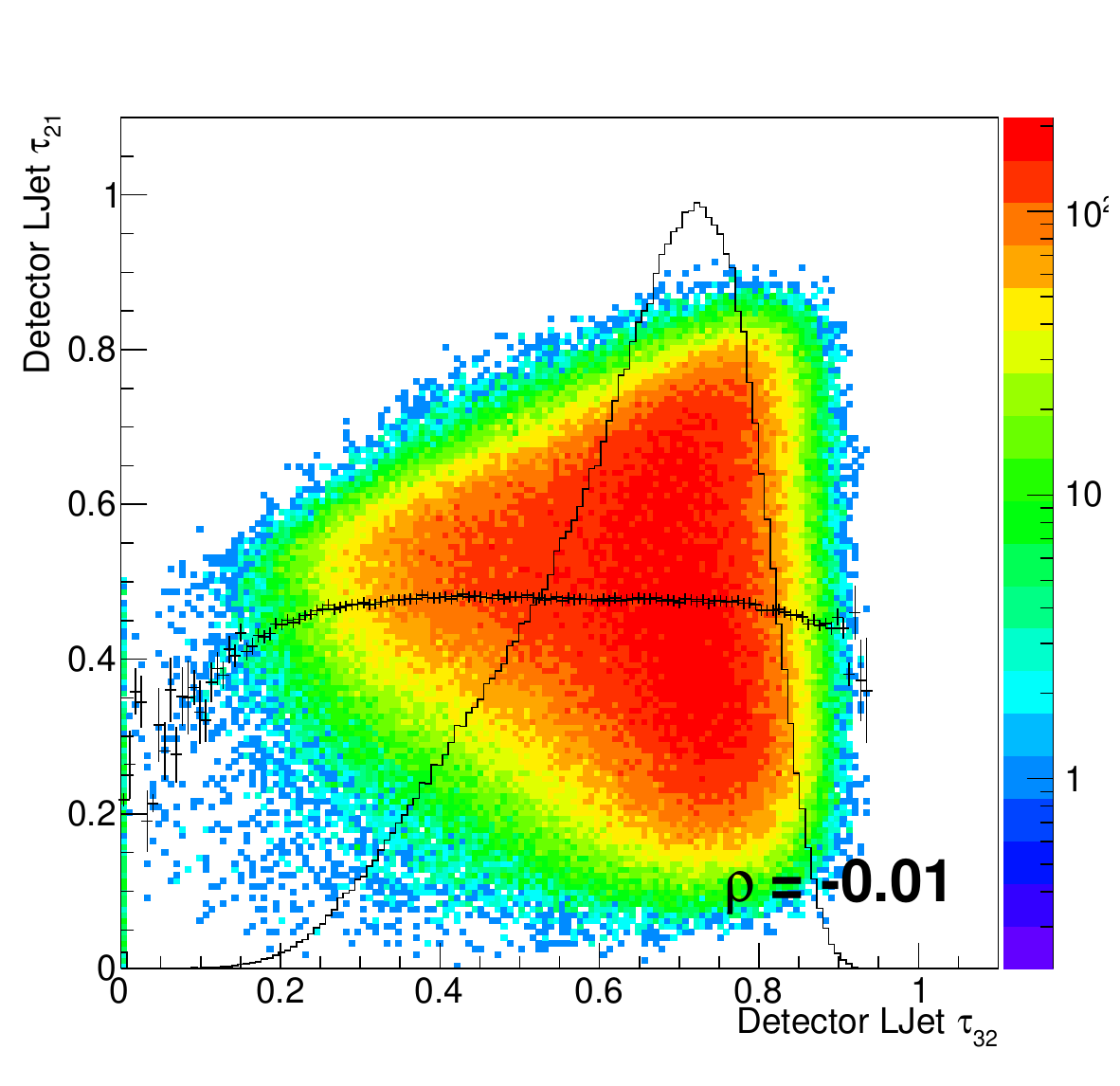} \\
\end{tabular}
\caption{The $\tau_{21}$ (left) and $\tau_{32}$ (middle) jet structure variables plotted versus the large-$R$ jet mass and the $\tau_{21}$ vs. $\tau_{32}$ for the $Z'$ sample with $m_{Z'} = 1\,$TeV at the particle (top) and detector (bottom) levels. The cross markers are the profile histogram while the solid line stands for the (scaled) projection of the histogram to the $x$ axis. The standard correlation factor~$\rho$ is also indicated.}
\label{fig_taus_vs_m}
\end{center}
\end{figure*}

The jet subjettiness variables \cite{Thaler:2010tr} $\tau_{k}$, $k=1,2,3$, and their ratios $\tau_{ij} \equiv \tau_{i}/\tau_{j}$ were employed as provided by the \Delphes{} jet properties together with a~simple selection on the large-$R$ jet mass to identify jets coming from the hadronic decays of the $W$ boson or a~top quark. The large-$R$ jets were tagged as
\begin{itemize}
\item $W$-jets if $ 0.10 < \tau_{21} < 0.60 \, \land \,   0.50 < \tau_{32} < 0.85 \, \land \,  m_J \in [70, 110] \,\mathrm{GeV}$;
\item top-jets if $ 0.30 < \tau_{21} < 0.70 \, \land \,   0.30 < \tau_{32} < 0.80 \, \land \,  m_J \in [140, 215] \,\mathrm{GeV}$.
\end{itemize}
As illustrated in~Figure~\ref{fig_taus_vs_m} for the $Z'$ sample with $m_{Z'} = 1\,$TeV, correlations can be found between the substructure variables and the large-$R$ jet mass, demonstrating not only the small correlation between $\tau_{21}$ and $\tau_{32}$ but also the worse jet mass resolution at the detector as compared to the particle level. 

The above definitions result in a~$W$-tagging efficiency of about $0.60$ in the jet \pt{} range of $[150,350]\,\mathrm{GeV}$ with mistag (fake) rate if about $0.15$; and to the top-tagging efficiency of about $0.55$ in the jet \pt{} range of $[300,700]\,\mathrm{GeV}$ with a~mistag rate below $0.10$; with real (fake) efficiencies evaluated using the \ttbar{} ($b\bar{b}+$jets) sample (see~\ref{app:Tag}). 
For the $Z'$ samples the $t$ and $W$ mistag rates are higher for $\pt < 600\,$GeV than in the \ttbar{} sample and falls of again towards the high $\pt$ due to the fact that a~simple angular matching of parton $t$ and $W$ particles defining the true jet labels is not suitable to obtain non-$t$/$W$ jets for highly boosted samples containing real $t$ and $W$. 

The $b$-jets at the detector level were used as provided by the tagging efficiency in the default \Delphes{} ATLAS card, while for studies at the particle level, custom tagging based on the presence of a~$B$~hadron angularly close to the particle jet was implemented, using $B$~hadrons with $\pt{} > 5\,$GeV. Particle-level jets were tagged as $b$-jets if such a~$B$~hadron was within $\Delta R = \sqrt{(\Delta\eta)^2 + (\Delta\phi)^2} < 0.4$. Furthermore, in order to account for the practically important case of falsely $b$-tagged jets, a~custom random mistagging with a~rate of 1\% was added for originally non-$b$-tagged jets.

\section{Event selection}
\label{sec:Selection}

When taking into account only the hadronic decays of the $\ttbar$ events, the $W^+ b W^- \bar{b}$ intermediate state can be experimentally observed, depending on the Lorentz boost of the parent top quarks, as two large-$R$ jets corresponding to the two highly-boosted top quarks; or as a~combination of only one such top-like jet accompanied by an isolated $b$-jet and a~large-$R$ jet corresponding to decay products of one of the hadronically decaying $W$ boson ($W$-like jet); or as a~final state with two $b$-jets and two $W$-like jets.
We did not pursue the resolved topology or events with only a~single $W$-jet as dedicated techniques are needed to study the combinatorics of up to $\geq 6$ small-$R$ jets and also because we do expect a~substantial boost of the top quarks when looking for signs of new physics with particles of large masses.

\begin{figure*}[!h]
\begin{center}
\begin{tabular}{ccc}
  \includegraphics[width=0.33\textwidth]{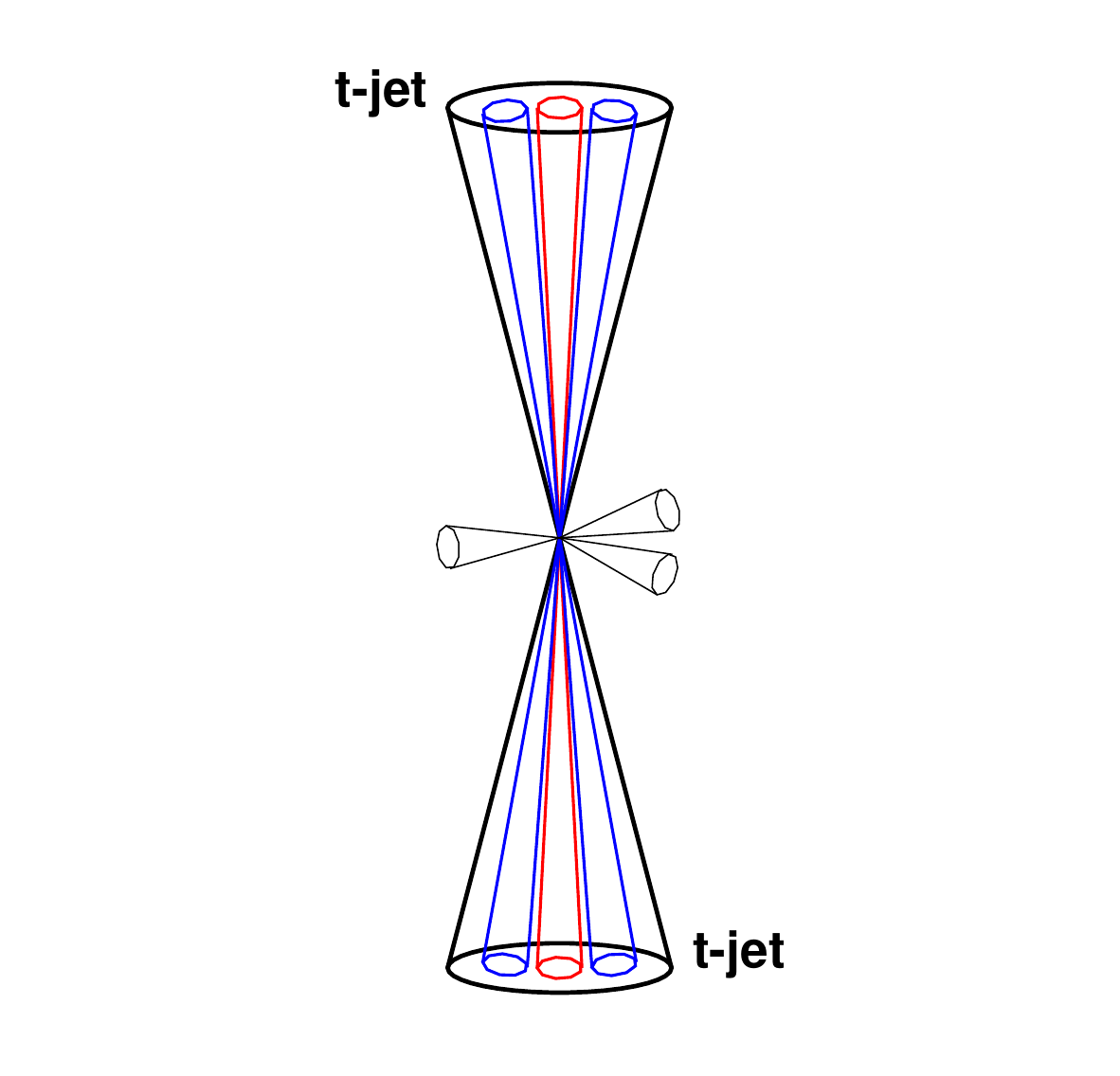} &
  \includegraphics[width=0.33\textwidth]{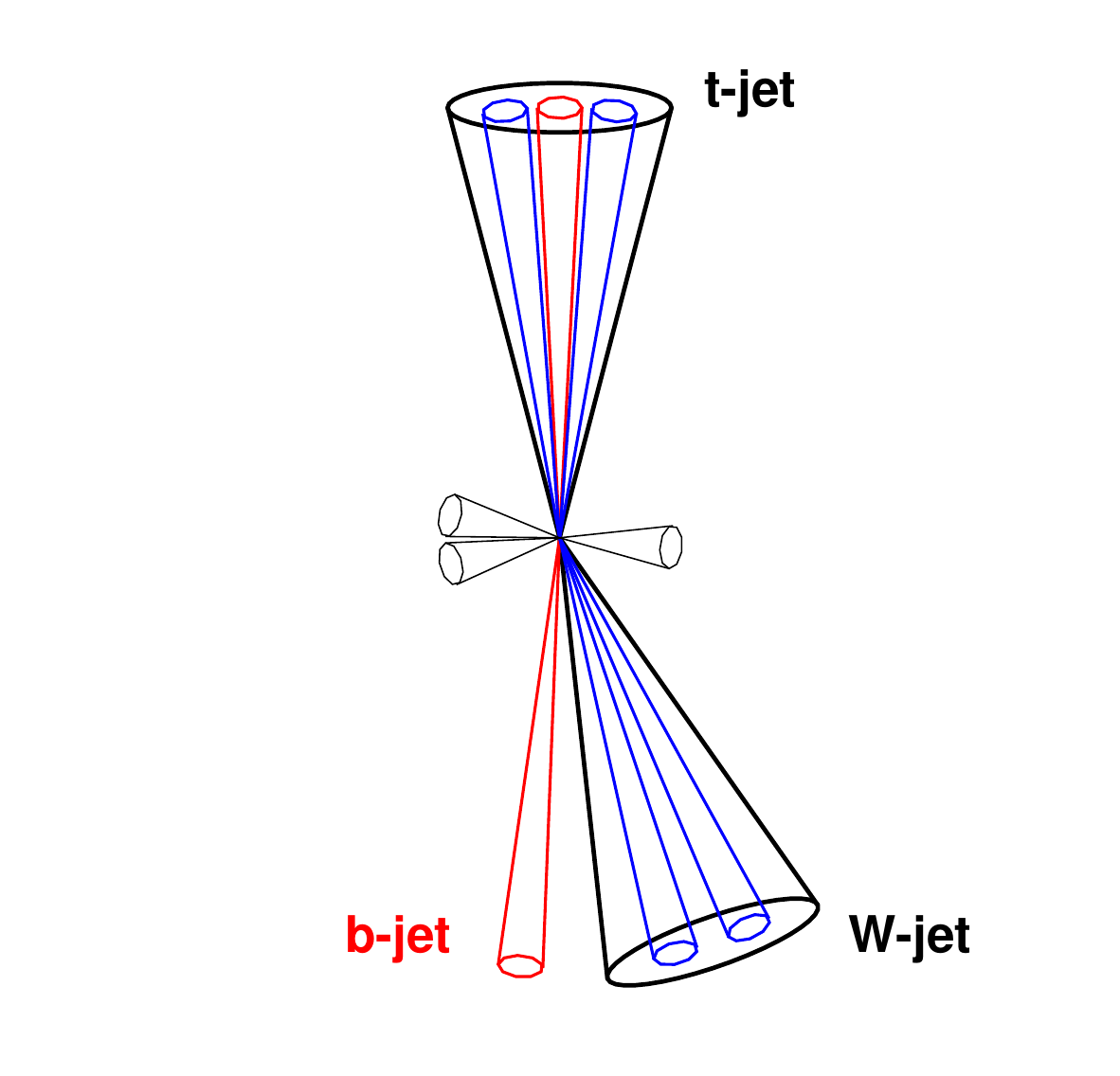} &
  \includegraphics[width=0.33\textwidth]{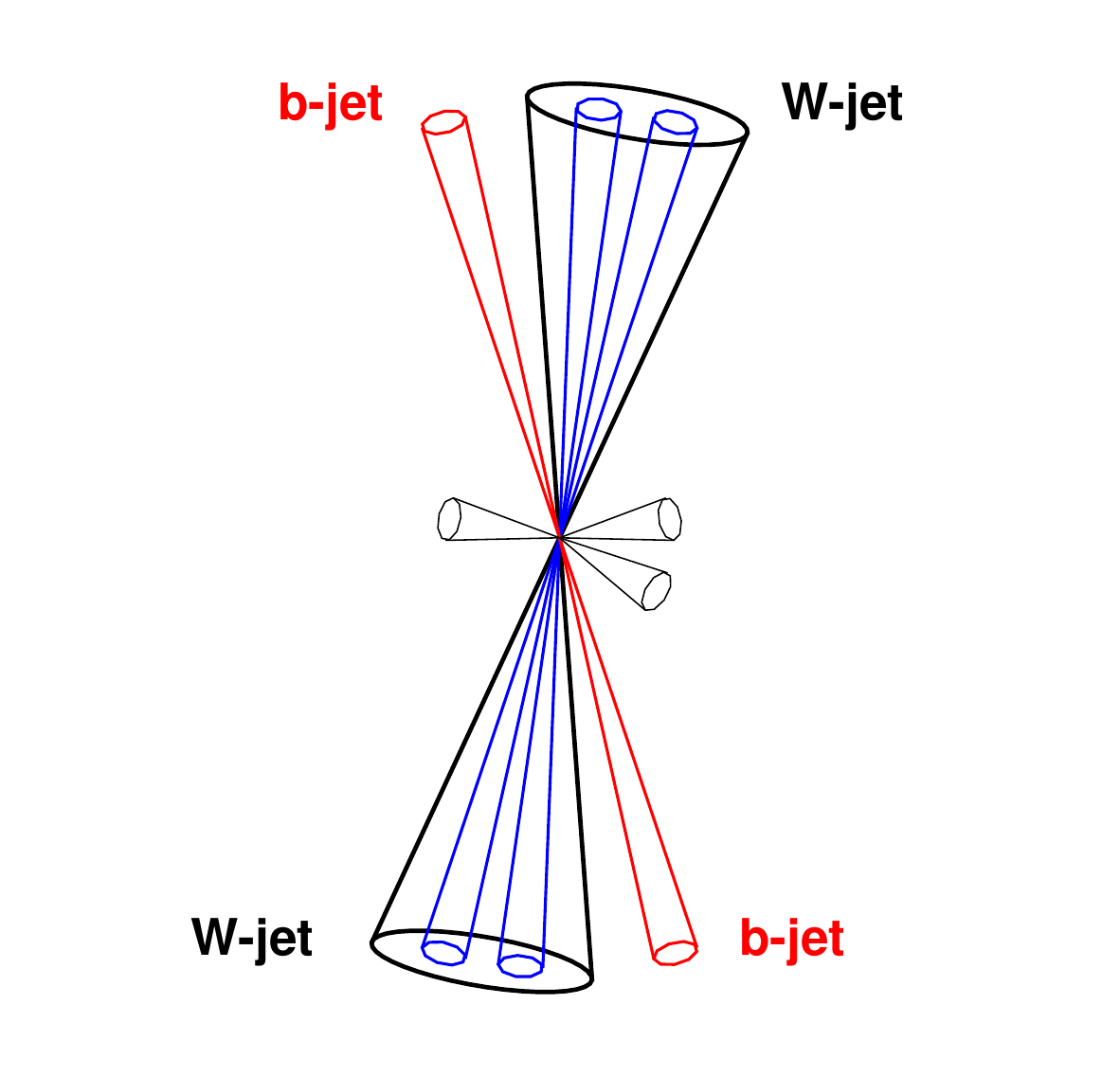} \\
\end{tabular}
\caption{Cartoons of the studied hadronic $\ttbar$ event topologies, from left to right: 2B0S (boosted-boosted), 1B1S (boosted-semiboosted) and 0B2S (semiboosted-semiboosted). The large-$R$ jets are shown as bold large black large cones, with sub-jets in blue while the $b$ sub-jets are in red. Additional jets are shown as black cones of small radii.}
\label{fig_cartoons}
\end{center}
\end{figure*}

\begin{figure*}[p]
\begin{center}
\begin{tabular}{cc}
  \includegraphics[width=0.47\textwidth]{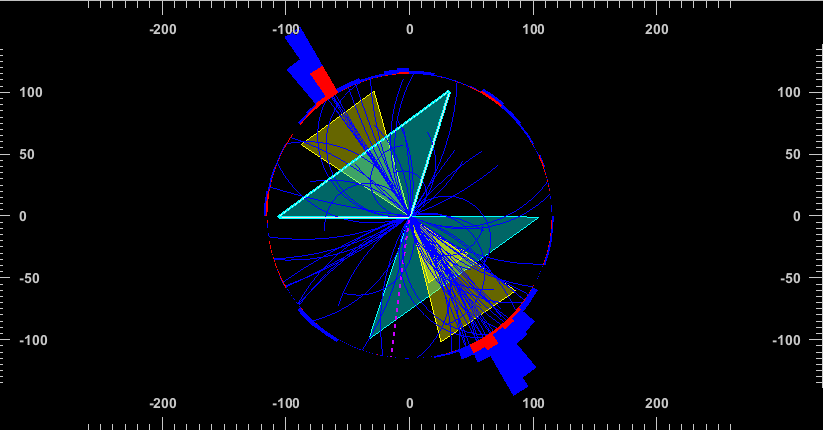} &
  \includegraphics[width=0.47\textwidth]{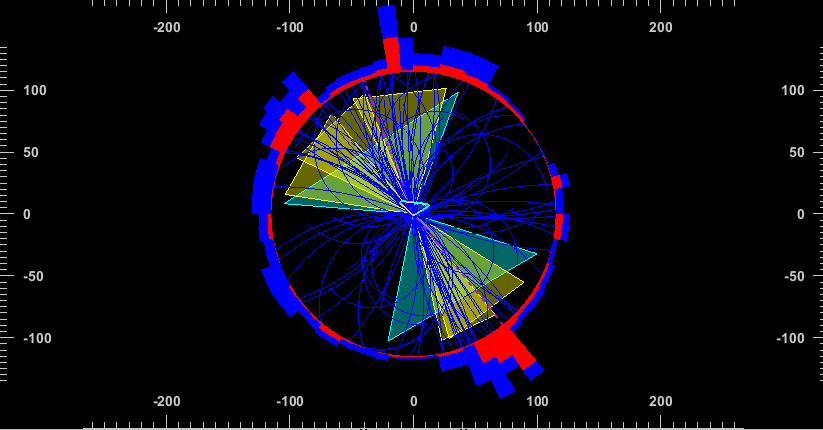} \\
  \includegraphics[width=0.47\textwidth]{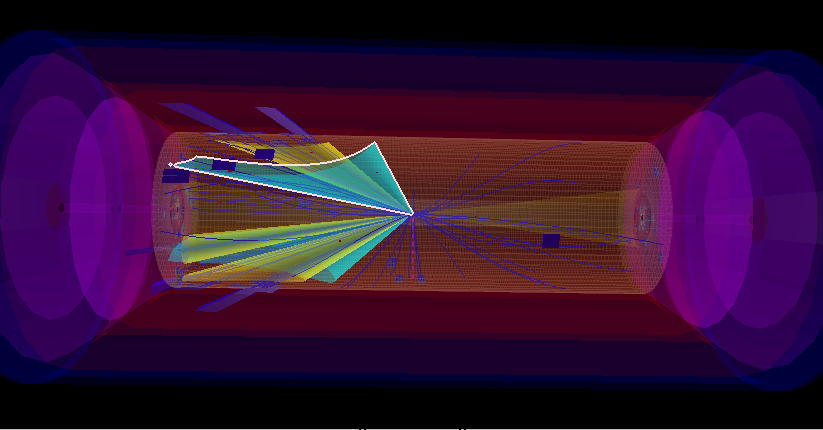} &
  \includegraphics[width=0.47\textwidth]{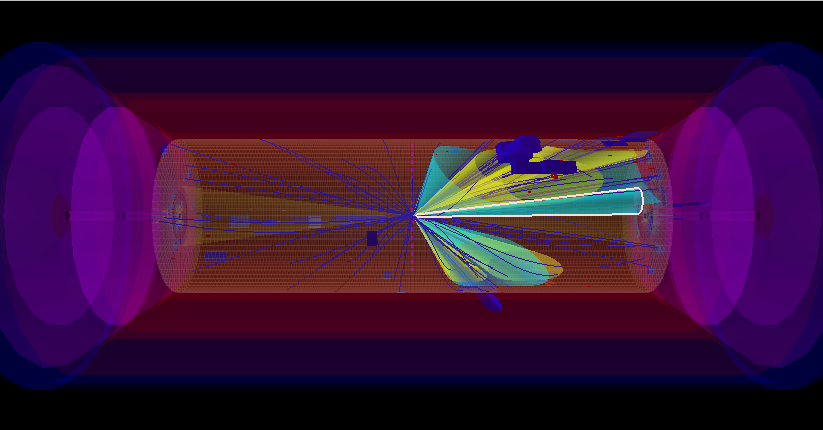}
\end{tabular}
\caption{The \Delphes{} event display of a~boosted hadronic \ttbar{} event (left) and a~more complex final state, namely additional jet activity (right), from the $Z'$ sample with $m_{Z'} = 1\,$TeV. Small-$R$ (large-$R$) jets are depicted as yellow (light blue) cones. Tracks of charged particles as well as electromagnetic (red) and hadronic (blue) calorimeter energy deposits are also shown.}
\label{fig_evt_displ_zp1000_3}
\end{center}
\end{figure*}

\begin{figure*}[p]
\begin{center}
\begin{tabular}{cc}
  \includegraphics[width=0.490\textwidth]{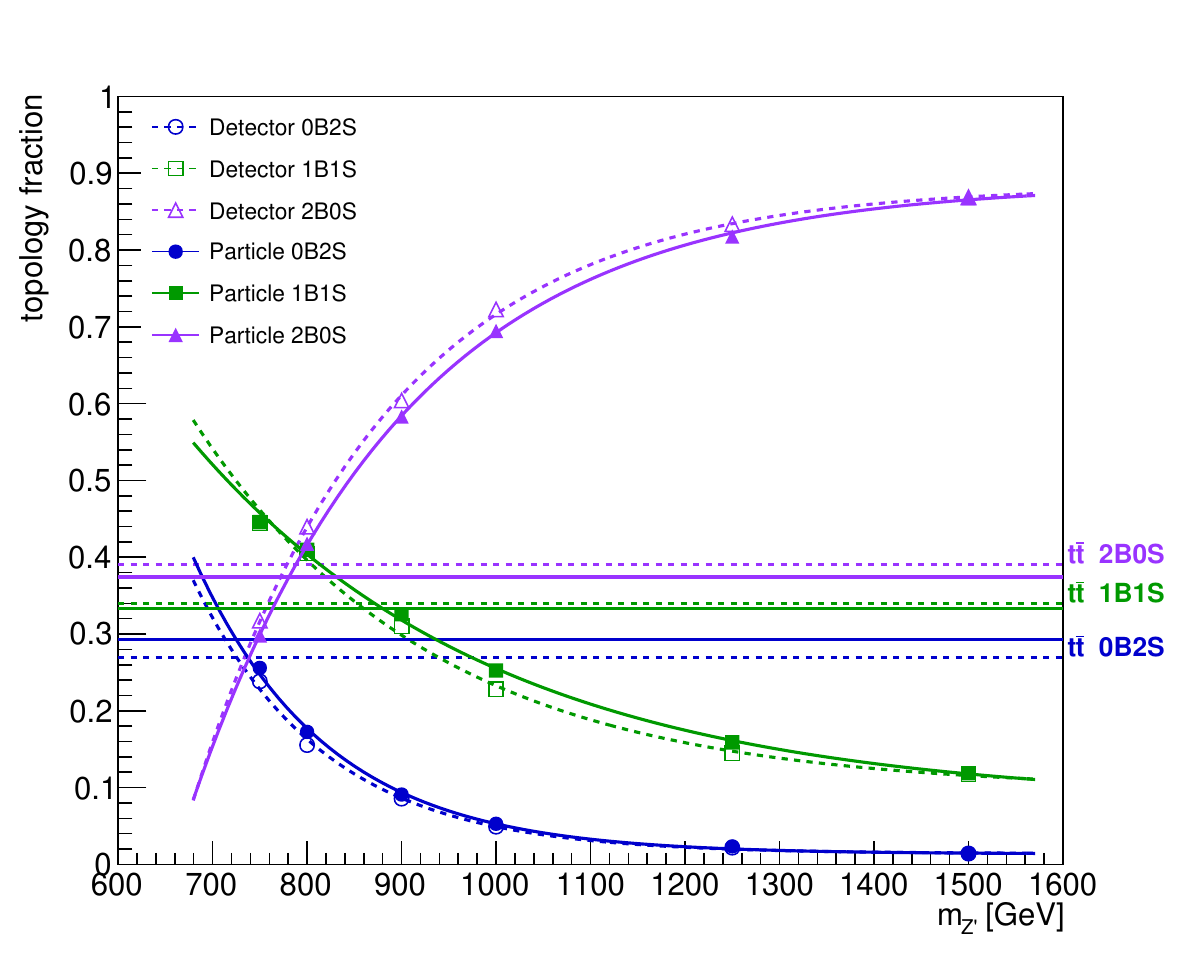} &
  \includegraphics[width=0.420\textwidth]{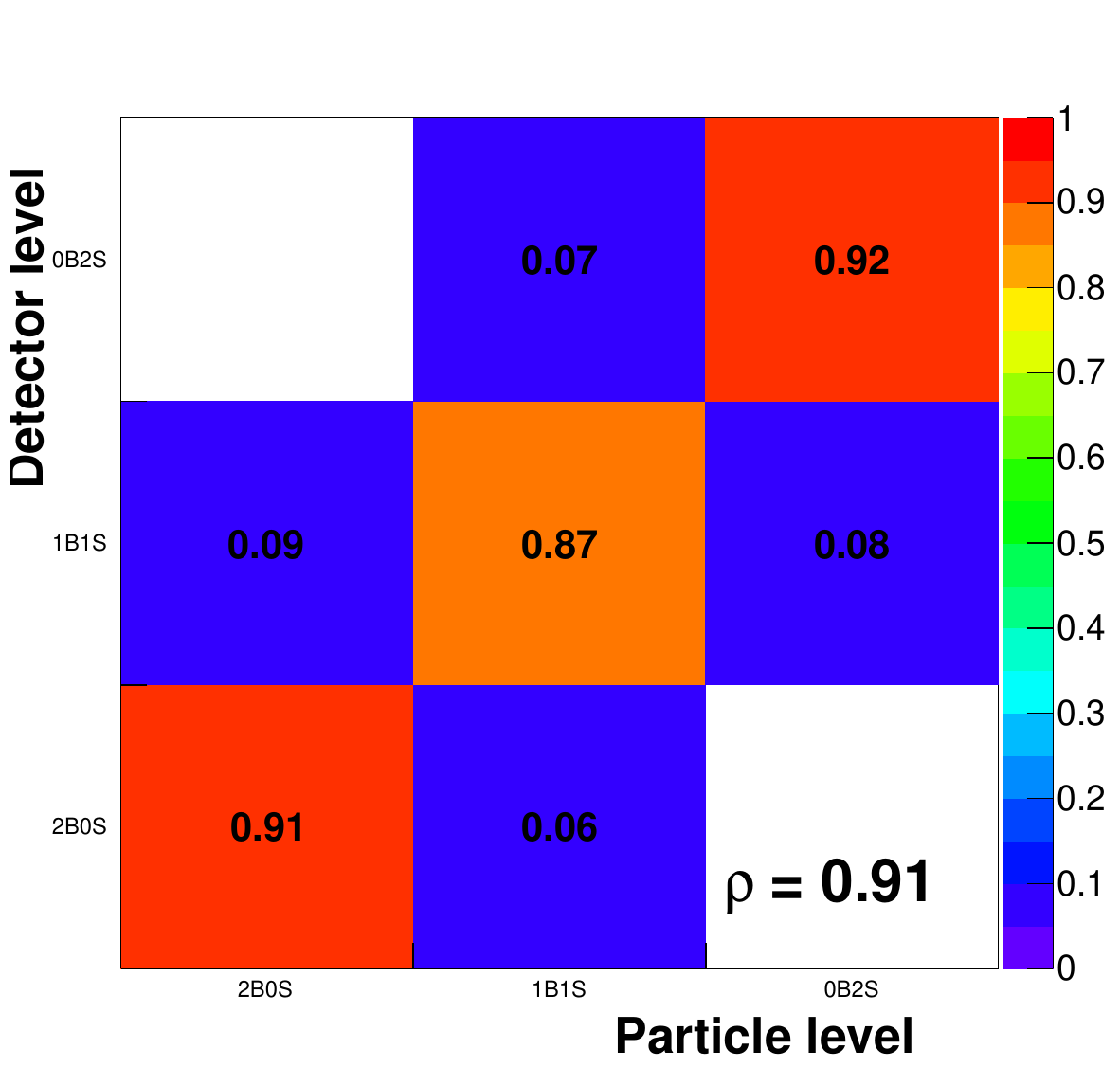} \\
\end{tabular}
\caption{Left: Fractions of the hadronic $\ttbar{}$ 2B0S, 1B1S and 0B2S topologies in various samples as function of the parent $Z'$ mass. The guiding lines are exponential fits and the horizontal lines mark the fractions for the \ttbar{} control sample. Both particle-level (solid lines, full markers) and \Delphes{} detector-level (dashed lines, open markers) fractions are plotted. Right: Migration between the selection topologies at the particle and \Delphes{} detector levels with the overall correlation coefficient $\rho$ indicated.}
\label{fig_topos_graph_migra}
\end{center}
\end{figure*}

We probed for two classes of boosted large-$R$-jets appear in events,
either as top-jets, \emph{i.e.} jets with a~mass and substructure consistent with
the hypothesis that the top quark decay products were reconstructed
in such a~large-$R$ jet; or as $W$-jets, \emph{i.e.} those consistent with the hadronic decay of a~$W$ boson. We denoted the top-like large-$R$ jets as
boosted (B), and the $W$-like jets as semiboosted (S). Based on their
numbers in the event, we define in total three topologies of the
all-hadronic $\ttbar{}$ final state. These are denoted as the boosted-boosted
(2B0S), boosted-semiboosted (1B1S) or semiboosted-semiboosted (0B2S), as
depicted in~Figure~\ref{fig_cartoons} or as \Delphes{} event display
examples in~Figure~\ref{fig_evt_displ_zp1000_3}.
The fraction of the considered topologies as a~function of the generated mass of the $Z'$
vector boson are shown in~Figure~\ref{fig_topos_graph_migra}, together
with the migration rate between the topologies at the particle
and detector levels, in order to check the stability of the topologies
concept at both levels. The migration is observed to be at the level of $10$\%
in terms of the fraction of events migrating to different
topologies. One can see that while the fractions of semiboosted
topologies (1B1S and 0B2S) decrease with the mass of the resonance, they
still constitute non-negligible fractions of 10--20\% at medium masses
of $Z'$, and about a~30\% fraction for the SM $\ttbar{}$ sample.

\noindent The same event selection applied at the particle and detector levels can be summarised as follows:
\begin{itemize}
  \item require at least two large-$R$ jets with $\pt > 80\,$GeV and $|\eta| < 2.0$;
  \item preselect small-$R$ jets with $\pt > 25\,$GeV and $|\eta| < 2.5$;
  \item require at least two small-$R$ jets tagged as $b$-jets;
  \item reject events with an isolated high-$\pt$ lepton above $25\,$GeV, in order to remove events from the $pp \rightarrow y_0 \rightarrow \ttbar$ for which, due to the triangle loop in the corresponding diagrams, the \ttbar{} decay mode could not be specified at the generator level.
  \item Finally, attempt, in this order, the boosted-boosted (2B0S), boosted-semiboosted (1B1S) or semiboosted-semiboosted (0B2S) topologies reconstruction of the $\ttbar{}$ final state by requiring the corresponding numbers of top-tagged and $W$-tagged jets.
\end{itemize}
\noindent Examples of the resulting detector-level yields for the $Z'$ model in the 1B1S topology are listed in~Table~\ref{tab:ex:yields_zp}.

\begin{table*}[!h]
\centering
\begin{tabular}{l|r}
 \hline
 1B1S topology &  Expected yields \\ \hline 
 $Z'\rightarrow t\bar{t}, m_{Z'} = 1 \,\mathrm{TeV} (\times 37.5\mathrm{k} \cdot 2^{2})$           & $293300 \pm 6200$  \\ 
 $t\bar{t}, p_\mathrm{T}^{j1} \geq 200 \,\mathrm{GeV} p_\mathrm{T}^{j2} \in (60,200) \,\mathrm{GeV}$  & $252300 \pm 2300$  \\ 
 $t\bar{t}, p_\mathrm{T}^{j1,j2} \geq 200 \,\mathrm{GeV}$       & $105400 \pm 730$  \\ 
 $t\bar{t}, p_\mathrm{T}^{j1,j2} \in (60, 200) \,\mathrm{GeV}$  & $232700 \pm 6400$  \\ 
 $WWbb, p_\mathrm{T}^{j1,j2} > 60 \,\mathrm{GeV}$               & $411700 \pm 7200$  \\ 
 $Wbbjj, p_\mathrm{T}^{j1,j2} > 60 \,\mathrm{GeV}$              & $25800 \pm 1600$  \\ 
 $bbjj, p_\mathrm{T}^{j1} \geq 200 \,\mathrm{GeV} p_\mathrm{T}^{j2} \in (60,200) \,\mathrm{GeV}$  & $1420600 \pm 45900$  \\ 
 $bbjj, p_\mathrm{T}^{j1,j2} \geq 200 \,\mathrm{GeV}$                                           & $1419700 \pm 19400$  \\ 
 \hline 
  pseudodata &  $4091100$    \\ 
  prediction & $4161600 \pm 51200$   \\ 
  pseudodata / prediction & 0.98  \\ 
  \hline 
 \end{tabular} 

    \caption{cross-section weighted event yields for the $\ttbar$ and non-$\ttbar$ background samples and the vector $Z'$ model with $m_{Z'} = 1000\,$GeV in the 1B1S topology.}
  \label{tab:ex:yields_zp}
\end{table*}

\section{Variables}

\subsection{Kinematic variables}

Typical variables studied in the $\ttbar{}$ system for SM precision differential cross-section measurements are the
transverse momenta of the two top quarks (both used in this study, with event weights of $0.5$), rapidity of the top quarks ($y^{t}$) and of the \ttbar{} system ($y^{\ttbar{}}$), and the mass ($m^{\ttbar{}}$) and transverse momentum ($\pttt$) of the \ttbar{} system; their 2D and 3D extensions and variables related to additional jet activity in the event. Further variables are composed from the two top quarks momenta: These are the $\cos\theta^*$ (angle between a~top quark and the $z$ axis in a~frame where the $\ttbar{}$ system has zero momentum along the $z$ axis); the laboratory opening angle between the two top quarks ($\delta_{\ttbar{}}$) and the $\Delta\phi$ between the two top quarks in the transverse plane ($\DeltaPhitt$).
Other observables studied are the out-of-plane momentum $\pout$ based on the direction of one of the top quarks defining a~plane together with the beam ($z$ axis) direction $\hat{z}$, to which the momentum of the other top quark is projected; and the $\Yboost$ and $\Chittbar$ variables, defined as
\begin{eqnarray}
  \pout     &\equiv&  \vec{p}^{\,t2} \cdot \frac{\vec{p}^{\,t1} \times \hat{z}}{|\vec{p}^{\,t1}\times \hat{z}|}   \,, \quad \mathrm{and\,\, 1 \leftrightarrow 2} \\
  \Yboost   &\equiv& \frac12 \left| y^{t2} + y^{t1} \right| \\
  \Chittbar &\equiv& \exp \left| y^{t2} - y^{t1} \right| \,.
\end{eqnarray}
These are sensitive to final state radiation, the boost of the $\ttbar{}$ system and thus also to parton distribution functions; and to new physics via their sensitivity to the production angle in the central-mass-system. We also studied the ratio of the two top quark transverse momenta, $\Rtt{}$, the ratio of the sub-leading to leading top quark \pt{}. One can also use event shape variables (originally studied in QCD jet physics) based on the normalized momentum tensor defined using the three-momenta of objects $\mathcal{O}$ in the event (usually jets) as
\begin{eqnarray}
  \mathcal{M}_{ij} \equiv \frac{\sum\limits_\mathcal{O} p^\mathcal{O}_i p^\mathcal{O}_j}{\sum\limits_\mathcal{O}  \left(\vec{p}^\mathcal{O}\right)^2}
  \end{eqnarray}
where $i,j$ run over the three space indices. The matrix has eigenvalues $\lambda_1 \leq \lambda_2 \leq \lambda_3$
used to define aplanarity $\mathcal{A} \equiv \frac32 \lambda_1$ and sphericity $\mathcal{S} \equiv \frac32 (\lambda_1 + \lambda_2)$.
Corresponding aplanarity and sphericity for the small-$R$ jets are labelled as $\mathcal{A}^j$ and~$\mathcal{S}^j$ in plots.
Other global variables were also studied, namely:
\begin{itemize}
   \item the sum of \pt{} of small-$R$ jets $H_\mathrm{T}^{j}$;
   \item missing transverse energy $E_\mathrm{T}^\mathrm{miss}$; and the sum of $H_\mathrm{T}^{j} + E_\mathrm{T}^\mathrm{miss}$;
   \item the sum of masses of large-$R$ jets $\sum\limits_J m_J$, labelled also as Sum $m_J$ in plots;
   \item and the invariant masses of the 4-vector sum of large-$R$ jets 4-momenta $m^\mathrm{vis}_{\sum J}$, labelled also as $m^\mathrm{vis}_{\mathrm{Sum}J}$ in plots.
\end{itemize}
These were motivated by the fact that invisible BSM particles carrying significant transverse momentum can exhibit tails in $\Etmiss$, $\pout$, or contribute to higher jet activity in the event due to a~larger energy scale of events with production of heavy particles. We held onto the global event shape variables like aplanarity, sphericity; and retained $H_\mathrm{T}$ only for small-$R$ jets while keeping the jet-mass related quantities for larger-$R$ jets only.

\noindent In addition, dimensionless variables with the potential of possessing smaller experimental uncertainties in absolute jet energy scale calibration were studied by constructing ratios relative to $\mttbar$ or to the geometric mean of the two top quark transverse momenta $\sqrt{p_\mathrm{T}^{t1} \, p_\mathrm{T}^{t2}}$.

We construct variables ratios which naturally encode more information from the event, and we further construct two-dimensional (2D) variables from the above list of 23 variables, studying 253 variables in total, see Section~\ref{subsect:var_cmp} for their performance.

\subsection{Shape comparison}

\begin{figure*}
\begin{center}
\begin{tabular}{c}
\includegraphics[width=0.90\textwidth]{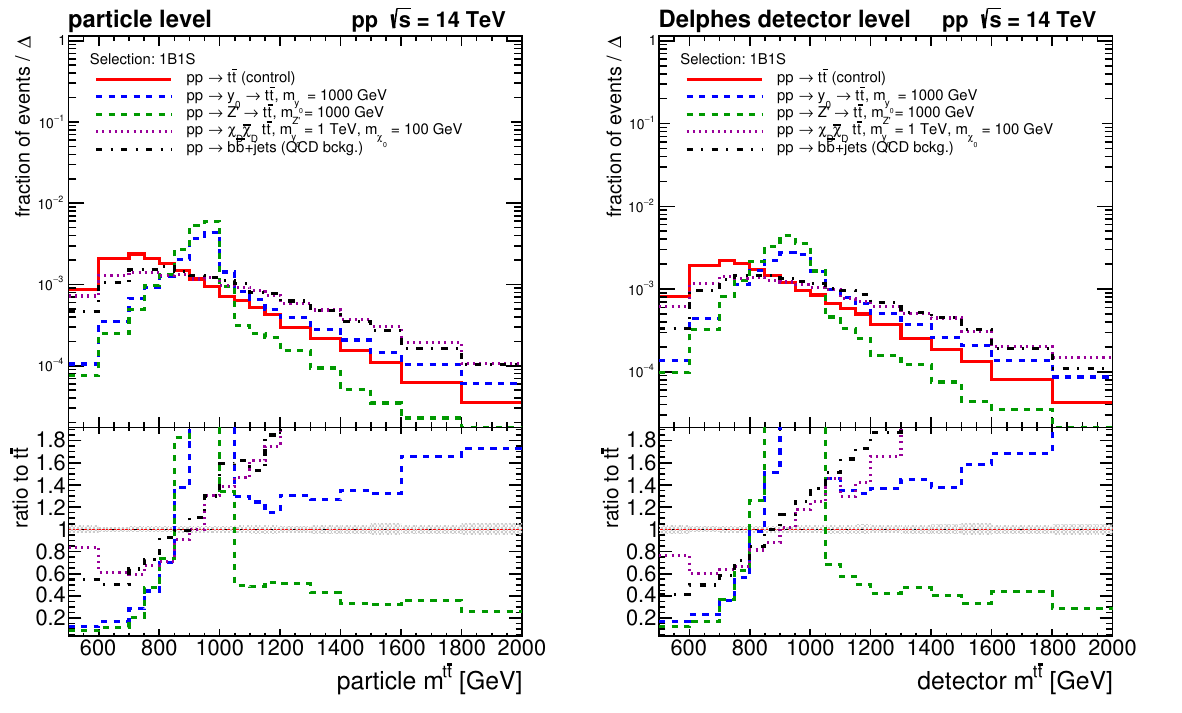}\\
\end{tabular}
\caption{Shape comparison for the invariant mass of the $\ttbar$ pair in the boosted-semiboosted topology for the SM and BSM $\ttbar$ signal processes as well as for the SM multijet (QCD) background. Left: particle level, right: \Delphes{} ATLAS detector level.}
\label{fig_shapes_mtt_1B1S}
\end{center}
\end{figure*}

\begin{figure*}
\begin{center}
\begin{tabular}{c}
\includegraphics[width=0.90\textwidth]{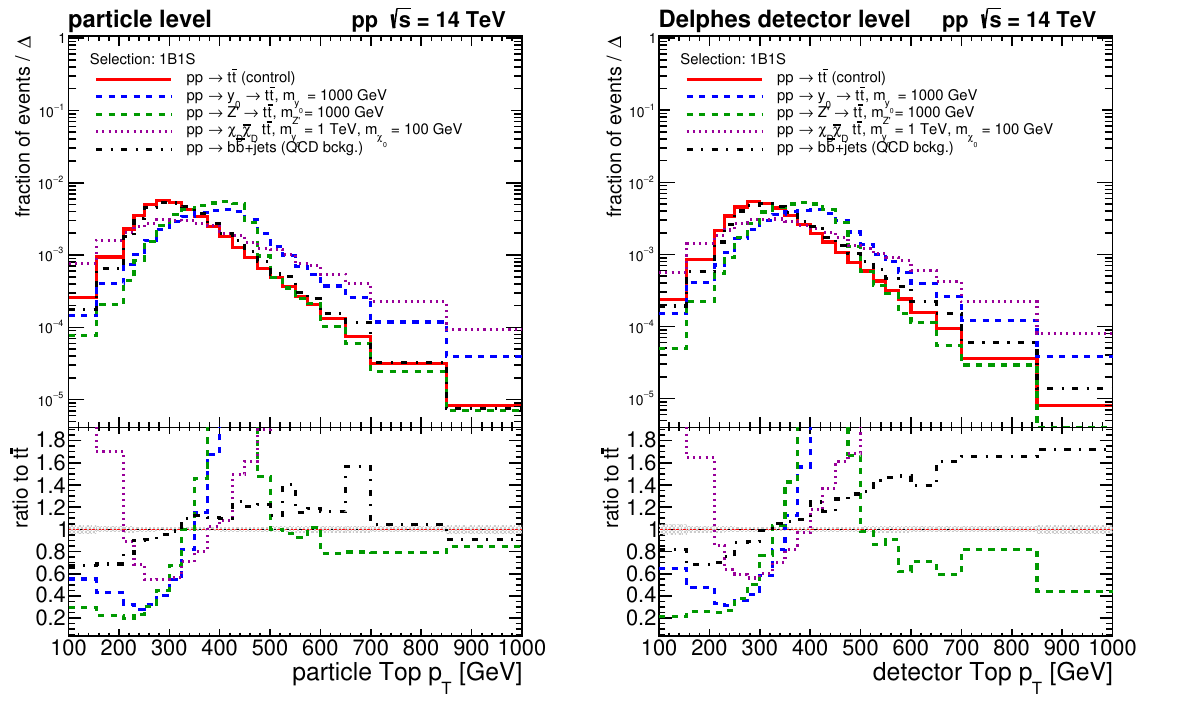}\\
\end{tabular}
\caption{Shape comparison for the $\pt$ of the hadronic top quark candidate in the boosted-semiboosted topology for the SM and BSM $\ttbar$ signal processes as well as for the SM multijet (QCD) background. Left: particle level, right: \Delphes{} ATLAS detector level.}
\label{fig_shapes_pTt_1B1S}
\end{center}
\end{figure*}

\begin{figure*}
\begin{center}
\begin{tabular}{c}
\includegraphics[width=0.90\textwidth]{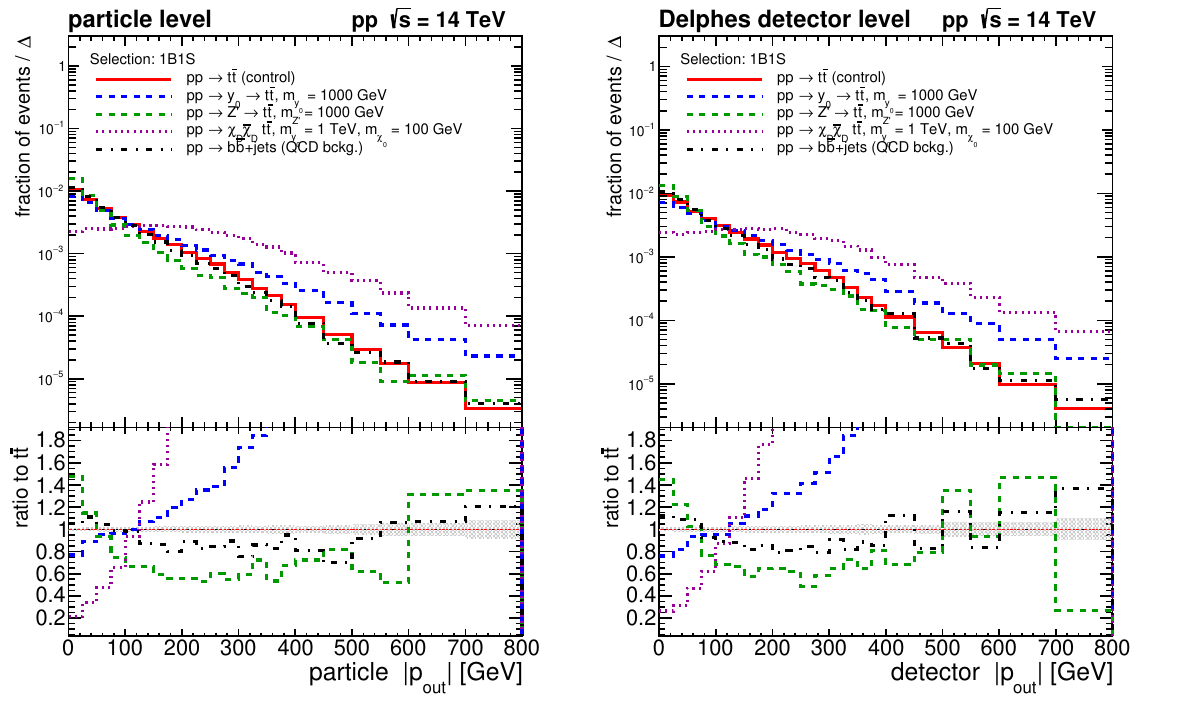}\\
\end{tabular}
\caption{Shape comparison for the absolute value of the out-of-plane momentum $|p_\mathrm{out}|$ in the boosted-semiboosted topology for the SM and BSM $\ttbar$ signal processes as well as for the SM multijet (QCD) background. Left: particle level, right: \Delphes{} ATLAS detector level.}
\label{fig_shapes_Pout_denser_1B1S}
\end{center}
\end{figure*}

\begin{figure*}
\begin{center}
\begin{tabular}{c}
  \includegraphics[width=0.90\textwidth]{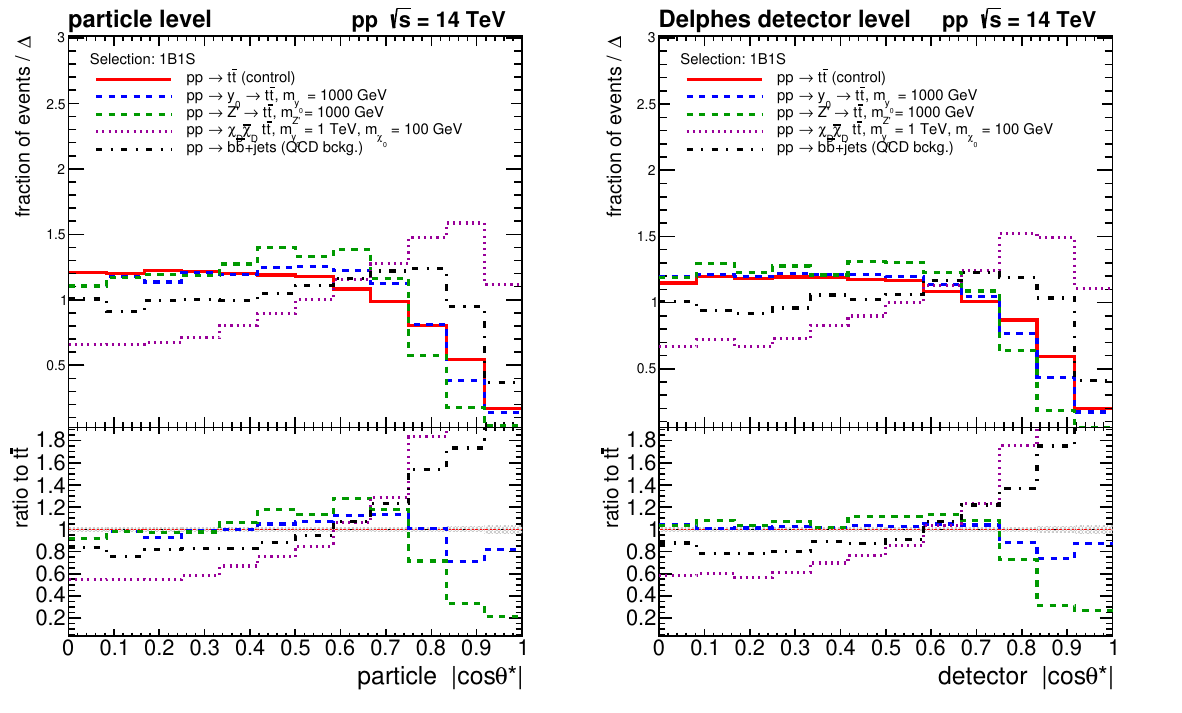} \\
\end{tabular}
\caption{Shape comparison for the absolute value of the $\cos\theta^*$ variable in the boosted-semiboosted topology for the SM and BSM $\ttbar$ signal processes as well as for the SM multijet (QCD) background. Left: particle level, right: \Delphes{} ATLAS detector level.}
\label{fig_shapes_costhetastar_1B1S}
\end{center}
\end{figure*}

We present a~selection of variables shape comparisons for the 1B1S topology in this section.
The most useful variable for searching for a~resonant $\ttbar{}$ production is naturally the invariant mass of the $\ttbar{}$ system, $m^{\ttbar}$. The shape comparison show a~clear peak for the vector ($Z'$) as well as the scalar ($y_0$) models. See Figure~\ref{fig_shapes_mtt_1B1S}.
But shapes of other variables are also modified for such a~signal, \emph{e.g.} the top quark transverse momentum in~Figure~\ref{fig_shapes_pTt_1B1S}, exhibits a~broader peak rather than just a~slope change.
In contrast, the DM model leading to the associated production of a~$\ttbar{}$ pair and a~pair of DM particles ($\chi_D \bar{\chi}_D$) leads not only to an enhancement of the missing transverse energy where such a~signal is often looked for, but also to a~more prominent tail in the out-of-plane momentum distribution ($\pout$). See~Figure~\ref{fig_shapes_Pout_denser_1B1S}.
The spin of the resonance decaying to a~pair of top quarks can be determined from the $\cos\theta^*$ distribution, as provided in~Figure~\ref{fig_shapes_costhetastar_1B1S}, as it leads to a~slightly flatter spectrum for the scalar $y_0$ case and a~more peaked shape for the case of a~$Z'$ production.
Similar observations were found also in the 2B0S and 0B2S event topologies.



\subsection{Correlations between variables}

\begin{figure*}
\begin{center}
   \begin{tabular}{c}
  \includegraphics[width=0.68\textwidth]{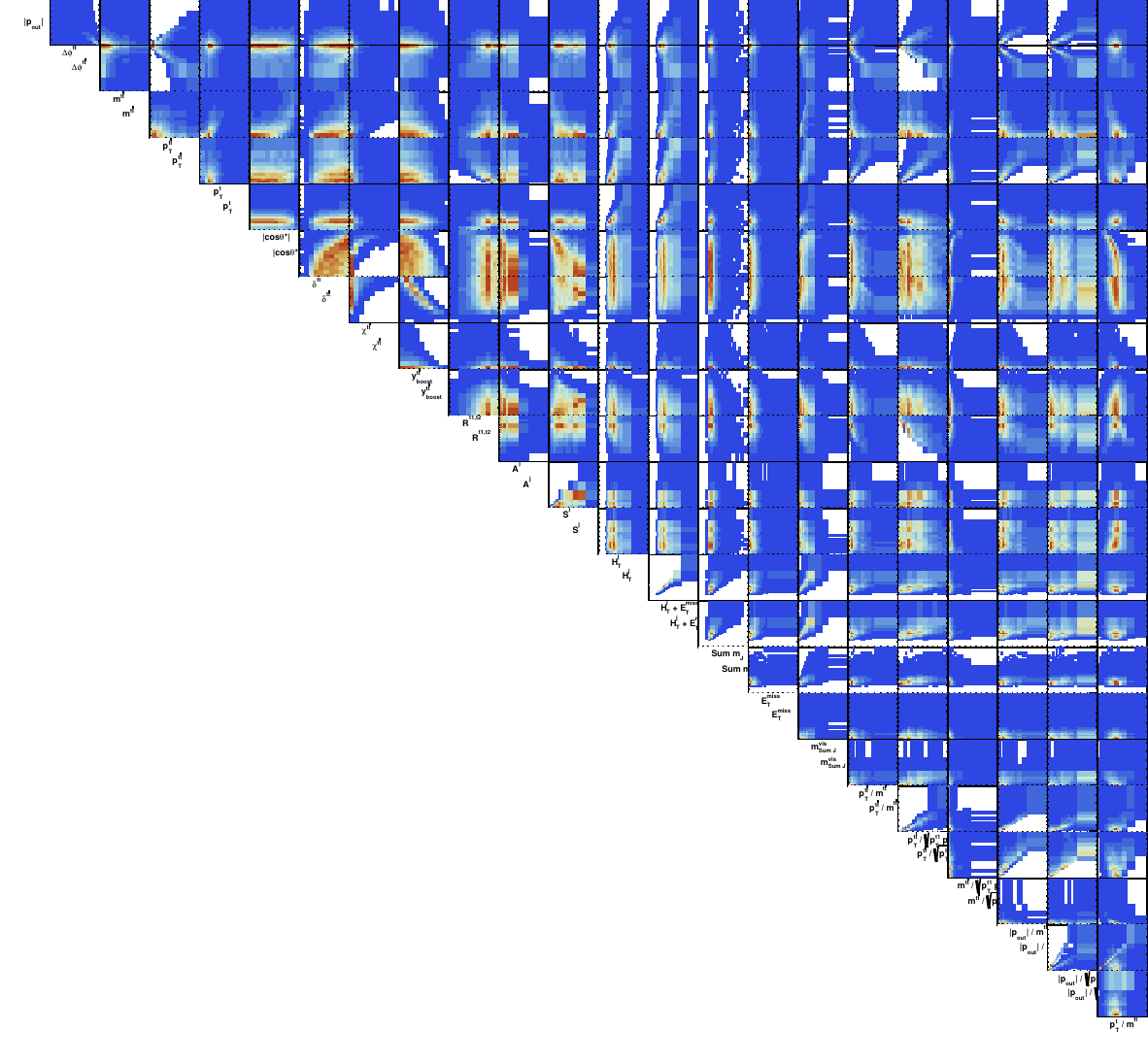} \\ 
  \includegraphics[width=0.81\textwidth]{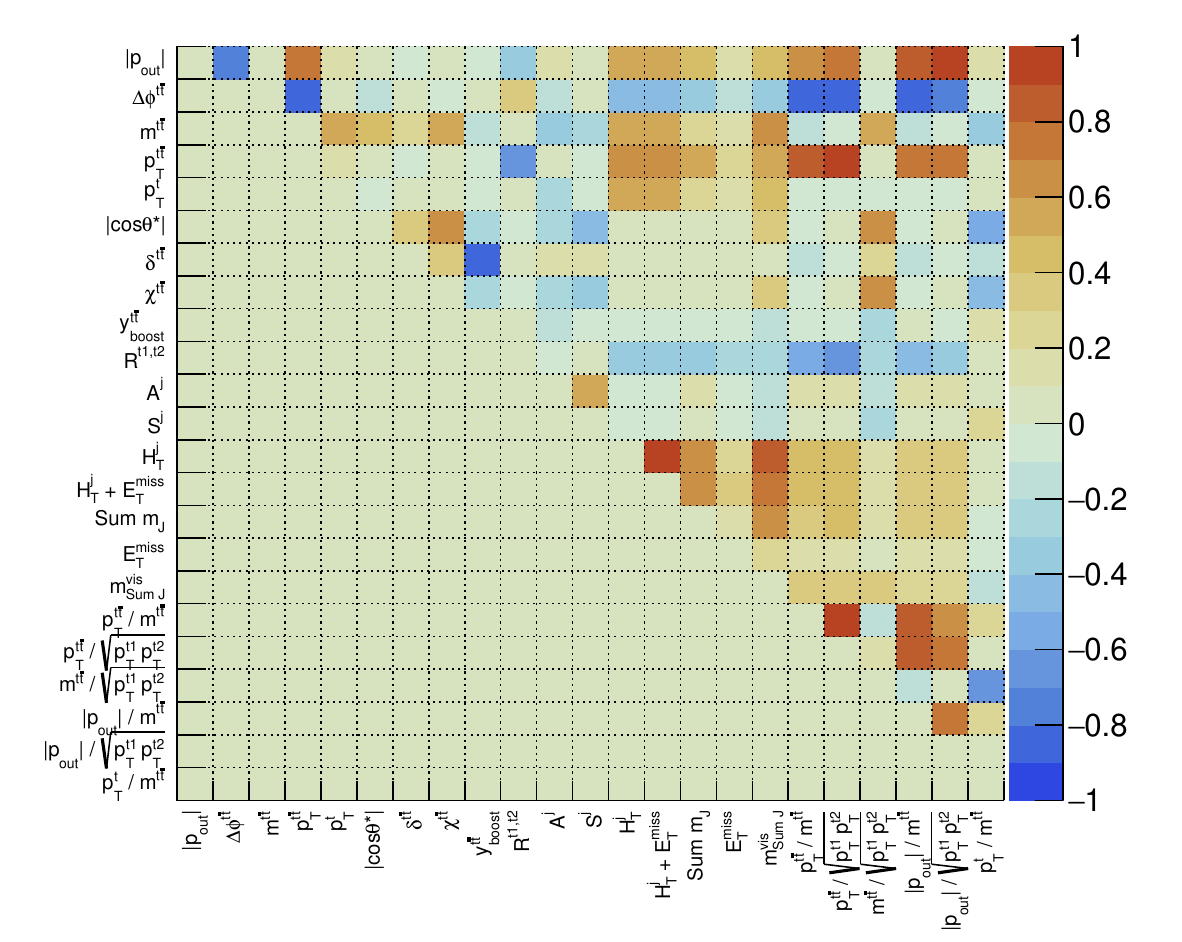} \\
\end{tabular}
\caption{Example of correlations between bins of considered kinematic variables (top) as well as the correlation factor between the variables (bottom) at the detector level in the boosted-semiboosted (1B1S) topology evaluated on the $\ttbar{}$ control sample. The same order of variables is preserved at both axes and on both plots.}
\label{fig_corr_vars}
\end{center}
\end{figure*}

Correlation between variables were studied with the motivation of verifying later whether higher or lower correlation (see Figure~\ref{fig_corr_vars}) lead to the expected change in the power of the variables to separate signal from background.
Besides the obvious anti-correlations of variables to relative quantities where the former is in the denominator, and the negative correlation of $\DeltaPhitt$ to $\pttt$ and $\pout$ by construction,
one can make several non-trivial observations. Out of all the variables, $\pout$ (or an absolute value of it) is especially a useful one as it has only a~small correlation to $\mtt$ (we observe 0.07 for the $\ttbar$ control sample, see also Figure~\ref{fig_corr_vars}) while it exhibits a~large shape change especially in tails for models with additional DM particles produced, see~Figure~\ref{fig_shapes_Pout_denser_1B1S}. It has also been used for MC generators tuning as it is sensitive to the initial and final state radiation~\cite{ATL-PHYS-PUB-2016-020} but caution should be taken in order not to tune to possible new physics effects.

We kept variables even with large absolute correlations for detailed study of their performance in terms of a~signal significance.
Both $\ptt$ and $\pttt$ have large correlation to variables quantifying the jet $\pt$ activity in the event, like $\HTj$ and the total large-$R$ jets visible mass $m^\mathrm{vis}_{\sum J}$. 
There are also large negative correlations of sphericity and $\Yboost$ to $\mtt / \sqrt{p_\mathrm{T}^{t1} \, p_\mathrm{T}^{t2}}$, of $\Rtt$ to $\pttt$ and $|\pout|$; and of $|\cos\theta^*|$ and $\Chittbar$ to $\ptt / \mtt$.

\clearpage
\section{Results}


\subsection{Pseudo-data study}

For the purpose of creating a~realistic mixed sample of selected SM backgrounds (associated production of a~$W$ bosons and $b$-jets), the $\ttbar{}$ sample and a~new physics signal, detector-level distributions were created by mixing the samples based on their cross section $\sigma$ using weights $w = \sigma \, L/ N_\mathrm{gen}$ (with $L$ being the target luminosity and $N_\mathrm{gen}$ the number of generated events with non-zero matching weights), with an additional scaling of the signal in order to make it more prominent so as to study its shape and significance. Such mixed samples were used throughout the study to asses the performance of the variables proposed in the previous section.

For the purpose of creating stacked plots consisting of physics and $\ttbar$ background and a~selected signal model, the scalar $y_0$ model with the same mass was scaled by a~factor of 30 and additional factors of $2^{3/2}$ ($2^{4}$) in the 1B1S (0B2S) topologies.
Similarly, and in order to approximately match the $y_0$ signal yield, the $Z'$ model with $m_{Z'} = 1000\,$GeV was scaled, due to its smaller cross-section, by a~factor of 37.5k, with additional $2^{1/2}$, $2^2$ and $2^4$ factors in the 2B0S, 1B1S and 0B2S topologies, respectively. The larger $y_0$ production cross section is partially due to the higher gluon-gluon parton luminosity in the production mechanism.
Finally, the sample of the DM pair creation with an associated $\ttbar{}$ pair was scaled by a~factor of 3k and an additional factor of $2^{\frac12}$ in the 0B2S topology.
The purpose of the additional topology-dependent scalings is to reach a~similar level of signal significance across the topologies.

For the case of 1D plots, the target luminosity was set to $1\,\mathrm{ab}^{-1}$, being about $1/3$ of the design High Luminosity LHC programme.
For the \BumpHunter{} studies, the target luminosity used is $100\times$ smaller for the following reasons: the signal scale factors and therefore signal yields were set large enough for the 1D stack comparison in order to see where signal accumulates for each variable. But for a search exercise such amount of signal would be too large and pseudo-data compatibility to the background-only hypothesis would be negligible. Therefore, either smaller signal scale-factors would have to be applied to a larger luminosity, or one can perform the variables comparison at lower luminosity. The latter approach was adopted also for the following reason: the Monte-Carlo luminosities of the generated samples correspond roughly to the case of $0.01\,\mathrm{ab}^{-1}$. Corresponding statistical fluctuations are imprinted on the binned observables. Scaling to an arbitrarily high luminosity would mean that the fluctuations would be taken as a serious disagreement between pseudo-data and prediction, which would not be fair. Thus, a trade-off between the luminosity and signal strength has to be resolved maintaining optimal $S/B$ ratio for the \BumpHunter{} study to perform nontrivially, \emph{i.e.} not giving extreme results towards either of the background-only or signal-saturated conclusions so that variables performance can be compared over an intermediate region of a~modest signal strength, allowing a noticeable separation between less and more powerful observables.

\begin{figure*}
\begin{center}
\begin{tabular}{ccc}
\includegraphics[width=0.33\textwidth]{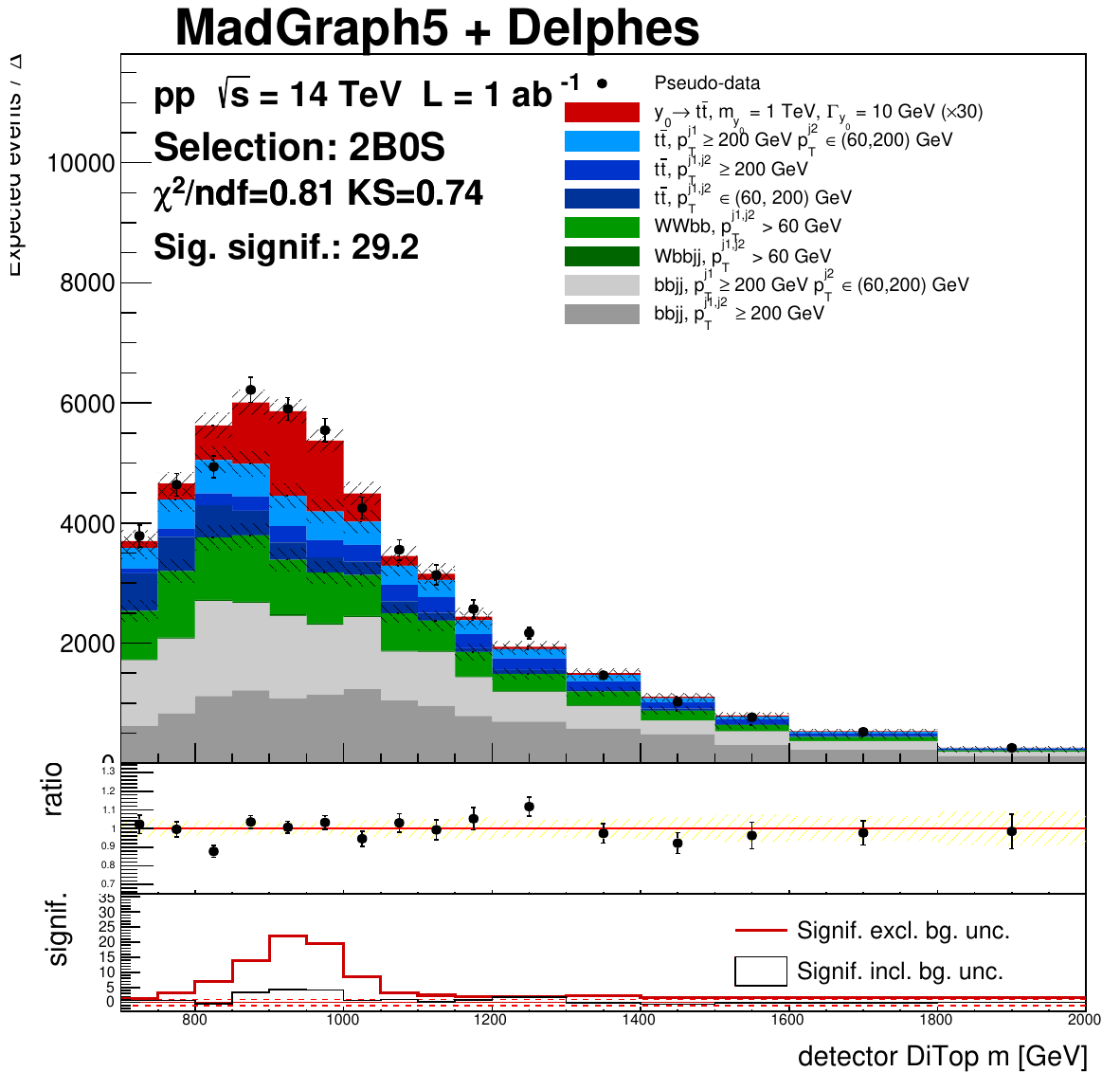} &
\includegraphics[width=0.33\textwidth]{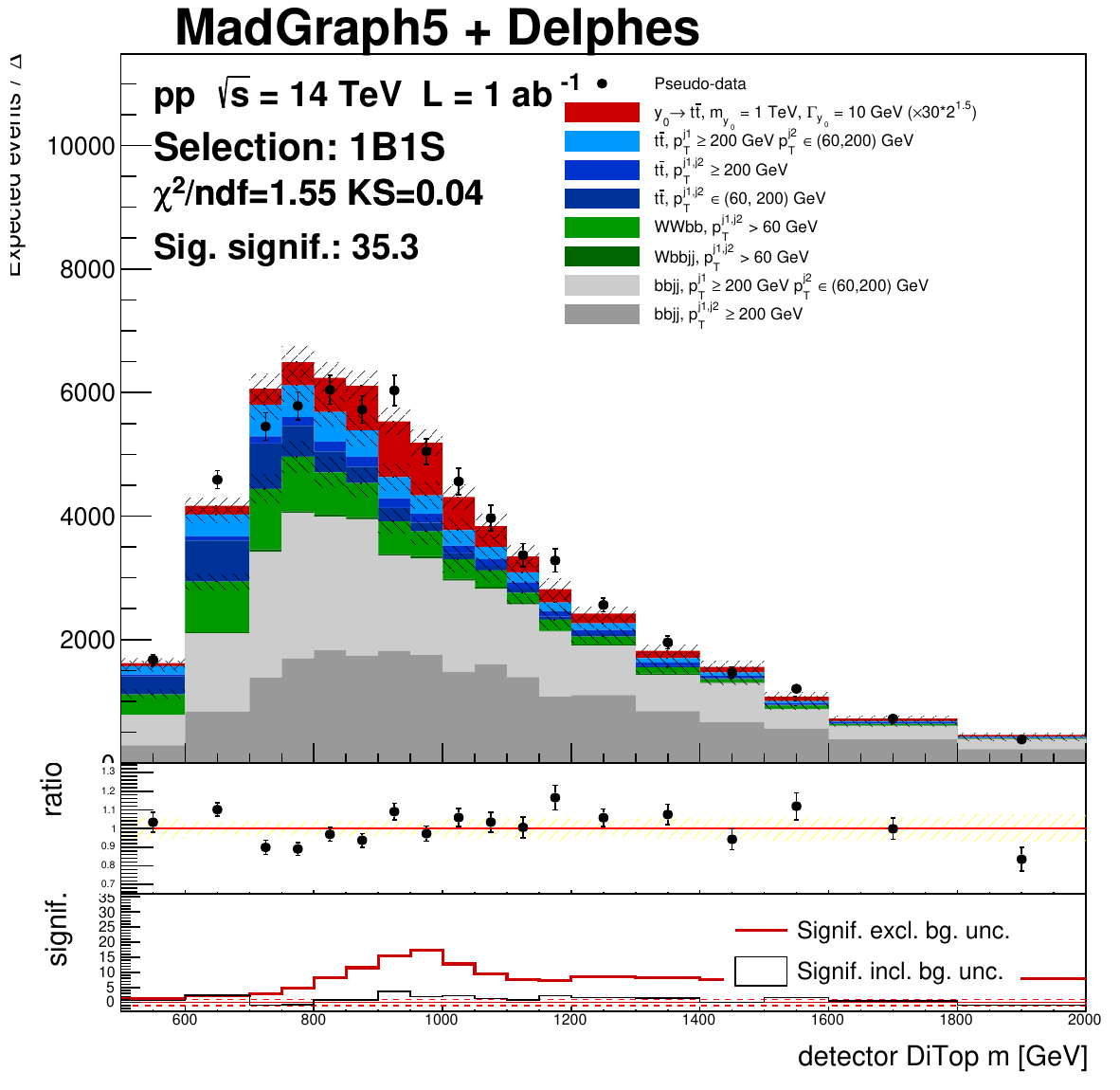} &
\includegraphics[width=0.33\textwidth]{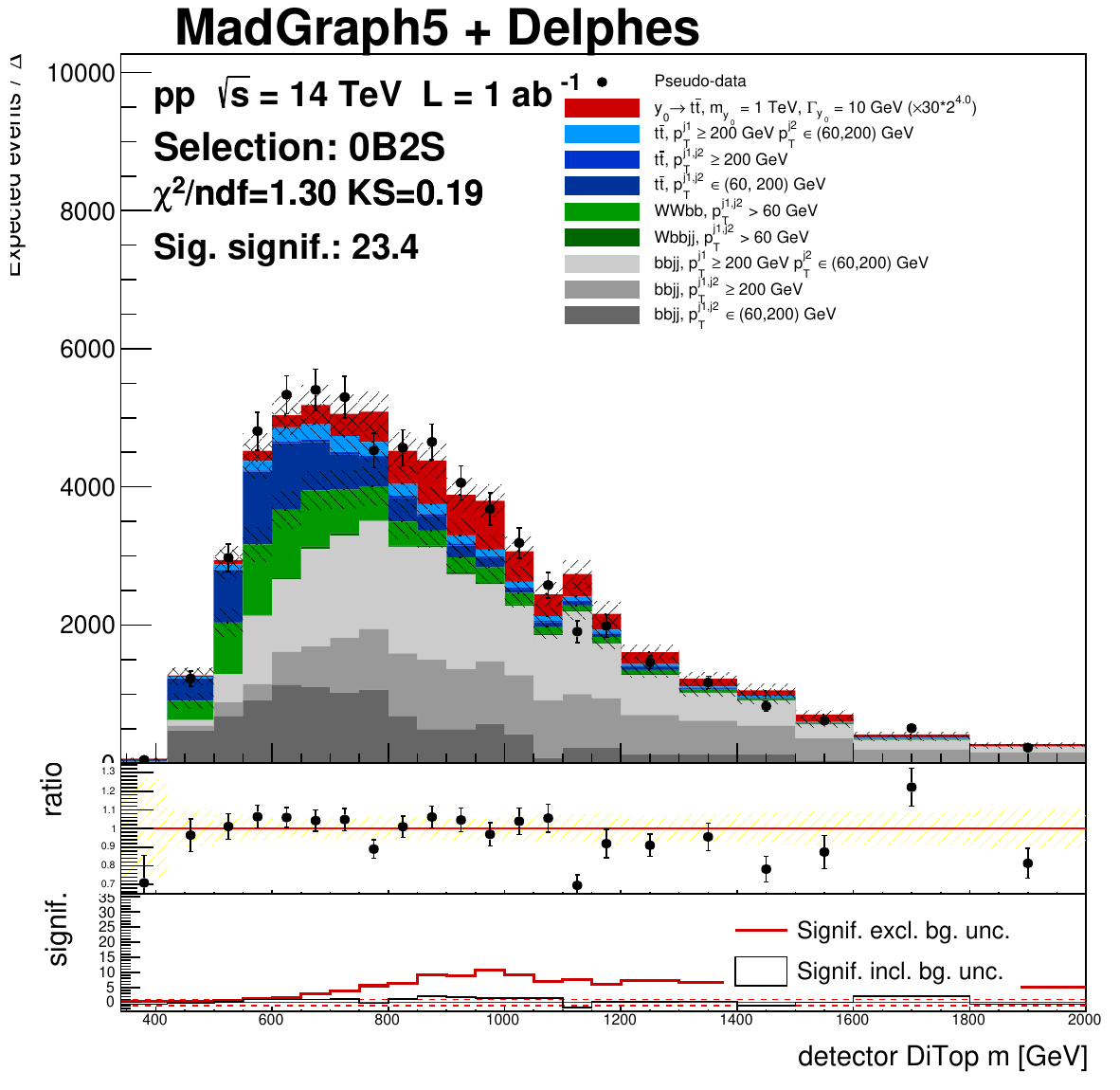} \\
\includegraphics[width=0.33\textwidth]{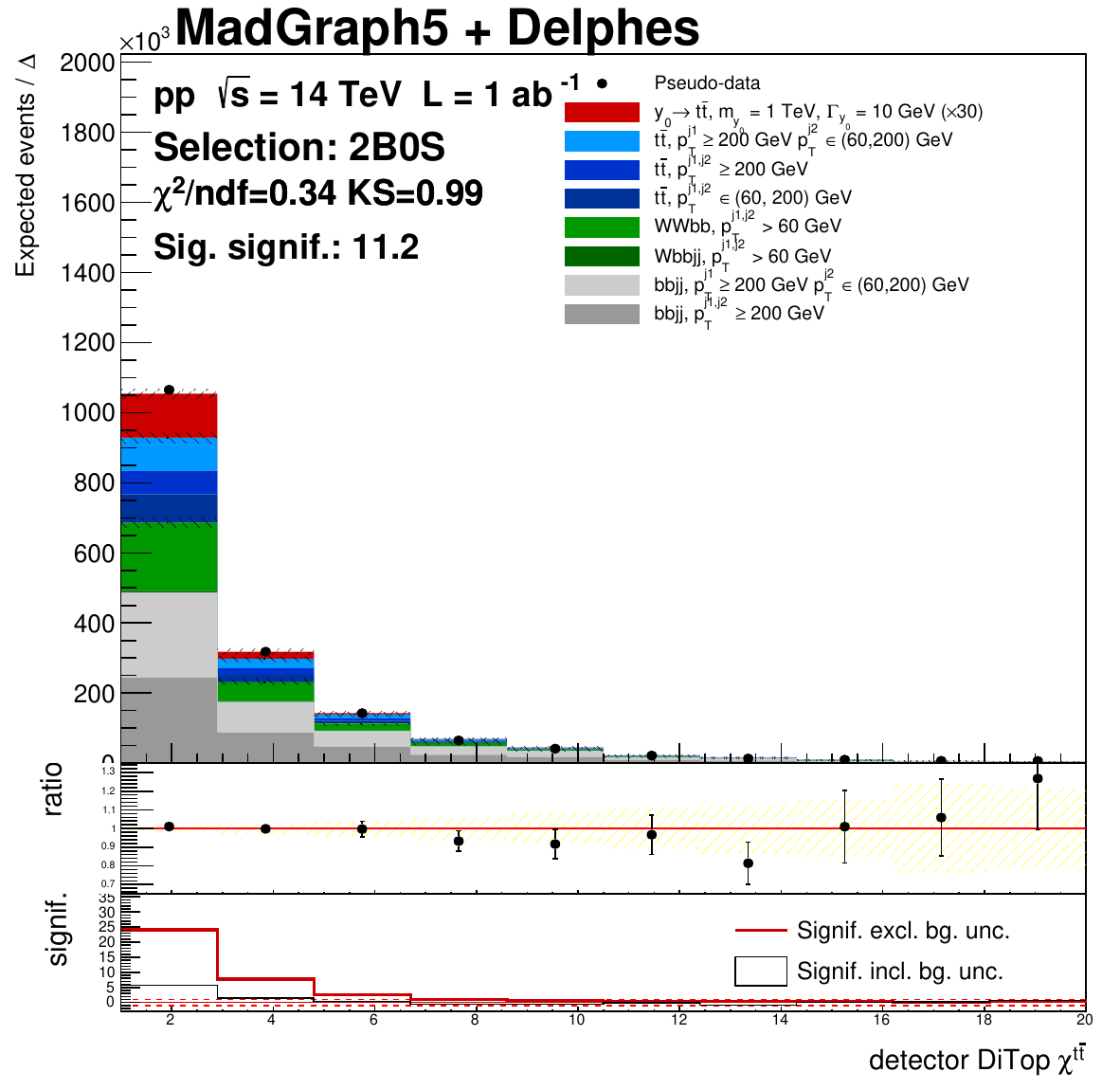} &
\includegraphics[width=0.33\textwidth]{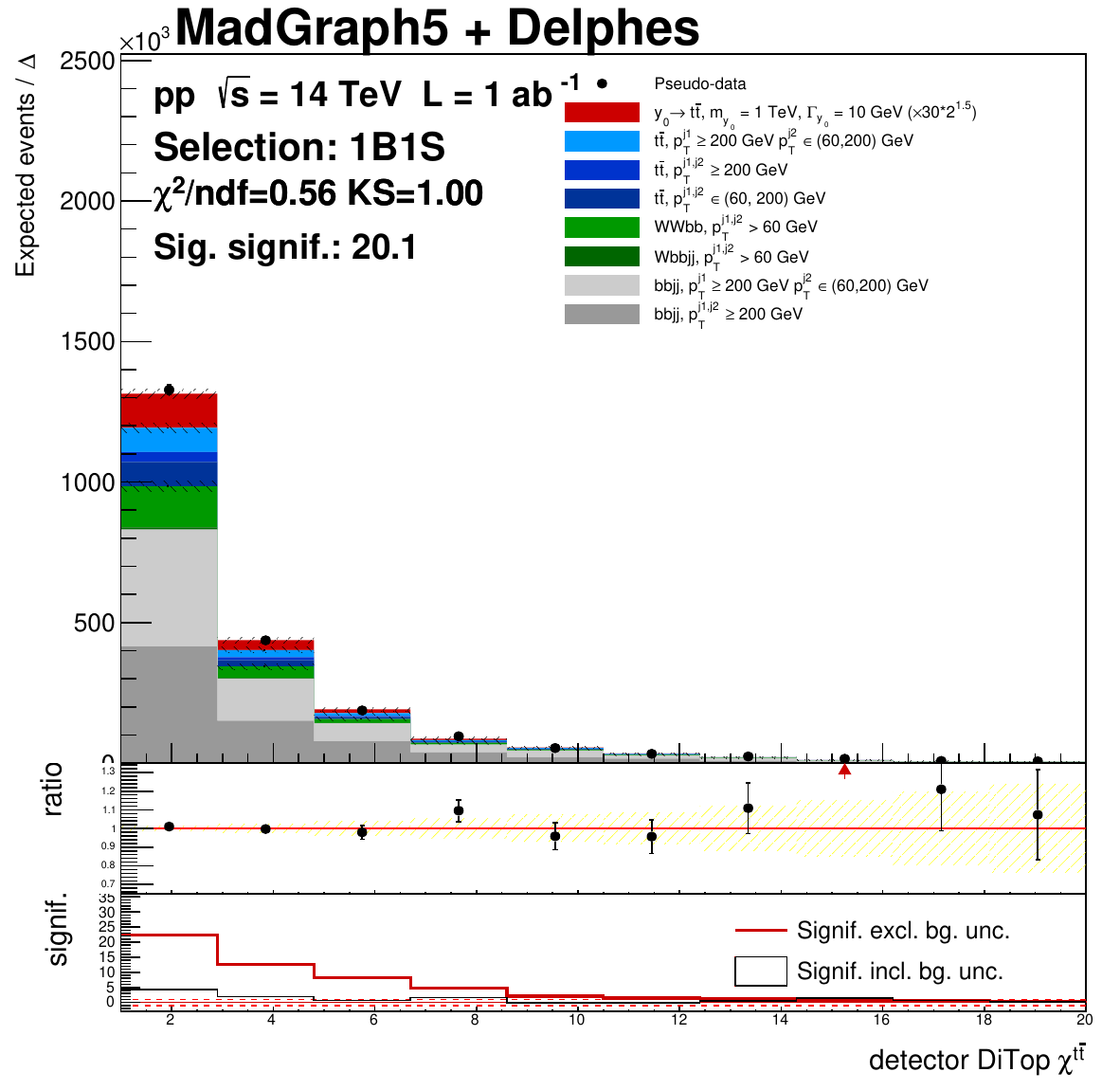} &
\includegraphics[width=0.33\textwidth]{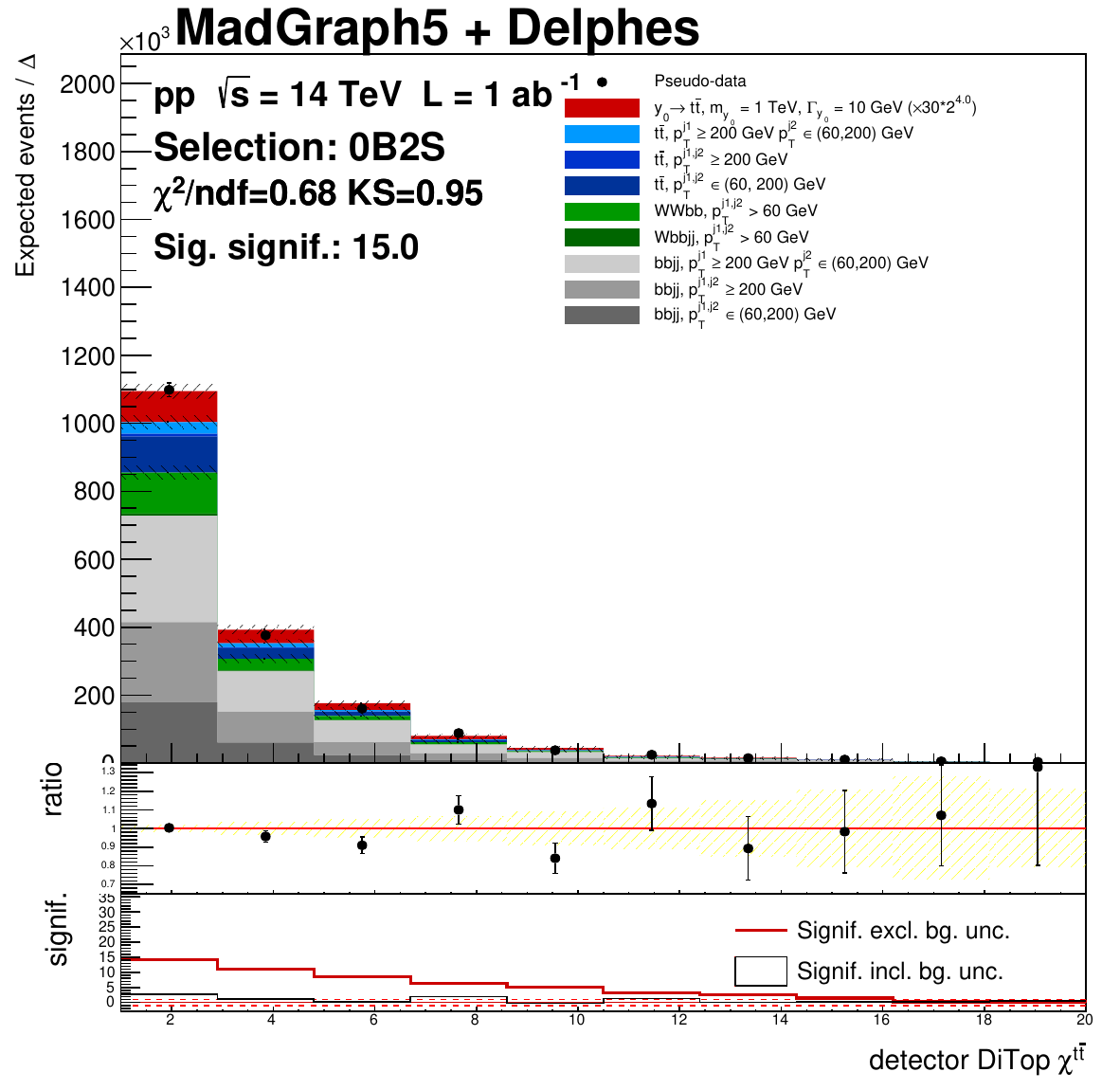} \\
2B0S  &
1B1S  &
0B2S  \\
\end{tabular}
\caption{Stacked invariant mass of the $\ttbar$ pair (top) and the $\Chittbar$ variable (bottom) at the \Delphes{} detector level for the scaled $y_0$ model with $m_{y_0} = 1\,$TeV, with both the pseudo-data/prediction ratio and the signal significance shown in lower pads, for the boosted-boosted (2B0S, left), boosted-semiboosted (1B1S, middle) and semiboosted-semiboosted (0B2S, right) topologies. The $\chi^2$ and Kolmogorov-Smirnov (KS) test statistics between the pseudo-data and predictions as provided by ROOT~\protect\cite{Antcheva:2009zz} and the total signal significance are also indicated.
}
\label{fig_stack_Detector_model_y0_compact_RS}
\end{center}
\end{figure*}
\begin{figure*}
\begin{center}
\begin{tabular}{ccc}
\includegraphics[width=0.33\textwidth]{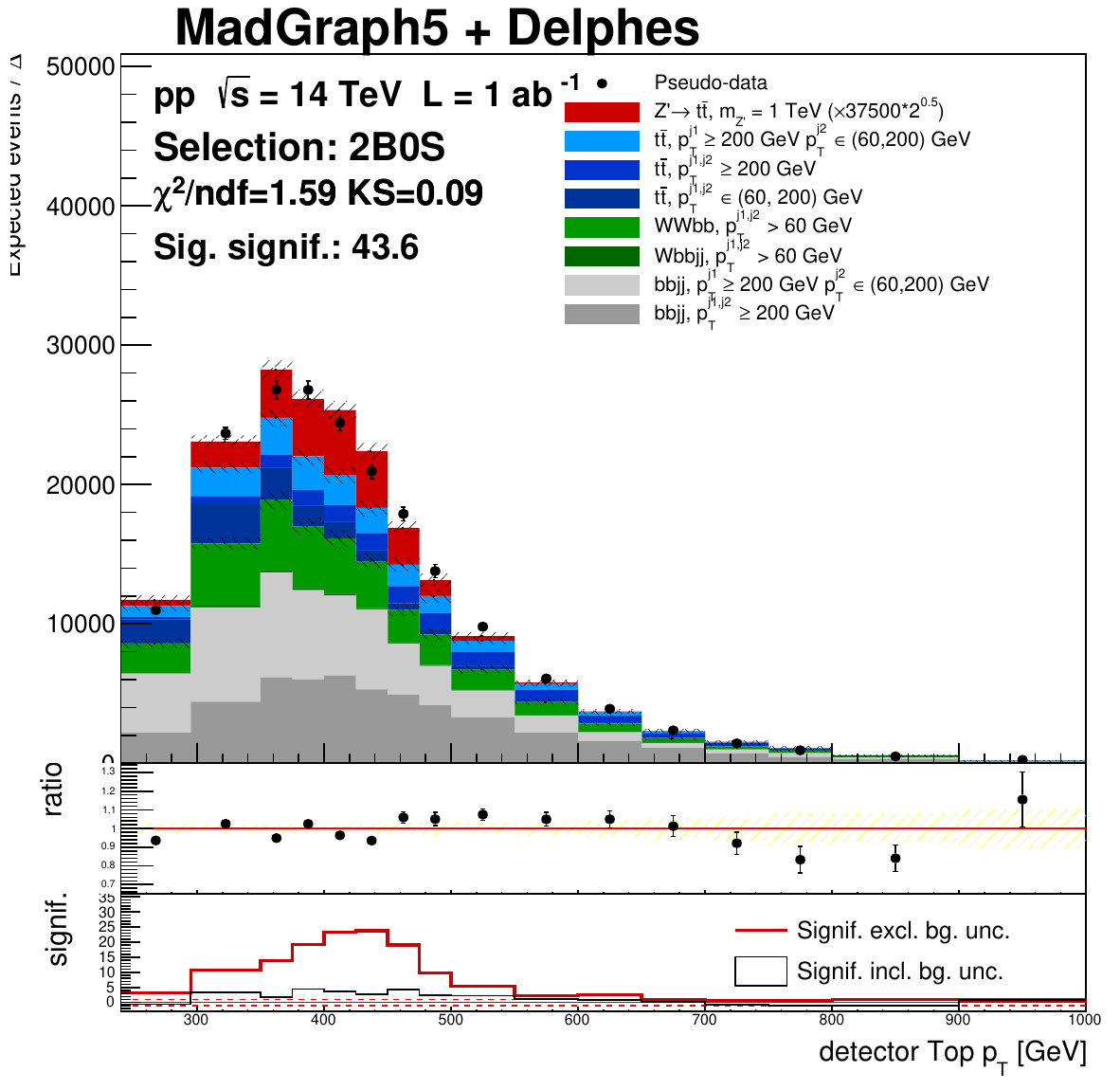} &
\includegraphics[width=0.33\textwidth]{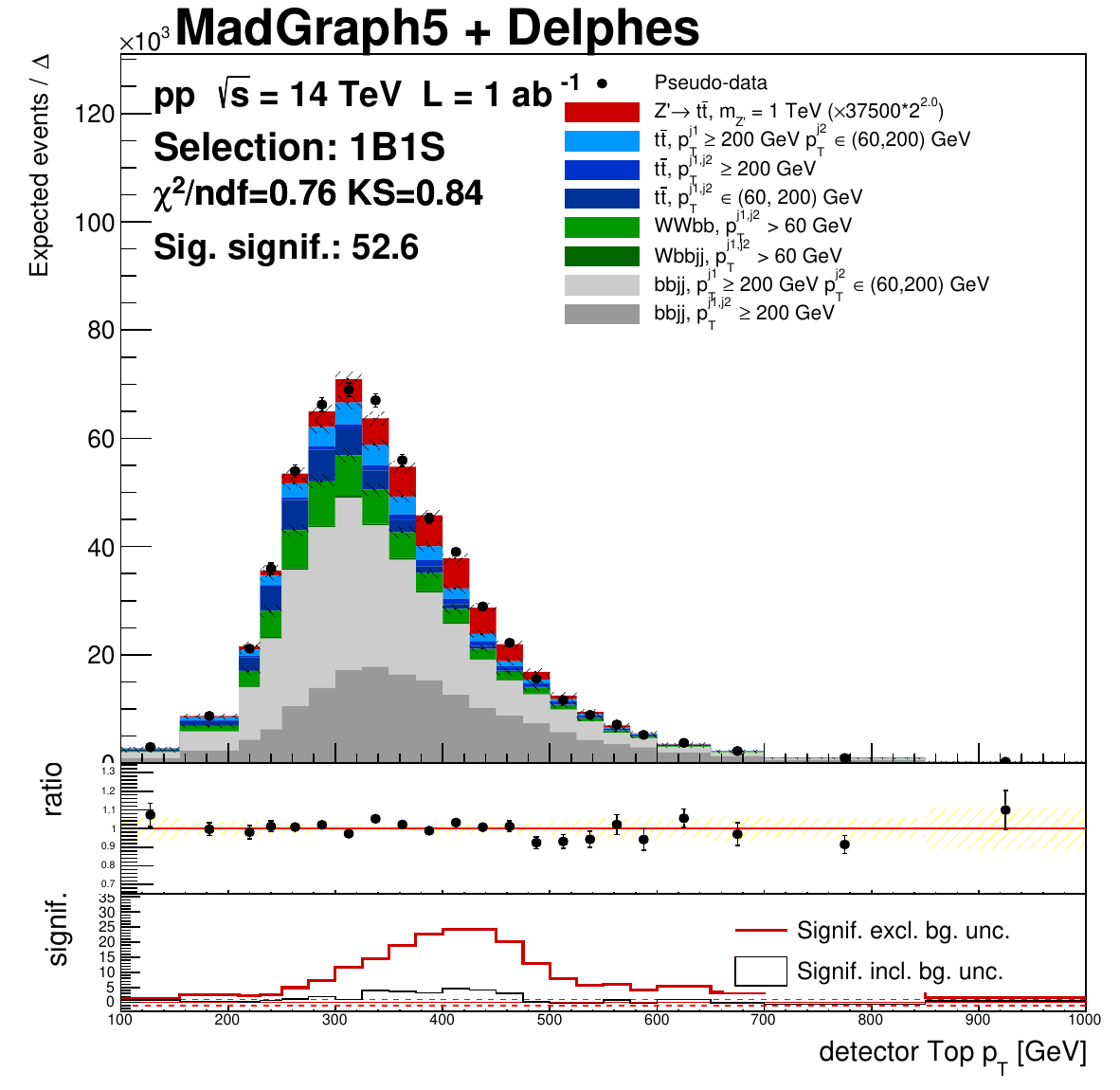} &
\includegraphics[width=0.33\textwidth]{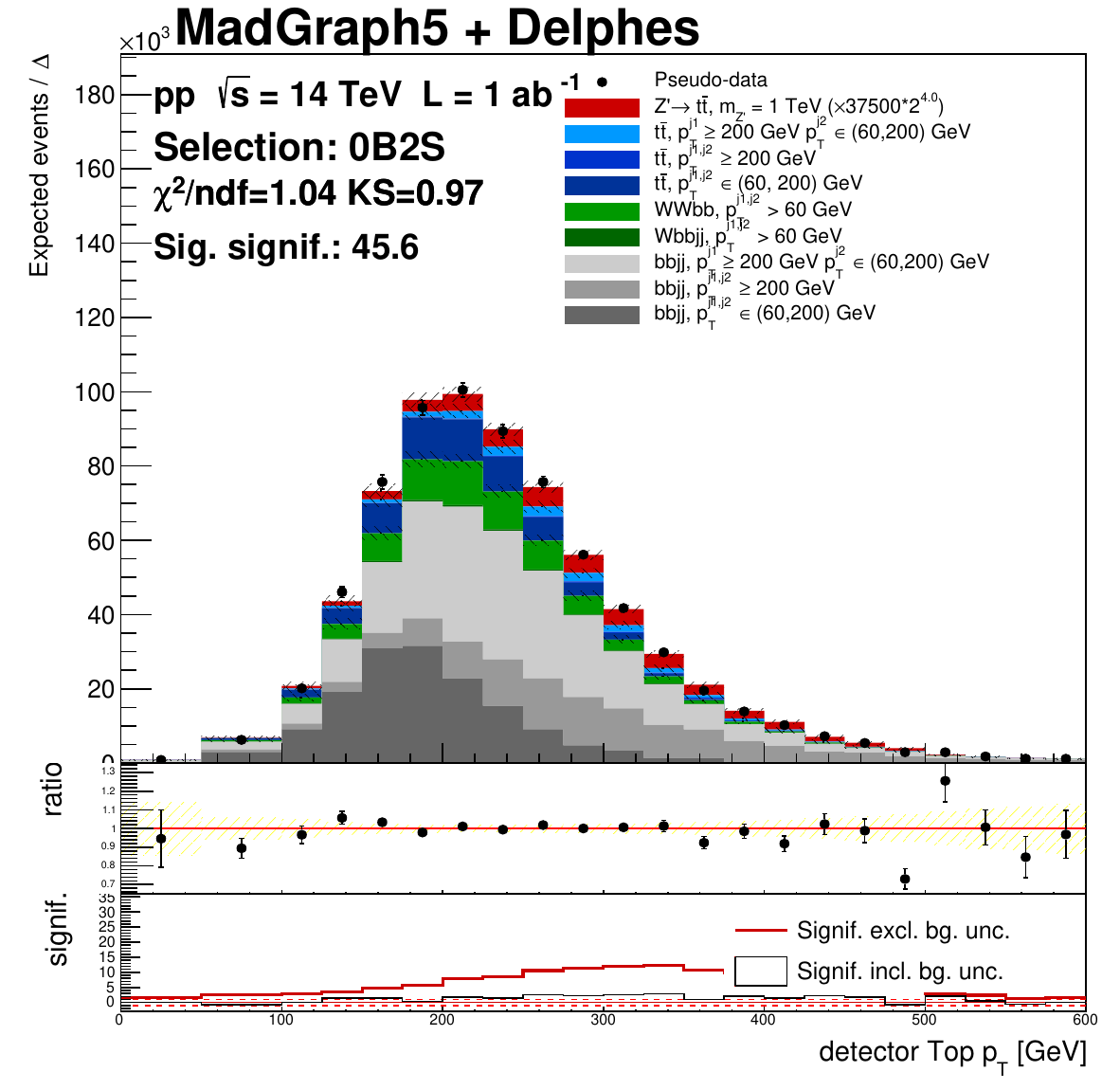} \\
\includegraphics[width=0.33\textwidth]{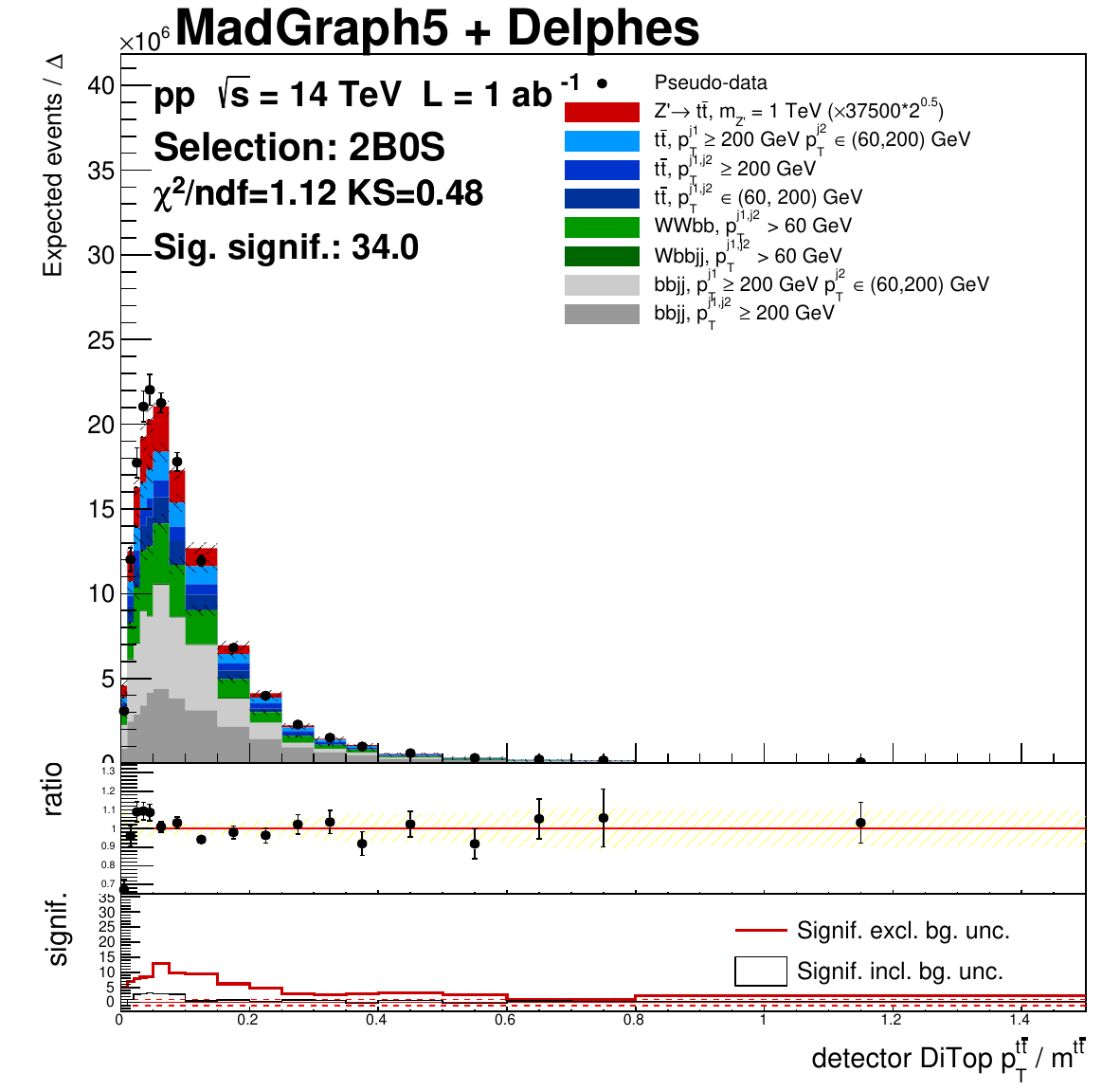} &
\includegraphics[width=0.33\textwidth]{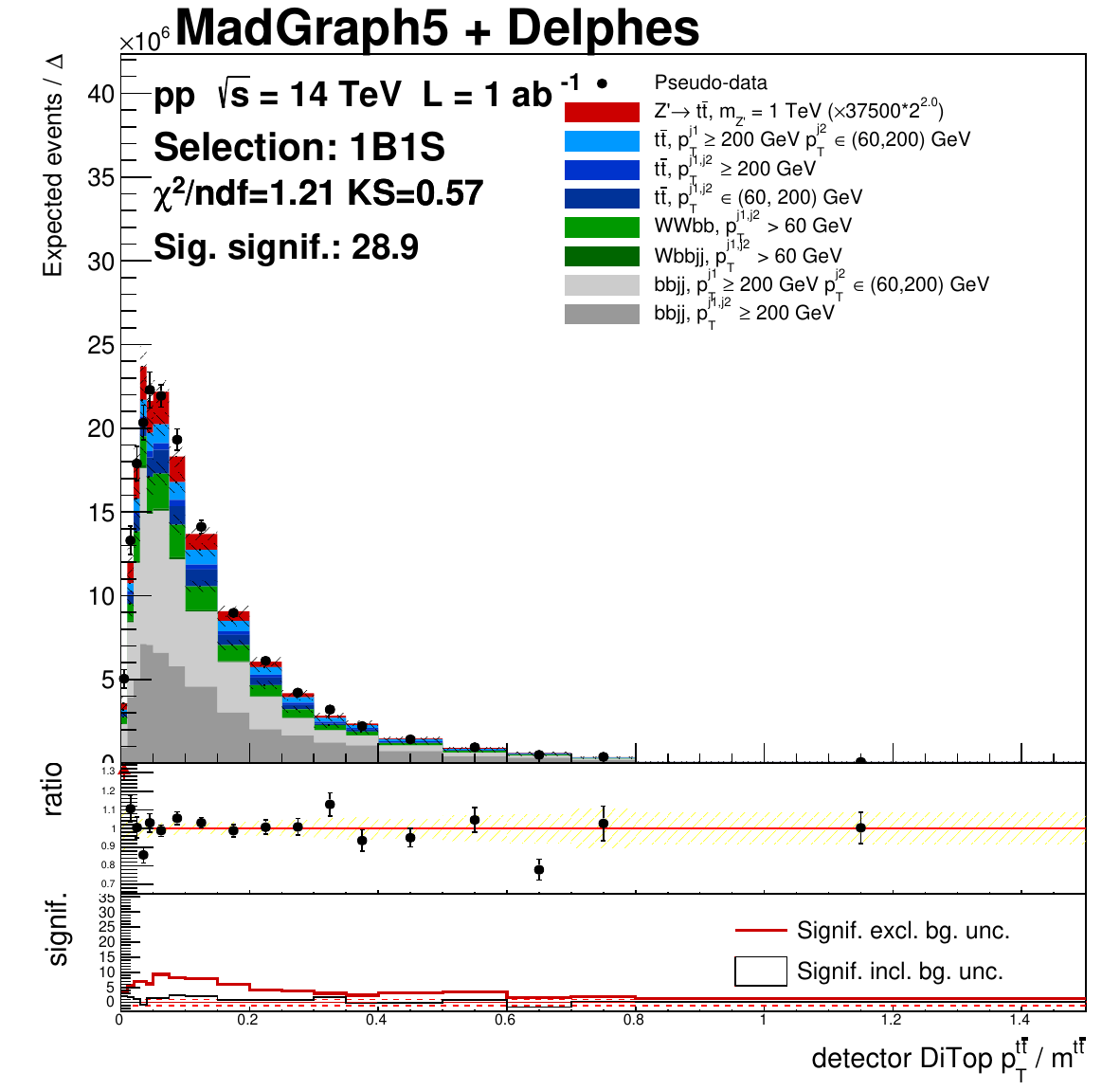} &
\includegraphics[width=0.33\textwidth]{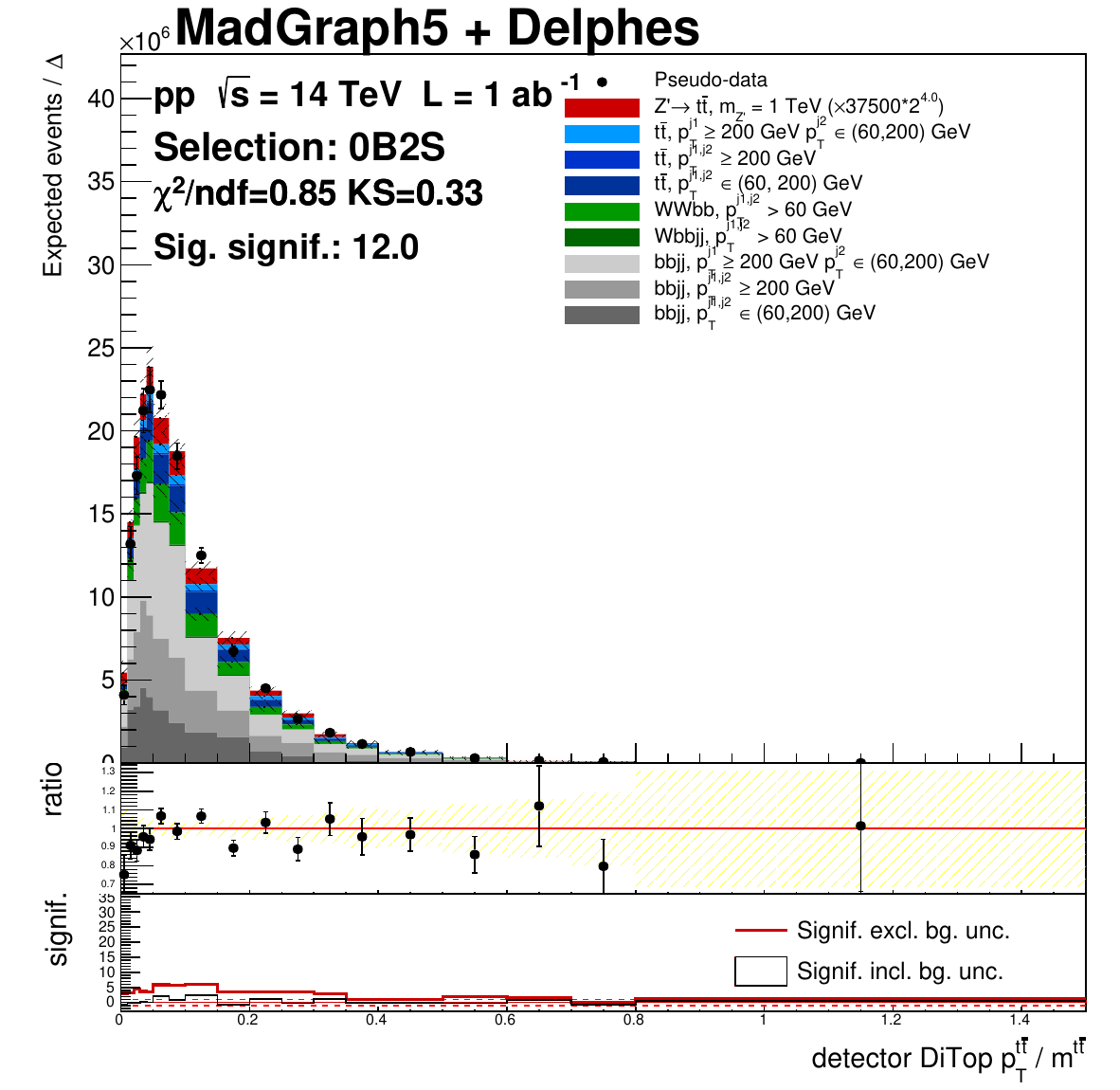} \\
2B0S  &
1B1S  &
0B2S  \\
\end{tabular}
\caption{ Stacked $\pt{}$ of the reconstructed top quarks (top) and the $\ttbar$ pair $\pt$ relative to the pair mass (bottom) at the \Delphes{} detector level for the scaled $Z'$ model with $m_{Z'}=1\,$TeV with both the pseudo-data/prediction ratio and the signal significance shown in lower pads, for the boosted-boosted (2B0S, left), boosted-semiboosted (1B1S, middle) and semiboosted-semiboosted (0B2S, right) topologies.
 The $\chi^2$ and Kolmogorov-Smirnov (KS) test statistics between the pseudo-data and predictions as provided by ROOT~\protect\cite{Antcheva:2009zz} and the total signal significance are also indicated.
}
\label{fig_stack_Detector_model_zp_compact_RS}
\end{center}
\end{figure*}
\begin{figure*}
\begin{center}
\begin{tabular}{ccc}
\includegraphics[width=0.33\textwidth]{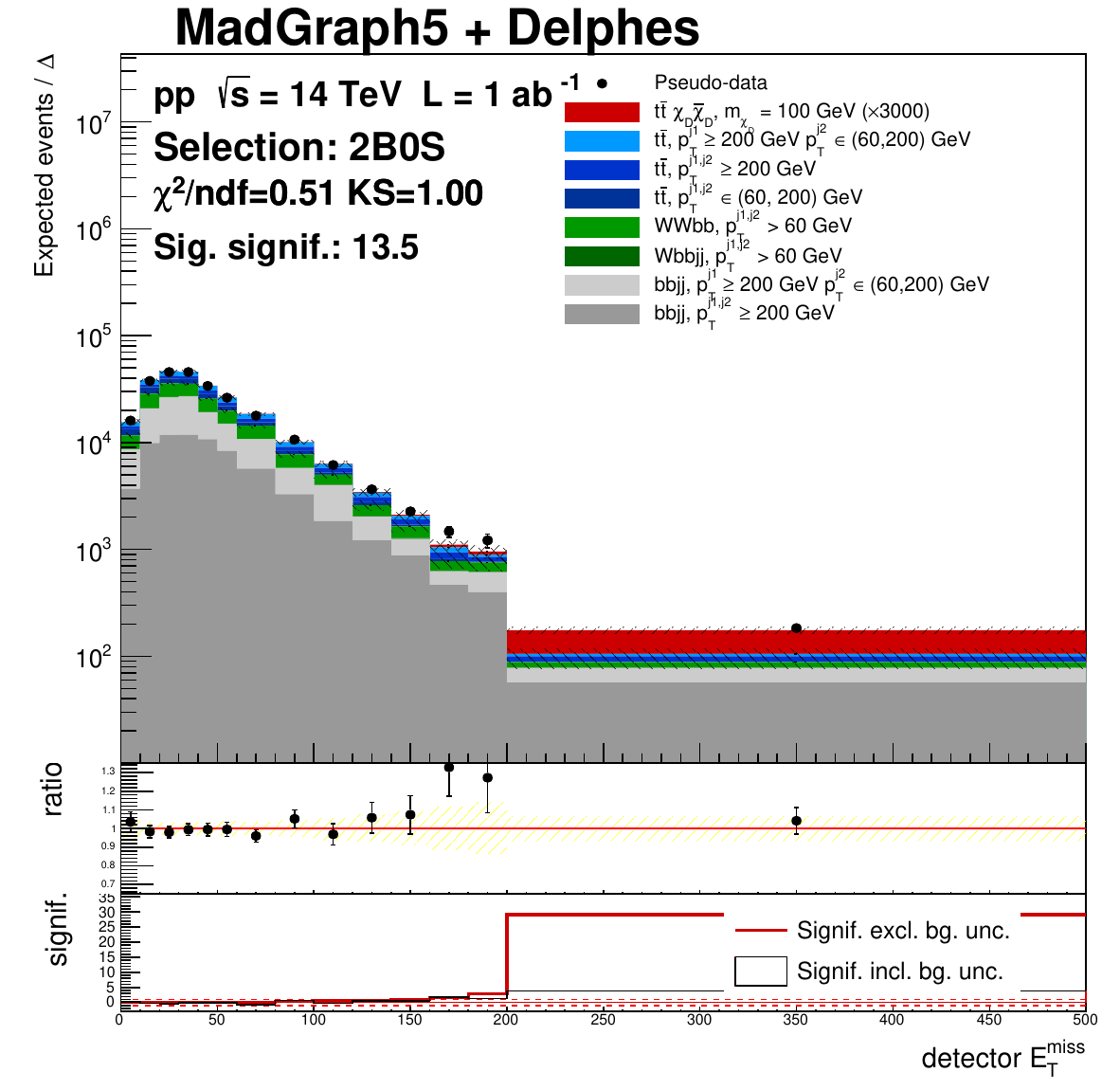} &
\includegraphics[width=0.33\textwidth]{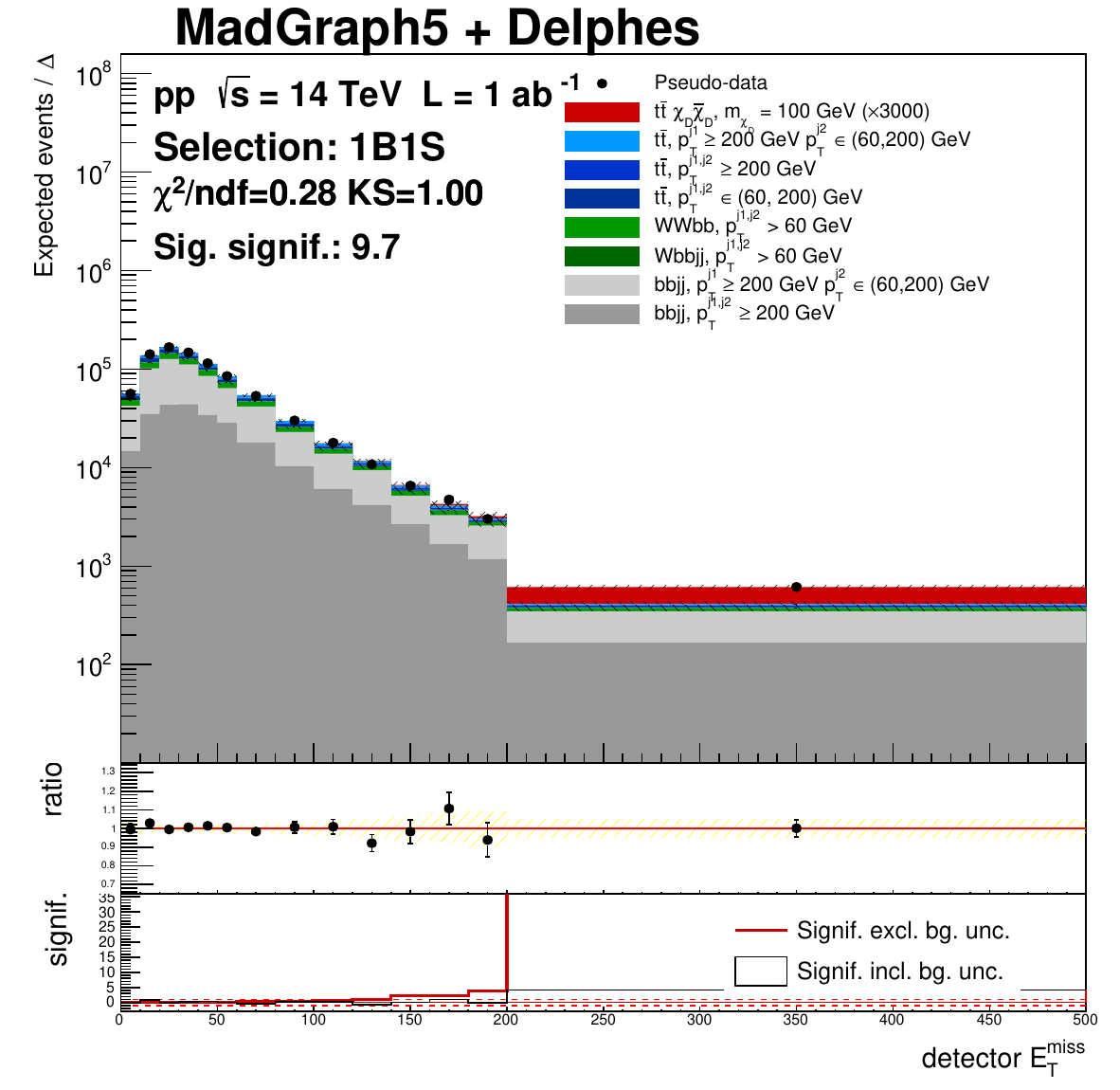} &
\includegraphics[width=0.33\textwidth]{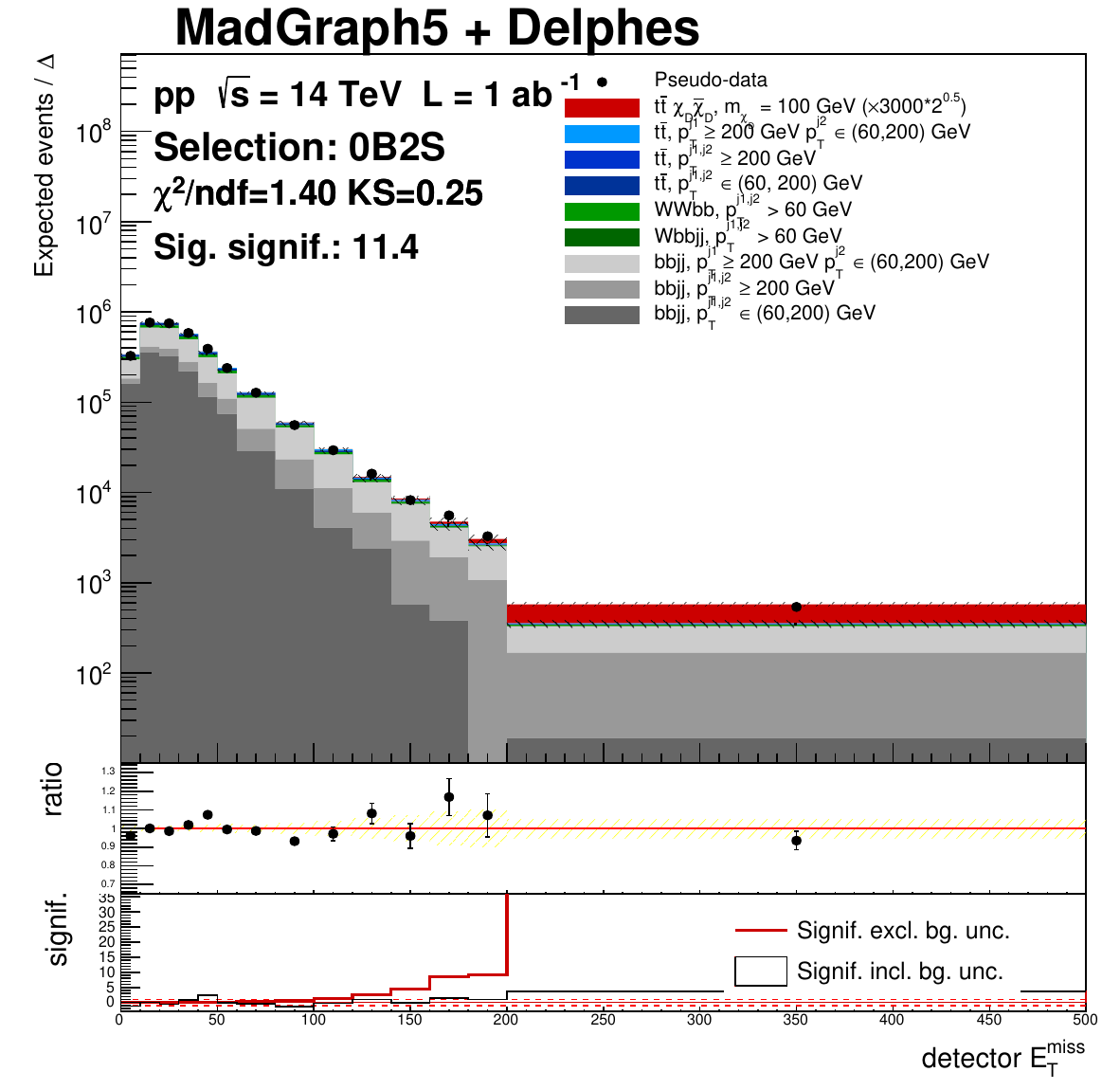} \\
2B0S  &
1B1S  &
0B2S  \\
\end{tabular}
\caption{ Stacked absolute value of the missing transverse energy at the \Delphes{} detector level for the scaled $y_0 \rightarrow \chi_D \bar{\chi}_{D},\, m_{\chi_D} = 100\,$ GeV model with both the pseudo-data/prediction ratio and the signal significance shown in lower pads, for the boosted-boosted (2B0S, left), boosted-semiboosted (1B1S, middle) and semiboosted-semiboosted (0B2S, right) topologies. The $\chi^2$ and Kolmogorov-Smirnov (KS) test statistics between the pseudo-data and predictions as provided by ROOT~\protect\cite{Antcheva:2009zz} and the total signal significance are also indicated.
}
\label{fig_stack_Detector_models_Met_compact_RS}
\end{center}
\end{figure*}

Here we present examples of stacked samples forming predictions compared to the pseudo-data from the same but statistically independent samples. One can observe peaked $\mtt$ and the $\Chittbar$ variable distributions of signal for the $y_0$ model (Figure~\ref{fig_stack_Detector_model_y0_compact_RS}), a~broader enhancement in terms of the top quark transverse momentum \ptt{} and $\pttt / \mtt$ for the $Z'$ model (Figure~\ref{fig_stack_Detector_model_zp_compact_RS}), together with the large signal in the tail of the $\Etmiss$ distribution for the DM pair production (Figure~\ref{fig_stack_Detector_models_Met_compact_RS}).

\subsection{Statistically varied replicas}

The technique of bootstrapping~\cite{Bohm:389738} is employed in order to construct statistically varied pseudo-independent versions (replicas) of 1D as well as 2D histograms, by repeatedly filling 100 replicas by events weighted by random weights drawn from the Poisson distribution of mean of $1$, \emph{i.e.} as $w \sim \mathrm{Poisson}(1)$.
\noindent This ensures that every event is used in each replica once on average and enables one to construct statistically correlated replicas across arbitrary histograms (projections) of any dimension over the events. The main aim is to evaluate signal excess over the replicas, thus obtaining a~more robust estimator of the signal significance for each 1D or 2D spectrum in each topology and model.

\subsection{\BumpHunter{} algorithm extension}

The binned 1D significance of a~signal over the total $\ttbar$ and non-$\ttbar$ background is evaluated at the detector level for scalar and vector resonant production of the \ttbar{} pair as well as for the model of paired DM and $\ttbar{}$ production as $N_\mathrm{sig} / \sqrt{\mathrm{Var}_\mathrm{\,bg} + \mathrm{Var}_\mathrm{\,sig}}$.  This is shown in Figures~\ref{fig_stack_Detector_model_y0_compact_RS}--\ref{fig_stack_Detector_models_Met_compact_RS} as the red (black) curve in the lowest pad without (with) taking into account the uncertainties of the background.

In order to quantify the signal excess and find the most sensitive bin area, the 1D \BumpHunter{} (BH) algorithm~\cite{Choudalakis:2011qn} has been employed. It finds the 1D window of a~predefined size in which the probability of the data compatibility with the background-only hypothesis ($p$-value, $p_\mathrm{val}$) is the smallest, constructing a~tests statistics  $t \equiv - \log p_\mathrm{val}^\mathrm{min}$ which grows with the discrepancy. The $p$-value for the given area can be computed in detail using the normalized incomplete gamma function as $\Gamma(d,b)$ for the case when the number of data events $d$ is larger than the expected number of background events $b$, or as $1 - \Gamma(d+1,b)$ otherwise, with
\begin{eqnarray}
  \Gamma(x,y) \equiv \frac{1}{\Gamma(x)} \int\limits_{0}^{y} \zeta^{x-1} \mathrm{e}^{-\zeta} \,\mathrm{d}\zeta\,,\quad \Gamma(x) \equiv \int\limits_{0}^{\infty} \zeta^{x-1} \mathrm{e}^{-\zeta} \,\mathrm{d}\zeta \,.
  \end{eqnarray}
The function appears in a~useful replacement for the sum of Poisson weight factors determining the probability of observing $d$ events or more as the sum of Poisson terms for a~distribution with mean $b$ as \footnote{There seems to be a~typesetting mistake in Eq.~18 in~\cite{Choudalakis:2011qn} which should read $\sum\limits_{n=d}^{\infty} \frac{b^n}{n!}\mathrm{e}^{-b}$.}
\begin{eqnarray}
  \sum\limits_{k=d}^{\infty} \frac{b^k}{k!}\mathrm{e}^{-b} =  \Gamma(d,b) \,.
\end{eqnarray}

In our adaptation of the 1D \BumpHunter{} algorithm, the following constrains on the 1D area of interest are imposed, with the algorithm details outlined as follows:
\begin{itemize}
  \item The initial 1D window width is 2 bins, starting with the window placed at the left of the spectrum.
  \item Compute the \BumpHunter{} test statistics~$t$.
  \item Add the left- or right-neighbouring bin to the interval based on larger resulting~$t$.
  \item Continue as long as the number of bins $N$ of the interval under test in 1D fulfills $ N \leq \frac23 \,  n^\mathrm{bins}$ where $ n^\mathrm{bins}$ is the numbers of non-empty bins and as long as $t$ grows.
  \item Start again with a~shifted search window by one bin from left to right.
  \item Try all such resulting 1D intervals, return the one with the best BH score.
\end{itemize}

We propose a~natural 2D extension of the \BumpHunter{} algorithm following these considerations:
\begin{itemize}
  \item The initial 2D window width is $2\times 2$ bins, placing the window at the bottom left of the 2D histogram.
  \item Compute the \BumpHunter{} test statistics~$t$.
  \item Add a~directly neighbouring bin to the current 2D area (excluding diagonal neighbours) which leads to the largest positive change in~$t$.
  \item Continue as long as $t$ grows and while the number of bins $N$ of the 2D area under test with its widths in $x$ and $y$ ($w_x$ and $w_y$) fulfills $N < n^\mathrm{bins} \, \land \,   w_x \leq \frac23 \,  n^\mathrm{bins}_x  \, \land \,   w_y \leq \frac23 \,  n^\mathrm{bins}_y$, where $n^\mathrm{bins}$ is the number of non-empty bins and $ n^\mathrm{bins}_{x,y}$ are the numbers of non-empty bins of the 1D histogram projections along the $x,y$ axes.
  \item Start again shifting the search window gradually by one bin in $x$ and $y$ directions from left to right.
  \item Try all such 2D areas, return the one with the best BH score.
\end{itemize}

For presentational purposes we further define the BH score as a~logarithm of its test statistics, noting that a~small shift in the score means a~large shift in the actual \BumpHunter{} test statistics.


\subsection{Variables performance comparison}
\label{subsect:var_cmp}

Histograms of the 1D and 2D \BumpHunter{} (BH) scores $\log t$ as well as the 2D histograms themselves for selected variables are shown in~Figures~\ref{fig_bestVars_y0}--\ref{fig_bestVars_xdxd}, with the 1D and 2D best BH areas also indicated.
The results are obtained by repeating the \BumpHunter{} procedure over the 100 statistically pseudo-independent replicas of the 1D and 2D spectra using the stacked backgrounds (including the $\ttbar{}$ samples) and the different signal models in the pseudo-data.

The best discriminating variables in terms of the \BumpHunter{} score for the three benchmark models are shown in~Tables~\ref{tab:bestBHscores_y0}--\ref{tab:bestBHscores_xdxd}.
Variables are compared to the best performing one in terms of the significance of the difference in their scores, taking into account both the variance of the best variable score as well as that of the variable-under-test. Variables scoring within 1-$\sigma$ are included in the tables with those scoring within 1-$\sigma$ marked in bold.
In the case of the  $\ttbar{} + \chi_D \bar{\chi}_D$ signal model, although with only a small overall signal-to-background ratio, the best performing variable, $H_\mathrm{T}^{j} + E^\mathrm{miss}_\mathrm{T}$ vs. $H_\mathrm{T}^{j}$, performs so outstandingly that the $\sigma$ range has been extended to allow for more variables appear in the corresponding Table~\ref{tab:bestBHscores_xdxd}. Also, for this model and in the 1B1S topology, this variable saturates the score, resulting in its zero variance, with a similar behaviour in the other two topologies (scores having about a half of the standard deviation compared to other models and variables) which was another motivation for increasing the significance width to judge more variables performance in this case.
It has been observed that a~single 1D variable never scores better than 2D variables.

\begin{figure*}
  \begin{center}
    \includegraphics[width=0.99\textwidth]{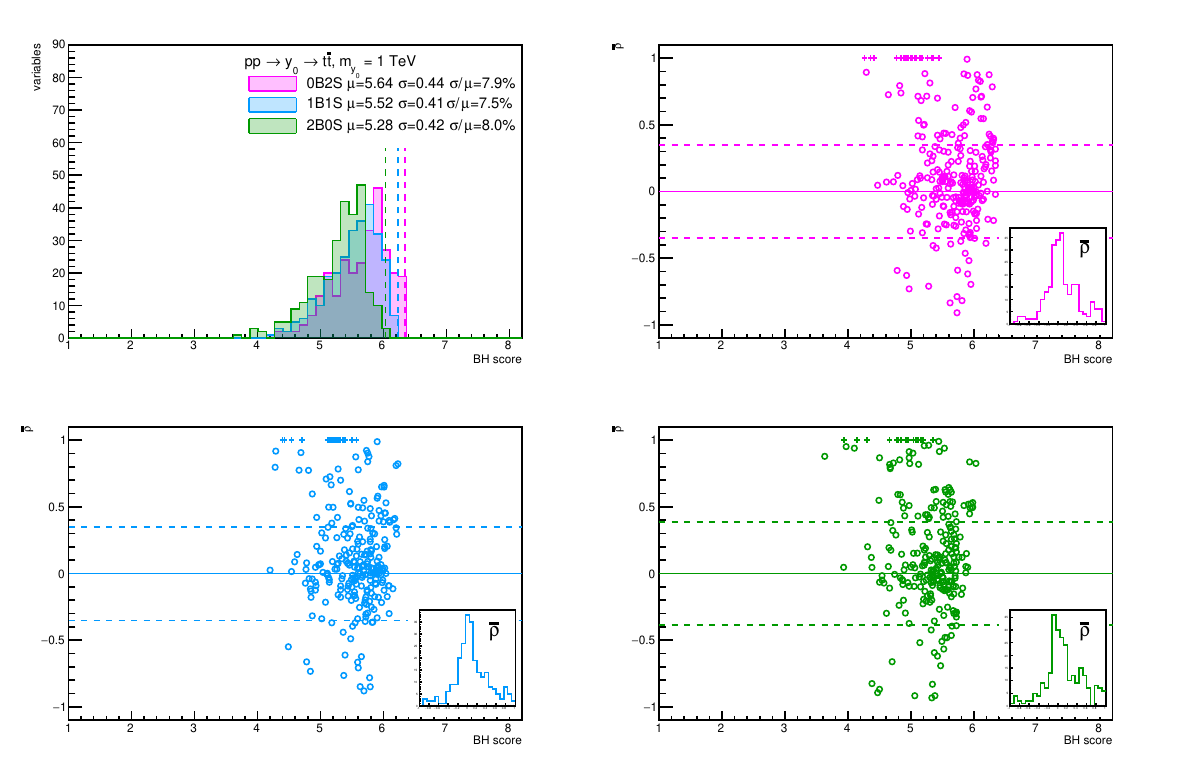} \\
     \caption{Mean BH scores over the variables for the scalar $y_0$ model with $m_{y_0} = 1000\,$GeV (top left). The vertical dashed lines indicate the BH score for the best variable. In the 2D plots with the BH score on the $x$-axis and the correlation coefficient between two variables on the $y$-axis, the circle markers show the correlation vs. the 2D BH results while the crosses represent the 1D BH scores plotted at $\bar{\rho} = 1$. Green: the 2B0S, blue: 1B1S and pink: 0B2S topologies. The insets show the correlations distribution and the dashed horizontal guiding lines are drawn along $ \pm \sigma_{\bar{\rho}}$, the correlations standard deviation over the 2D variables studied.}
\label{fig_bestVars_y0_mean}
\end{center}
\end{figure*}

\begin{figure*}
\begin{center}
    \includegraphics[width=0.99\textwidth]{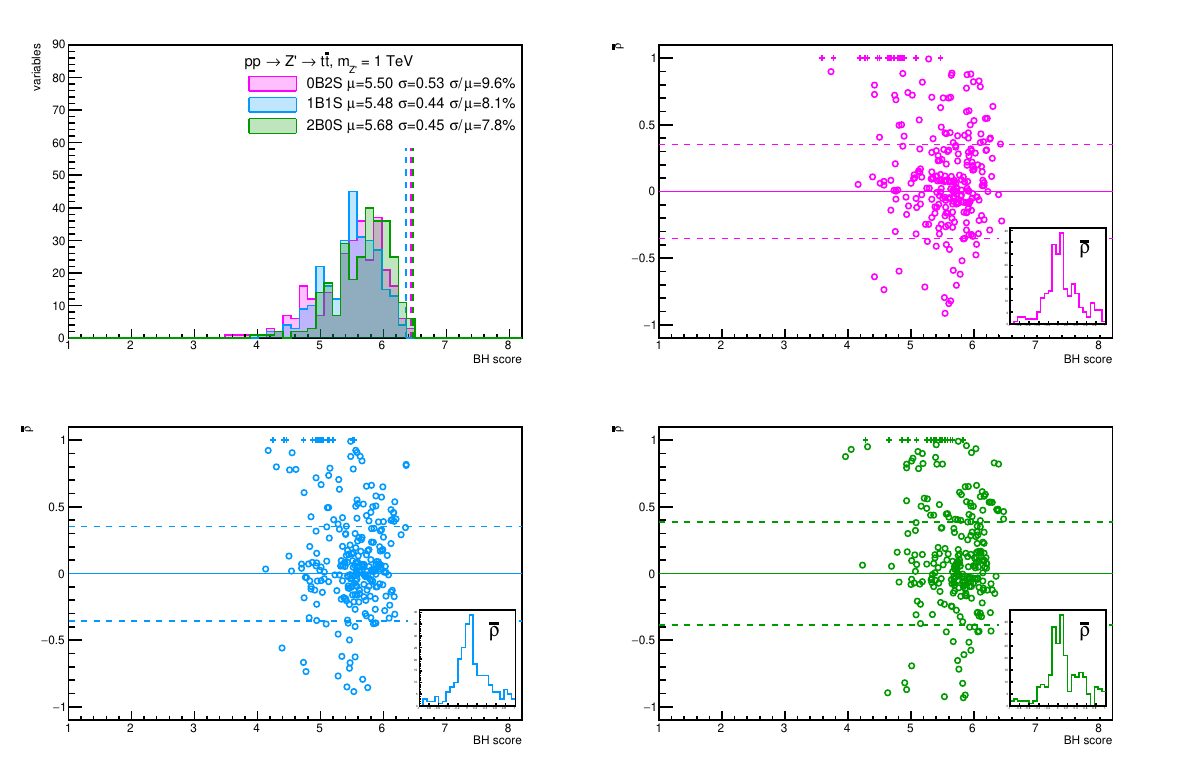} \\
     \caption{Mean BH scores over the variables for the vector $Z'$ model with $m_{Z'} = 1000\,$GeV (top left). The vertical dashed lines indicate the BH score for the best variable. In the 2D plots with the BH score on the $x$-axis and the correlation coefficient between two variables on the $y$-axis, the circle markers show the correlation vs. the 2D BH results while the crosses represent the 1D BH scores plotted at $\bar{\rho} = 1$. Green: the 2B0S, blue: 1B1S and pink: 0B2S topologies. The insets show the correlations distribution and the dashed horizontal guiding lines are drawn along $ \pm \sigma_{\bar{\rho}}$, the correlations standard deviation over the 2D variables studied.}
\label{fig_bestVars_zp_mean}
\end{center}
\end{figure*}

\begin{figure*}
\begin{center}
     \includegraphics[width=0.99\textwidth]{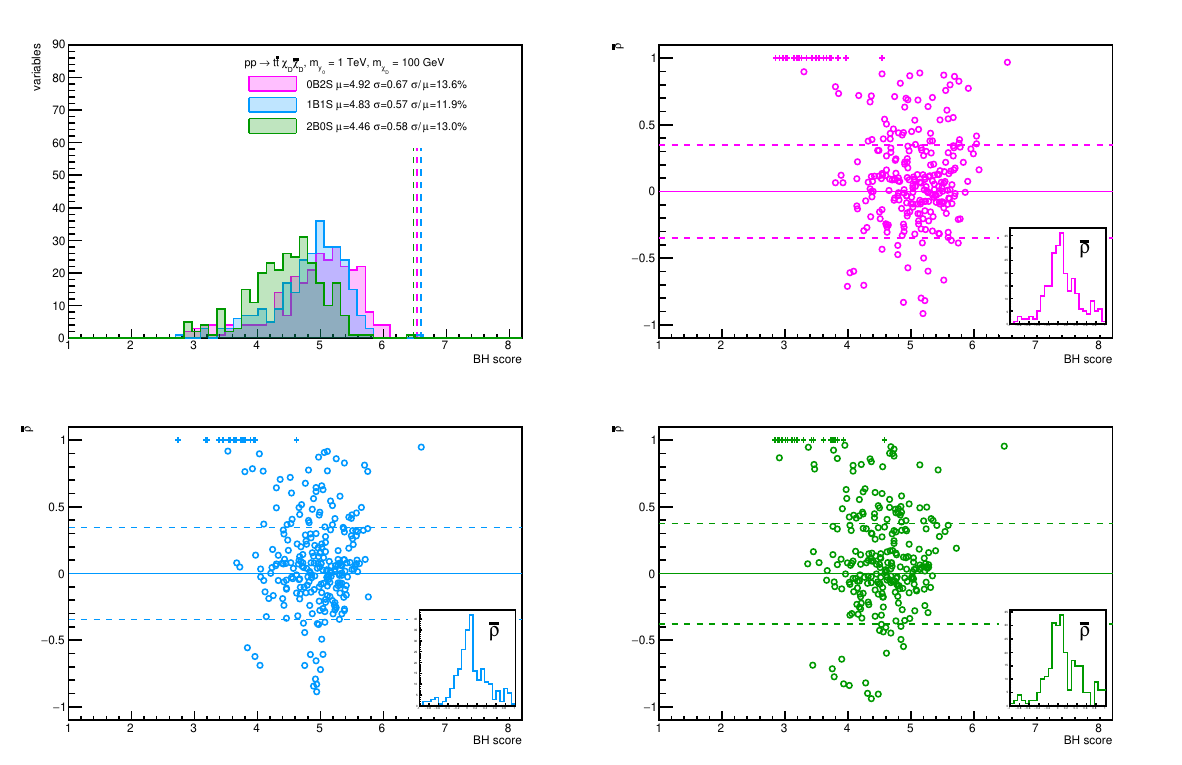}
 \caption{Mean BH scores over the variables for the associated production of $\ttbar{}$ with a~pair of DM particles with $m_{\chi_D} = 100\,$GeV (top left). The vertical dashed lines indicate the BH score for the best variable. In the 2D plots with the BH score on the $x$-axis and the correlation coefficient between two variables on the $y$-axis, the circle markers show the correlation vs. the 2D BH results while the crosses represent the 1D BH scores plotted at $\bar{\rho} = 1$. Green: the 2B0S, blue: 1B1S and pink: 0B2S topologies. The insets show the correlations distribution and the dashed horizontal guiding lines are drawn along $ \pm \sigma_{\bar{\rho}}$, the correlations standard deviation over the 2D variables studied.}
 \label{fig_bestVars_xdxd_mean}
\end{center}
\end{figure*}

Histograms of the score over all the variables are shown as 1D plots in~Figures~\ref{fig_bestVars_y0_mean}--\ref{fig_bestVars_xdxd_mean}. From within the 2D plots, the correlation coefficient for the 253 studied 2D spectra built from 23 kinematic variables plotted as function of the best BH score, one can observe a~denser population of points within approximately $|\bar{\rho}| \lesssim 0.2$, but with a~large spread. The $\bar{\rho}$ is the correlation coefficient averaged over 100 pseudo-experiments, computed using the $\ttbar$ control sample. One can conclude a~small correlation is not a~necessary condition for two variables to possibly provide a~stronger 2D signal significance than the individual 1D variables. The best variables are a~mixture of those with small, large as well as negative correlations, which are sometimes negligible for SM $\ttbar$ but can be largely negative for signal. There are also cases of only a~smaller positive correlation for signal.

Interestingly, the $\mtt$ variable is not the best standalone variable for the $y_0$ and $Z'$ models but it is often preferred when paired with another variable, \emph{e.g.} with $\HT$, $\Etmiss$, or also with jets sphericity or relative variables like  $\pttt / \mtt$. Another important observation is that the top quark $\pt$ does not seem to help in finding a~signal much, which may be regarded as good news to worries of possibly tuning new physics into physics models using this variable, while the transverse momentum of the $\ttbar$ system seems to be very sensitive (and harder to model in general). Other powerful variables turn out to be the sum of the large-$R$ jet masses and the invariant mass of the 4-vector sum of large-$R$ jets  $m^\mathrm{vis}_{\sum J}$, which has correlation of about $0.60$ to $\mttbar$. Sphericity and aplanarity variables also appear in the tables.

For the DM signal case, the best variable turns out to be $\Etmiss$ in combination, as a 2D variable, with $\HT + \Etmiss$. Although both variables are highly correlated ($\sim 1$ for the $\ttbar$ and $\sim 0.9$ for the signal samples), the 2D histogram actually helps to separate the signal by moving it off the diagonal, leading to a~large significance and a~large (2D) BH score. The signal scale factor for this model was chosen so that the 2D \BumpHunter{} significance for this variable is substantial, which effectively reduces the significances in all other variables to much smaller values.
It was found that besides the missing transverse energy, the best performing variables for the DM signal are similar to those for the $Z'$ model, \emph{e.g.} $\pttt / \mtt$, $\pttt / \sqrt{p_\mathrm{T}^{t1} \, p_\mathrm{T}^{t2}}$, $H_\mathrm{T}^j$. The $\pout$ variable, which has a~large correlation to $\pttt$ ($\rho=0.86$) but did not score prominently for the $y_0$ or $Z'$ models is also relevant for the DM signal case. One can thus think of the $\pout$ variable as a~useful tuning instead of a~search tool, or a~proxy to $\pttt$.


\begin{figure*}
\begin{center}
  \begin{tabular}{cccc}
    \input{bestfigs_2D_y0_1000GeV_2B0S}
   \end{tabular}
  \begin{tabular}{cccc}
    \input{bestfigs_2D_y0_1000GeV_1B1S}
   \end{tabular}
   \begin{tabular}{cccc}
     \input{bestfigs_2D_y0_1000GeV_0B2S}
   \end{tabular}
   \caption{Left: a histogram of 1D (orange, purple) and 2D (green) \BumpHunter{} scores over pseudo-experiments. Best variables for the scalar $y_0$ model with $m_{y_0} = 1000\,$GeV in the 2B0S (top), 1B1S (middle) and 0B2S (bottom) topologies.
Right: the green boxes indicate the best 2D ($xy$) \BumpHunter{} area while the vertical purple and horizontal orange bars indicate the bin ranges of the best 1D \BumpHunter{} areas for variables on the $y$ and $x$ axes, respectively. The blue (red) boxes area is proportional to the number of total signal$+$background (signal) events.
   }
\label{fig_bestVars_y0}
\end{center}
\end{figure*}
\begin{figure*}
\begin{center}
  \begin{tabular}{cccc}
    \input{bestfigs_2D_zp_1000GeV_2B0S}
   \end{tabular}
  \begin{tabular}{cccc}
    \input{bestfigs_2D_zp_1000GeV_1B1S}
   \end{tabular}
   \begin{tabular}{cccc}
     \input{bestfigs_2D_zp_1000GeV_0B2S}
   \end{tabular}
   \caption{Left: a histogram of 1D (orange, purple) and 2D (green) \BumpHunter{} scores over pseudo-experiments. Best variables for the vector $Z'$ model with $m_{Z'} = 1000\,$GeV in the 2B0S (top), 1B1S (middle) and 0B2S (bottom) topologies.
   Right: the green boxes indicate the best 2D ($xy$) \BumpHunter{} area while the vertical purple and horizontal orange bars indicate the bin ranges of the best 1D \BumpHunter{} areas for variables on the $y$ and $x$ axes, respectively. The blue (red) boxes area is proportional to the number of total signal$+$background (signal) events.
}
\label{fig_bestVars_zp}
\end{center}
\end{figure*}
\begin{figure*}
\begin{center}
  \begin{tabular}{cccc}
    \input{bestfigs_2D_xdxdtt_100GeV_2B0S}
   \end{tabular}
  \begin{tabular}{cccc}
    \input{bestfigs_2D_xdxdtt_100GeV_1B1S}
   \end{tabular}
   \begin{tabular}{cccc}
     \input{bestfigs_2D_xdxdtt_100GeV_0B2S}
   \end{tabular}
   \caption{Left: a histogram of 1D (orange, purple) and 2D (green) \BumpHunter{} scores over pseudo-experiments. The best variable for the associated production of a $\ttbar{}$ pair with a pair of DM particles with $m_{\chi_D} = 100\,$GeV in the 1B1S topology (the picture is very similar in the other topologies).
   Right: the green boxes indicate the best 2D ($xy$) \BumpHunter{} area while the vertical purple and horizontal orange bars indicate the bin ranges of the best 1D \BumpHunter{} areas for variables on the $y$ and $x$ axes, respectively. The blue (red) boxes area is proportional to the number of total signal$+$background (signal) events.
}
\label{fig_bestVars_xdxd}
\end{center}
\end{figure*}

\clearpage
\clearpage

\begin{table*}[p]
  \centering
\resizebox*{0.31\textheight}{!}{
     \begin{tabular}{l|lll}
\hline 
variable & 2B0S & 1B1S & 0B2S \\ \hline 
  $E^\mathrm{miss}_\mathrm{T}$ vs. $H_\mathrm{T}^{j} + E^\mathrm{miss}_\mathrm{T}$ &    &   ${6.0} \pm 0.2$  &   $\mathbf{6.24} \pm 0.23$  \\ 
  $E^\mathrm{miss}_\mathrm{T}$ vs. $H_\mathrm{T}^{j}$                    &   ${5.78} \pm 0.22$  &   ${6.02} \pm 0.21$  &   $\mathbf{6.34} \pm 0.21$  \\ 
  $E^\mathrm{miss}_\mathrm{T}$ vs. $\sum m^{J}$                          &    &    &   ${6.14} \pm 0.2$  \\ 
  $E^\mathrm{miss}_\mathrm{T}$ vs. $m^{t\bar{t}}$                        &   ${5.76} \pm 0.23$  &   ${5.93} \pm 0.22$  &   ${6.18} \pm 0.2$  \\ 
  $E^\mathrm{miss}_\mathrm{T}$ vs. jets Sphericity                       &    &    &   ${6.11} \pm 0.2$  \\ 
  $H_\mathrm{T}^{j} + E^\mathrm{miss}_\mathrm{T}$ vs. $H_\mathrm{T}^{j}$ &    &   ${5.9} \pm 0.23$  &    \\ 
  $H_\mathrm{T}^{j} + E^\mathrm{miss}_\mathrm{T}$ vs. $R^{t2,t1}$        &   ${5.72} \pm 0.23$  &    &    \\ 
  $H_\mathrm{T}^{j} + E^\mathrm{miss}_\mathrm{T}$ vs. $\chi^{t\bar{t}}$  &    &   ${5.92} \pm 0.24$  &    \\ 
  $H_\mathrm{T}^{j} + E^\mathrm{miss}_\mathrm{T}$ vs. $m^{t\bar{t}}$     &   $\mathbf{5.97} \pm 0.18$  &   ${6.03} \pm 0.23$  &   ${6.15} \pm 0.21$  \\ 
  $H_\mathrm{T}^{j} + E^\mathrm{miss}_\mathrm{T}$ vs. $p_\mathrm{T}^{t\bar{t}}$ &    &   ${5.91} \pm 0.22$  &    \\ 
  $H_\mathrm{T}^{j} + E^\mathrm{miss}_\mathrm{T}$ vs. $y_\mathrm{boost}^{t\bar{t}}$ &    &   $\mathbf{6.09} \pm 0.24$  &    \\ 
  $H_\mathrm{T}^{j} + E^\mathrm{miss}_\mathrm{T}$ vs. jets Sphericity    &   $\mathbf{5.9} \pm 0.24$  &   ${5.99} \pm 0.24$  &    \\ 
  $H_\mathrm{T}^{j}$ vs. $\chi^{t\bar{t}}$                               &    &   ${5.93} \pm 0.24$  &    \\ 
  $H_\mathrm{T}^{j}$ vs. $\delta^{t\bar{t}}$                             &    &   ${5.93} \pm 0.27$  &    \\ 
  $H_\mathrm{T}^{j}$ vs. $m^{t\bar{t}}$                                  &   $\mathbf{5.98} \pm 0.19$  &   ${6.02} \pm 0.23$  &   ${6.11} \pm 0.23$  \\ 
  $H_\mathrm{T}^{j}$ vs. $y_\mathrm{boost}^{t\bar{t}}$                   &    &   $\mathbf{6.08} \pm 0.23$  &    \\ 
  $H_\mathrm{T}^{j}$ vs. jets Aplanarity                                 &   ${5.72} \pm 0.22$  &    &    \\ 
  $H_\mathrm{T}^{j}$ vs. jets Sphericity                                 &   $\mathbf{5.87} \pm 0.24$  &   ${5.99} \pm 0.23$  &   ${6.08} \pm 0.22$  \\ 
  $R^{t2,t1}$ vs. $m^{t\bar{t}}$                                         &    &    &   ${6.13} \pm 0.22$  \\ 
  $R^{t2,t1}$ vs. $y_\mathrm{boost}^{t\bar{t}}$                          &    &   ${5.92} \pm 0.19$  &    \\ 
  $\delta^{t\bar{t}}$ vs. $m^{t\bar{t}}$                                 &    &    &   ${6.01} \pm 0.27$  \\ 
  $\sum m^{J}$ vs. $H_\mathrm{T}^{j} + E^\mathrm{miss}_\mathrm{T}$       &    &   ${5.97} \pm 0.23$  &   ${6.11} \pm 0.22$  \\ 
  $\sum m^{J}$ vs. $H_\mathrm{T}^{j}$                                    &    &   ${6.01} \pm 0.25$  &   ${6.12} \pm 0.24$  \\ 
  $\sum m^{J}$ vs. $m^{t\bar{t}}$                                        &   ${5.72} \pm 0.21$  &    &    \\ 
  $m^{t\bar{t}} / \sqrt{p_\mathrm{T}^{t1} p_\mathrm{T}^{t2}}$ vs. $H_\mathrm{T}^{j} + E^\mathrm{miss}_\mathrm{T}$ &    &   ${5.91} \pm 0.22$  &    \\ 
  $m^{t\bar{t}} / \sqrt{p_\mathrm{T}^{t1} p_\mathrm{T}^{t2}}$ vs. $H_\mathrm{T}^{j}$ &    &   ${5.92} \pm 0.19$  &    \\ 
  $m^{t\bar{t}}$ vs. $\Delta\phi^{t\bar{t}}$                             &    &    &   ${6.07} \pm 0.21$  \\ 
  $m_{\sum J}^\mathrm{vis}$ vs. $E^\mathrm{miss}_\mathrm{T}$             &   ${5.72} \pm 0.23$  &   ${6.06} \pm 0.2$  &   ${6.13} \pm 0.2$  \\ 
  $m_{\sum J}^\mathrm{vis}$ vs. $H_\mathrm{T}^{j} + E^\mathrm{miss}_\mathrm{T}$ &   $\textcolor{red}{\mathbf{6.03}} \pm 0.23$  &   $\mathbf{6.2} \pm 0.24$  &   ${6.1} \pm 0.22$  \\ 
  $m_{\sum J}^\mathrm{vis}$ vs. $H_\mathrm{T}^{j}$                       &   $\mathbf{5.93} \pm 0.24$  &   $\textcolor{red}{\mathbf{6.23}} \pm 0.25$  &   ${6.07} \pm 0.28$  \\ 
  $m_{\sum J}^\mathrm{vis}$ vs. $R^{t2,t1}$                              &   ${5.72} \pm 0.24$  &    &    \\ 
  $m_{\sum J}^\mathrm{vis}$ vs. $\Delta\phi^{t\bar{t}}$                  &    &   ${5.9} \pm 0.23$  &    \\ 
  $m_{\sum J}^\mathrm{vis}$ vs. $\delta^{t\bar{t}}$                      &    &   ${6.02} \pm 0.26$  &    \\ 
  $m_{\sum J}^\mathrm{vis}$ vs. $\sum m^{J}$                             &    &   ${6.01} \pm 0.22$  &    \\ 
  $m_{\sum J}^\mathrm{vis}$ vs. $m^{t\bar{t}}$                           &    &   ${6.04} \pm 0.22$  &   $\mathbf{6.21} \pm 0.22$  \\ 
  $m_{\sum J}^\mathrm{vis}$ vs. $p_\mathrm{T}^{t\bar{t}}$                &    &   ${5.93} \pm 0.23$  &    \\ 
  $m_{\sum J}^\mathrm{vis}$ vs. $y_\mathrm{boost}^{t\bar{t}}$            &    &   $\mathbf{6.15} \pm 0.21$  &    \\ 
  $m_{\sum J}^\mathrm{vis}$ vs. jets Aplanarity                          &   ${5.87} \pm 0.2$  &    &    \\ 
  $m_{\sum J}^\mathrm{vis}$ vs. jets Sphericity                          &   ${5.82} \pm 0.23$  &   ${6.01} \pm 0.19$  &    \\ 
  $p_\mathrm{T}^{t\bar{t}} / \sqrt{p_\mathrm{T}^{t1} p_\mathrm{T}^{t2}}$ vs. $E^\mathrm{miss}_\mathrm{T}$ &    &   ${6.02} \pm 0.17$  &   $\mathbf{6.34} \pm 0.2$  \\ 
  $p_\mathrm{T}^{t\bar{t}} / \sqrt{p_\mathrm{T}^{t1} p_\mathrm{T}^{t2}}$ vs. $H_\mathrm{T}^{j} + E^\mathrm{miss}_\mathrm{T}$ &   $\mathbf{5.91} \pm 0.25$  &   $\mathbf{6.09} \pm 0.27$  &   $\mathbf{6.31} \pm 0.18$  \\ 
  $p_\mathrm{T}^{t\bar{t}} / \sqrt{p_\mathrm{T}^{t1} p_\mathrm{T}^{t2}}$ vs. $H_\mathrm{T}^{j}$ &   ${5.84} \pm 0.27$  &   $\mathbf{6.1} \pm 0.27$  &   $\mathbf{6.31} \pm 0.2$  \\ 
  $p_\mathrm{T}^{t\bar{t}} / \sqrt{p_\mathrm{T}^{t1} p_\mathrm{T}^{t2}}$ vs. $\delta^{t\bar{t}}$ &    &    &   ${6.15} \pm 0.27$  \\ 
  $p_\mathrm{T}^{t\bar{t}} / \sqrt{p_\mathrm{T}^{t1} p_\mathrm{T}^{t2}}$ vs. $\sum m^{J}$ &    &   ${5.96} \pm 0.21$  &   $\mathbf{6.26} \pm 0.19$  \\ 
  $p_\mathrm{T}^{t\bar{t}} / \sqrt{p_\mathrm{T}^{t1} p_\mathrm{T}^{t2}}$ vs. $m^{t\bar{t}}$ &    &   ${6.06} \pm 0.19$  &   $\textcolor{red}{\mathbf{6.34}} \pm 0.19$  \\ 
  $p_\mathrm{T}^{t\bar{t}} / \sqrt{p_\mathrm{T}^{t1} p_\mathrm{T}^{t2}}$ vs. $m_{\sum J}^\mathrm{vis}$ &   $\mathbf{5.94} \pm 0.24$  &   $\mathbf{6.2} \pm 0.23$  &   $\mathbf{6.21} \pm 0.22$  \\ 
  $p_\mathrm{T}^{t\bar{t}} / \sqrt{p_\mathrm{T}^{t1} p_\mathrm{T}^{t2}}$ vs. $p_\mathrm{T}^{t\bar{t}} / m^{t\bar{t}}$ &    &    &   ${6.05} \pm 0.27$  \\ 
  $p_\mathrm{T}^{t\bar{t}} / \sqrt{p_\mathrm{T}^{t1} p_\mathrm{T}^{t2}}$ vs. $y_\mathrm{boost}^{t\bar{t}}$ &    &   ${5.99} \pm 0.22$  &    \\ 
  $p_\mathrm{T}^{t\bar{t}} / \sqrt{p_\mathrm{T}^{t1} p_\mathrm{T}^{t2}}$ vs. jets Sphericity &   ${5.81} \pm 0.26$  &   ${5.99} \pm 0.2$  &   $\mathbf{6.21} \pm 0.23$  \\ 
  $p_\mathrm{T}^{t\bar{t}} / m^{t\bar{t}}$ vs. $E^\mathrm{miss}_\mathrm{T}$ &    &   ${6.02} \pm 0.18$  &   $\mathbf{6.34} \pm 0.19$  \\ 
  $p_\mathrm{T}^{t\bar{t}} / m^{t\bar{t}}$ vs. $H_\mathrm{T}^{j} + E^\mathrm{miss}_\mathrm{T}$ &   $\mathbf{5.98} \pm 0.23$  &   $\mathbf{6.18} \pm 0.24$  &   $\mathbf{6.29} \pm 0.2$  \\ 
  $p_\mathrm{T}^{t\bar{t}} / m^{t\bar{t}}$ vs. $H_\mathrm{T}^{j}$        &   $\mathbf{5.95} \pm 0.24$  &   $\mathbf{6.16} \pm 0.23$  &   $\mathbf{6.3} \pm 0.21$  \\ 
  $p_\mathrm{T}^{t\bar{t}} / m^{t\bar{t}}$ vs. $\delta^{t\bar{t}}$       &    &    &   ${6.06} \pm 0.22$  \\ 
  $p_\mathrm{T}^{t\bar{t}} / m^{t\bar{t}}$ vs. $\sum m^{J}$              &    &   ${5.93} \pm 0.2$  &   ${6.18} \pm 0.22$  \\ 
  $p_\mathrm{T}^{t\bar{t}} / m^{t\bar{t}}$ vs. $m^{t\bar{t}}$            &   ${5.8} \pm 0.22$  &   $\mathbf{6.09} \pm 0.24$  &   $\mathbf{6.31} \pm 0.22$  \\ 
  $p_\mathrm{T}^{t\bar{t}} / m^{t\bar{t}}$ vs. $m_{\sum J}^\mathrm{vis}$ &   $\mathbf{5.93} \pm 0.22$  &   $\mathbf{6.21} \pm 0.23$  &   $\mathbf{6.26} \pm 0.23$  \\ 
  $p_\mathrm{T}^{t\bar{t}} / m^{t\bar{t}}$ vs. $y_\mathrm{boost}^{t\bar{t}}$ &    &   ${6.04} \pm 0.23$  &    \\ 
  $p_\mathrm{T}^{t\bar{t}} / m^{t\bar{t}}$ vs. jets Sphericity           &   $\mathbf{5.88} \pm 0.21$  &   ${5.96} \pm 0.23$  &   ${6.15} \pm 0.25$  \\ 
  $p_\mathrm{T}^{t\bar{t}}$ vs. $m^{t\bar{t}}$                           &    &   ${5.96} \pm 0.22$  &   ${6.11} \pm 0.22$  \\ 
  $y_\mathrm{boost}^{t\bar{t}}$ vs. $m^{t\bar{t}}$                       &    &   ${5.92} \pm 0.21$  &    \\ 
  $y_\mathrm{boost}^{t\bar{t}}$ vs. $p_\mathrm{T}^{t\bar{t}}$            &    &   ${5.98} \pm 0.2$  &    \\ 
  $|p_\mathrm{out}| / m^{t\bar{t}}$ vs. $E^\mathrm{miss}_\mathrm{T}$     &    &    &   $\mathbf{6.28} \pm 0.2$  \\ 
  $|p_\mathrm{out}| / m^{t\bar{t}}$ vs. $H_\mathrm{T}^{j} + E^\mathrm{miss}_\mathrm{T}$ &    &    &   $\mathbf{6.26} \pm 0.24$  \\ 
  $|p_\mathrm{out}| / m^{t\bar{t}}$ vs. $H_\mathrm{T}^{j}$               &    &    &   $\mathbf{6.27} \pm 0.24$  \\ 
  $|p_\mathrm{out}| / m^{t\bar{t}}$ vs. $R^{t2,t1}$                      &    &    &   ${6.19} \pm 0.21$  \\ 
  $|p_\mathrm{out}| / m^{t\bar{t}}$ vs. $\sum m^{J}$                     &    &    &   $\mathbf{6.27} \pm 0.19$  \\ 
  $|p_\mathrm{out}| / m^{t\bar{t}}$ vs. $m^{t\bar{t}} / \sqrt{p_\mathrm{T}^{t1} p_\mathrm{T}^{t2}}$ &    &   ${5.86} \pm 0.28$  &    \\ 
  $|p_\mathrm{out}| / m^{t\bar{t}}$ vs. $m^{t\bar{t}}$                   &    &    &   $\mathbf{6.22} \pm 0.2$  \\ 
  $|p_\mathrm{out}| / m^{t\bar{t}}$ vs. $m_{\sum J}^\mathrm{vis}$        &    &    &   ${6.19} \pm 0.21$  \\ 
  $|p_\mathrm{out}| / m^{t\bar{t}}$ vs. $p_\mathrm{T}^{t\bar{t}} / \sqrt{p_\mathrm{T}^{t1} p_\mathrm{T}^{t2}}$ &    &    &   $\mathbf{6.29} \pm 0.23$  \\ 
  $|p_\mathrm{out}| / m^{t\bar{t}}$ vs. $p_\mathrm{T}^{t\bar{t}} / m^{t\bar{t}}$ &    &    &   $\mathbf{6.24} \pm 0.24$  \\ 
  $|p_\mathrm{out}| / m^{t\bar{t}}$ vs. jets Aplanarity                  &    &    &   ${6.16} \pm 0.25$  \\ 
  $|p_\mathrm{out}| / m^{t\bar{t}}$ vs. jets Sphericity                  &    &   ${5.9} \pm 0.25$  &   $\mathbf{6.31} \pm 0.19$  \\ 
  jets Sphericity vs. $\chi^{t\bar{t}}$                                  &   ${5.72} \pm 0.21$  &    &    \\ 
  jets Sphericity vs. $m^{t\bar{t}}$                                     &    &   ${5.97} \pm 0.22$  &   ${6.04} \pm 0.24$  \\ 
     \hline
\end{tabular}

  }
  \caption{The best 1D and/or 2D \BumpHunter{} variables scores ($\log t$) and the statistical uncertainty over 100 pseudo-experiments for stacked backgrounds (including the $\ttbar{}$ samples) and the $y_0$ signal model with $m_{y_0} = 1\,$TeV in the pseudo-data over the topologies. Empty fields indicate that the variable did not score within $1\sigma$ to the best scoring variable in given topology (red) while those scoring within $0.5\sigma$ are marked in bold.}
  \label{tab:bestBHscores_y0}
\end{table*}

\clearpage


\begin{table*}[p]
    \centering
\resizebox*{0.48\textheight}{!}{
     \begin{tabular}{l|lll}
\hline 
variable & 2B0S & 1B1S & 0B2S \\ \hline 
  $E^\mathrm{miss}_\mathrm{T}$ vs. $m^{t\bar{t}}$                        &    &   ${6.06} \pm 0.23$  &    \\ 
  $H_\mathrm{T}^{j} + E^\mathrm{miss}_\mathrm{T}$ vs. $\chi^{t\bar{t}}$  &    &    &   ${6.13} \pm 0.26$  \\ 
  $H_\mathrm{T}^{j} + E^\mathrm{miss}_\mathrm{T}$ vs. $m^{t\bar{t}}$     &   ${6.29} \pm 0.17$  &   ${6.16} \pm 0.22$  &   ${6.18} \pm 0.25$  \\ 
  $H_\mathrm{T}^{j}$ vs. $\chi^{t\bar{t}}$                               &    &    &   ${6.16} \pm 0.26$  \\ 
  $H_\mathrm{T}^{j}$ vs. $m^{t\bar{t}}$                                  &   ${6.26} \pm 0.19$  &   ${6.16} \pm 0.21$  &   ${6.21} \pm 0.24$  \\ 
  $m^{t\bar{t}}$ vs. $\Delta\phi^{t\bar{t}}$                             &    &    &   ${6.15} \pm 0.26$  \\ 
  $m_{\sum J}^\mathrm{vis}$ vs. $E^\mathrm{miss}_\mathrm{T}$             &    &   ${6.12} \pm 0.21$  &    \\ 
  $m_{\sum J}^\mathrm{vis}$ vs. $H_\mathrm{T}^{j} + E^\mathrm{miss}_\mathrm{T}$ &   $\mathbf{6.39} \pm 0.16$  &   $\mathbf{6.36} \pm 0.21$  &    \\ 
  $m_{\sum J}^\mathrm{vis}$ vs. $H_\mathrm{T}^{j}$                       &   ${6.32} \pm 0.17$  &   $\textcolor{red}{\mathbf{6.36}} \pm 0.21$  &    \\ 
  $m_{\sum J}^\mathrm{vis}$ vs. $R^{t2,t1}$                              &   ${6.26} \pm 0.18$  &    &    \\ 
  $m_{\sum J}^\mathrm{vis}$ vs. $\Delta\phi^{t\bar{t}}$                  &   ${6.27} \pm 0.16$  &   ${6.09} \pm 0.22$  &    \\ 
  $m_{\sum J}^\mathrm{vis}$ vs. $\delta^{t\bar{t}}$                      &    &   ${6.03} \pm 0.26$  &    \\ 
  $m_{\sum J}^\mathrm{vis}$ vs. $m^{t\bar{t}}$                           &    &   ${6.18} \pm 0.2$  &   ${6.3} \pm 0.2$  \\ 
  $m_{\sum J}^\mathrm{vis}$ vs. $p_\mathrm{T}^{t\bar{t}}$                &    &   ${6.12} \pm 0.22$  &    \\ 
  $m_{\sum J}^\mathrm{vis}$ vs. $y_\mathrm{boost}^{t\bar{t}}$            &    &   ${6.14} \pm 0.2$  &    \\ 
  $m_{\sum J}^\mathrm{vis}$ vs. jets Aplanarity                          &   ${6.35} \pm 0.16$  &    &    \\ 
  $m_{\sum J}^\mathrm{vis}$ vs. jets Sphericity                          &   ${6.29} \pm 0.18$  &   ${6.12} \pm 0.18$  &    \\ 
  $p_\mathrm{T}^{t\bar{t}} / \sqrt{p_\mathrm{T}^{t1} p_\mathrm{T}^{t2}}$ vs. $H_\mathrm{T}^{j} + E^\mathrm{miss}_\mathrm{T}$ &   ${6.36} \pm 0.17$  &   ${6.12} \pm 0.26$  &   ${6.26} \pm 0.25$  \\ 
  $p_\mathrm{T}^{t\bar{t}} / \sqrt{p_\mathrm{T}^{t1} p_\mathrm{T}^{t2}}$ vs. $H_\mathrm{T}^{j}$ &   $\mathbf{6.38} \pm 0.17$  &   ${6.17} \pm 0.27$  &   ${6.25} \pm 0.27$  \\ 
  $p_\mathrm{T}^{t\bar{t}} / \sqrt{p_\mathrm{T}^{t1} p_\mathrm{T}^{t2}}$ vs. $m^{t\bar{t}}$ &   ${6.27} \pm 0.17$  &   ${6.17} \pm 0.23$  &   $\mathbf{6.39} \pm 0.21$  \\ 
  $p_\mathrm{T}^{t\bar{t}} / \sqrt{p_\mathrm{T}^{t1} p_\mathrm{T}^{t2}}$ vs. $m_{\sum J}^\mathrm{vis}$ &   $\textcolor{red}{\mathbf{6.47}} \pm 0.13$  &   $\mathbf{6.35} \pm 0.19$  &   $\mathbf{6.42} \pm 0.22$  \\ 
  $p_\mathrm{T}^{t\bar{t}} / \sqrt{p_\mathrm{T}^{t1} p_\mathrm{T}^{t2}}$ vs. jets Sphericity &    &    &   ${6.22} \pm 0.26$  \\ 
  $p_\mathrm{T}^{t\bar{t}} / m^{t\bar{t}}$ vs. $H_\mathrm{T}^{j} + E^\mathrm{miss}_\mathrm{T}$ &   $\mathbf{6.38} \pm 0.17$  &   ${6.2} \pm 0.21$  &   ${6.15} \pm 0.27$  \\ 
  $p_\mathrm{T}^{t\bar{t}} / m^{t\bar{t}}$ vs. $H_\mathrm{T}^{j}$        &   $\mathbf{6.38} \pm 0.17$  &   ${6.14} \pm 0.25$  &    \\ 
  $p_\mathrm{T}^{t\bar{t}} / m^{t\bar{t}}$ vs. $m^{t\bar{t}}$            &   ${6.33} \pm 0.18$  &   ${6.18} \pm 0.21$  &   $\textcolor{red}{\mathbf{6.44}} \pm 0.18$  \\ 
  $p_\mathrm{T}^{t\bar{t}} / m^{t\bar{t}}$ vs. $m_{\sum J}^\mathrm{vis}$ &   $\mathbf{6.47} \pm 0.13$  &   $\mathbf{6.28} \pm 0.21$  &   ${6.29} \pm 0.22$  \\ 
  $p_\mathrm{T}^{t\bar{t}} / m^{t\bar{t}}$ vs. jets Sphericity           &    &    &   ${6.16} \pm 0.29$  \\ 
  $p_\mathrm{T}^{t\bar{t}}$ vs. $m^{t\bar{t}}$                           &    &   ${6.08} \pm 0.24$  &   ${6.28} \pm 0.22$  \\ 
  $|p_\mathrm{out}| / m^{t\bar{t}}$ vs. $H_\mathrm{T}^{j} + E^\mathrm{miss}_\mathrm{T}$ &    &    &   ${6.2} \pm 0.25$  \\ 
  $|p_\mathrm{out}| / m^{t\bar{t}}$ vs. $H_\mathrm{T}^{j}$               &    &    &   ${6.19} \pm 0.27$  \\ 
  $|p_\mathrm{out}| / m^{t\bar{t}}$ vs. $m^{t\bar{t}}$                   &    &    &   ${6.16} \pm 0.24$  \\ 
  jets Sphericity vs. $m^{t\bar{t}}$                                     &    &   ${6.1} \pm 0.2$  &   ${6.25} \pm 0.23$  \\ 
     \hline
\end{tabular}

}
  \caption{The best 1D and/or 2D \BumpHunter{} variables scores ($\log t$) and the statistical uncertainty over 100 pseudo-experiments for stacked backgrounds (including the $\ttbar{}$ samples) and for the $Z'$ signal model with $m_{Z'} = 1\,$TeV in the pseudo-data over the topologies. Empty fields indicate that the variable did not score within $1\sigma$ to the best scoring variable in given topology (red) while those scoring within $0.5\sigma$ are marked in bold.}
  \label{tab:bestBHscores_zp}
\end{table*}

\clearpage

\begin{table*}[p]
    \centering
\resizebox*{0.48\textheight}{!}{
     \begin{tabular}{l|lll}
\hline 
variable & 2B0S & 1B1S & 0B2S \\ \hline 
  $E^\mathrm{miss}_\mathrm{T}$ vs. $H_\mathrm{T}^{j} + E^\mathrm{miss}_\mathrm{T}$ &   $\mathbf{5.59} \pm 0.26$  &   ${5.56} \pm 0.27$  &    \\ 
  $E^\mathrm{miss}_\mathrm{T}$ vs. $H_\mathrm{T}^{j}$                    &   $\mathbf{5.72} \pm 0.27$  &   $\mathbf{5.71} \pm 0.27$  &   $\mathbf{6.08} \pm 0.29$  \\ 
  $E^\mathrm{miss}_\mathrm{T}$ vs. $R^{t2,t1}$                           &   ${5.27} \pm 0.28$  &    &    \\ 
  $E^\mathrm{miss}_\mathrm{T}$ vs. $\chi^{t\bar{t}}$                     &   ${5.28} \pm 0.28$  &    &    \\ 
  $E^\mathrm{miss}_\mathrm{T}$ vs. $\sum m^{J}$                          &   ${5.23} \pm 0.29$  &    &    \\ 
  $E^\mathrm{miss}_\mathrm{T}$ vs. $p_\mathrm{T}^{t\bar{t}}$             &   $\mathbf{5.41} \pm 0.28$  &    &    \\ 
  $E^\mathrm{miss}_\mathrm{T}$ vs. jets Sphericity                       &   ${5.34} \pm 0.26$  &    &    \\ 
  $H_\mathrm{T}^{j} + E^\mathrm{miss}_\mathrm{T}$ vs. $H_\mathrm{T}^{j}$ &   $\textcolor{red}{\mathbf{6.48}} \pm 0.15$  &   $\textcolor{red}{\mathbf{6.6}} \pm 0.0$  &   $\textcolor{red}{\mathbf{6.53}} \pm 0.11$  \\ 
  $H_\mathrm{T}^{j} + E^\mathrm{miss}_\mathrm{T}$ vs. $\chi^{t\bar{t}}$  &   $\mathbf{5.14} \pm 0.38$  &   $\mathbf{5.57} \pm 0.33$  &    \\ 
  $H_\mathrm{T}^{j} + E^\mathrm{miss}_\mathrm{T}$ vs. $\delta^{t\bar{t}}$ &   ${5.21} \pm 0.28$  &    &    \\ 
  $H_\mathrm{T}^{j} + E^\mathrm{miss}_\mathrm{T}$ vs. $m^{t\bar{t}}$     &    &   ${5.5} \pm 0.28$  &    \\ 
  $H_\mathrm{T}^{j}$ vs. $\chi^{t\bar{t}}$                               &   ${5.14} \pm 0.3$  &    &    \\ 
  $H_\mathrm{T}^{j}$ vs. $\delta^{t\bar{t}}$                             &   ${5.25} \pm 0.29$  &    &    \\ 
  $H_\mathrm{T}^{j}$ vs. $m^{t\bar{t}}$                                  &    &   ${5.57} \pm 0.26$  &   ${5.59} \pm 0.37$  \\ 
  $H_\mathrm{T}^{j}$ vs. $p_\mathrm{T}^{t\bar{t}}$                       &    &   ${5.65} \pm 0.24$  &   ${5.67} \pm 0.34$  \\ 
  $H_\mathrm{T}^{j}$ vs. $y_\mathrm{boost}^{t\bar{t}}$                   &   ${5.31} \pm 0.28$  &    &    \\ 
  $H_\mathrm{T}^{j}$ vs. jets Sphericity                                 &   ${5.28} \pm 0.26$  &    &    \\ 
  $\chi^{t\bar{t}}$ vs. $m^{t\bar{t}}$                                   &   $\mathbf{4.66} \pm 0.52$  &    &    \\ 
  $\sum m^{J}$ vs. $H_\mathrm{T}^{j}$                                    &    &    &   ${5.34} \pm 0.52$  \\ 
  $m^{t\bar{t}} / \sqrt{p_\mathrm{T}^{t1} p_\mathrm{T}^{t2}}$ vs. $E^\mathrm{miss}_\mathrm{T}$ &   ${5.22} \pm 0.28$  &    &    \\ 
  $m^{t\bar{t}} / \sqrt{p_\mathrm{T}^{t1} p_\mathrm{T}^{t2}}$ vs. $H_\mathrm{T}^{j} + E^\mathrm{miss}_\mathrm{T}$ &    &   $\mathbf{5.62} \pm 0.32$  &    \\ 
  $m^{t\bar{t}} / \sqrt{p_\mathrm{T}^{t1} p_\mathrm{T}^{t2}}$ vs. $H_\mathrm{T}^{j}$ &    &   $\mathbf{5.55} \pm 0.3$  &    \\ 
  $m^{t\bar{t}} / \sqrt{p_\mathrm{T}^{t1} p_\mathrm{T}^{t2}}$ vs. $m_{\sum J}^\mathrm{vis}$ &    &   $\mathbf{5.51} \pm 0.32$  &    \\ 
  $m_{\sum J}^\mathrm{vis}$ vs. $E^\mathrm{miss}_\mathrm{T}$             &   $\mathbf{5.39} \pm 0.28$  &   ${5.58} \pm 0.27$  &    \\ 
  $m_{\sum J}^\mathrm{vis}$ vs. $H_\mathrm{T}^{j} + E^\mathrm{miss}_\mathrm{T}$ &   ${5.43} \pm 0.24$  &   $\mathbf{5.75} \pm 0.28$  &   ${5.62} \pm 0.35$  \\ 
  $m_{\sum J}^\mathrm{vis}$ vs. $H_\mathrm{T}^{j}$                       &    &   $\mathbf{5.7} \pm 0.3$  &   $\mathbf{5.76} \pm 0.39$  \\ 
  $m_{\sum J}^\mathrm{vis}$ vs. $\chi^{t\bar{t}}$                        &   ${4.85} \pm 0.41$  &   ${5.38} \pm 0.31$  &    \\ 
  $m_{\sum J}^\mathrm{vis}$ vs. $\delta^{t\bar{t}}$                      &    &   ${5.28} \pm 0.34$  &    \\ 
  $p_\mathrm{T}^{t\bar{t}} / \sqrt{p_\mathrm{T}^{t1} p_\mathrm{T}^{t2}}$ vs. $E^\mathrm{miss}_\mathrm{T}$ &   $\mathbf{5.54} \pm 0.23$  &    &    \\ 
  $p_\mathrm{T}^{t\bar{t}} / \sqrt{p_\mathrm{T}^{t1} p_\mathrm{T}^{t2}}$ vs. $H_\mathrm{T}^{j} + E^\mathrm{miss}_\mathrm{T}$ &    &    &   $\mathbf{6.04} \pm 0.28$  \\ 
  $p_\mathrm{T}^{t\bar{t}} / \sqrt{p_\mathrm{T}^{t1} p_\mathrm{T}^{t2}}$ vs. $H_\mathrm{T}^{j}$ &    &   ${5.68} \pm 0.23$  &   $\mathbf{6.04} \pm 0.29$  \\ 
  $p_\mathrm{T}^{t\bar{t}} / \sqrt{p_\mathrm{T}^{t1} p_\mathrm{T}^{t2}}$ vs. $m^{t\bar{t}}$ &    &    &   ${5.86} \pm 0.26$  \\ 
  $p_\mathrm{T}^{t\bar{t}} / \sqrt{p_\mathrm{T}^{t1} p_\mathrm{T}^{t2}}$ vs. jets Sphericity &   ${5.21} \pm 0.3$  &    &    \\ 
  $p_\mathrm{T}^{t\bar{t}} / m^{t\bar{t}}$ vs. $E^\mathrm{miss}_\mathrm{T}$ &   ${5.45} \pm 0.23$  &    &    \\ 
  $p_\mathrm{T}^{t\bar{t}} / m^{t\bar{t}}$ vs. $H_\mathrm{T}^{j} + E^\mathrm{miss}_\mathrm{T}$ &    &    &   ${5.77} \pm 0.3$  \\ 
  $p_\mathrm{T}^{t\bar{t}} / m^{t\bar{t}}$ vs. $H_\mathrm{T}^{j}$        &   ${5.32} \pm 0.26$  &   $\mathbf{5.75} \pm 0.27$  &   ${5.78} \pm 0.29$  \\ 
  $p_\mathrm{T}^{t\bar{t}} / m^{t\bar{t}}$ vs. jets Sphericity           &   ${5.28} \pm 0.26$  &    &    \\ 
  $|p_\mathrm{out}| / m^{t\bar{t}}$ vs. $E^\mathrm{miss}_\mathrm{T}$     &    &    &   ${5.83} \pm 0.26$  \\ 
  $|p_\mathrm{out}| / m^{t\bar{t}}$ vs. $H_\mathrm{T}^{j} + E^\mathrm{miss}_\mathrm{T}$ &    &    &   $\mathbf{5.95} \pm 0.29$  \\ 
  $|p_\mathrm{out}| / m^{t\bar{t}}$ vs. $H_\mathrm{T}^{j}$               &    &    &   $\mathbf{5.99} \pm 0.28$  \\ 
  $|p_\mathrm{out}| / m^{t\bar{t}}$ vs. $m^{t\bar{t}} / \sqrt{p_\mathrm{T}^{t1} p_\mathrm{T}^{t2}}$ &    &   $\mathbf{5.76} \pm 0.31$  &    \\ 
  $|p_\mathrm{out}| / m^{t\bar{t}}$ vs. $p_\mathrm{T}^{t\bar{t}} / \sqrt{p_\mathrm{T}^{t1} p_\mathrm{T}^{t2}}$ &    &    &   $\mathbf{5.91} \pm 0.33$  \\ 
  $|p_\mathrm{out}| / m^{t\bar{t}}$ vs. $p_\mathrm{T}^{t\bar{t}} / m^{t\bar{t}}$ &    &   ${5.25} \pm 0.34$  &   ${5.69} \pm 0.35$  \\ 
  $|p_\mathrm{out}| / m^{t\bar{t}}$ vs. jets Sphericity                  &    &   ${5.47} \pm 0.29$  &   ${5.9} \pm 0.25$  \\ 
     \hline
\end{tabular}

}
\caption{The best 1D and/or 2D \BumpHunter{} variable scores ($\log t$) and the statistical uncertainty over 100 pseudo-experiments for stacked backgrounds (including the $\ttbar{}$ samples) and for the DM \ttbar{}-associated pair production $\ttbar{} + \chi_D \bar{\chi}_D$ signal model with $m_{\chi_D} = 100\,$GeV in the pseudo-data over the topologies. Empty fields indicate that the variable did not score within $2.5\sigma$ ($4\sigma$) in the 0B2S (1B1S and 2B0S) topologies to the best scoring variable in given topology (red) while those scoring within $2.2\sigma$ ($3.5\sigma$) are marked in bold.}
  \label{tab:bestBHscores_xdxd}
\end{table*}

\clearpage


\subsection{Comparison to other methods}

In order to evaluate the possible benefits of the 2D version of the \BumpHunter{} algorithm in terms of the corresponding BH score, comparisons to several other established statistical tests have been performed.

\begin{enumerate}

    \item The consistency of the pseudo-data with the background-only hypothesis based on a~$\chi^2$ test between the pseudo-data and the summed background distribution was evaluated. A~corresponding null hypothesis probability ($p_0$ value) was computed by the ROOT \texttt{TMath::Prob} function. This probability was then converted into a background-only hypothesis ``Bg'' score by computing a~logarithm of the negative $p_0$ value, giving the Bg score $t_0 = \log (-p_0)$ for the 1D and 2D variables studied.

    \item A likelihood fit was performed to extract the signal strength $\mu$ by minimizing the negative logarithm of the likelihood  function
      $$ \mathcal{L}(\mu|\mathbf{d}, \mathbf{s}, \mathbf{b}) \equiv  \prod_{i \in \mathrm{bins}} \mathcal{P}(d_i| \mu s_i + b_i)\,, $$
      where $d_i$ is the number of pseudo-data events and $s_i$ ($b_i$) numbers of the expected signal (background) events in $i$-th bin of a~1D or 2D distribution under study.
      As customary, the constant data factorial term is conveniently dropped in the logarithm of the assumed Poisson probability density
 $$   \log \mathcal{P}(d_i| \nu_i) = d_i \log \nu_i - \nu_i\,, \quad \mathrm{with} \quad \nu_i \equiv \mu s_i + b_i \,.$$
      In order to extract a~quantity comparable to the BH or Bg score, the compatibility of the fitted $\mu$ with the zero value was computed as a~simple $\chi^2$ test of one degree of freedom, and by computing the corresponding $p^\mu_0$ value and a~related $\mu$ score  $t^\mu_0 = \log (-p^\mu_0)$.

    \item As a~last cross-check, a~similar likelihood fit is performed but on the bin range determined by the BH procedure, leading to a~fitted $\mu_\mathrm{BH}$ and corresponding $\mu_\mathrm{BH}$ score $t^{\mu_\mathrm{BH}}_0 = \log (-p^{\mu_\mathrm{BH}}_0)$.
      
\end{enumerate}

\begin{figure*}
  \begin{center}
     \begin{tabular}{cc}
       \includegraphics[width=0.45\textwidth]{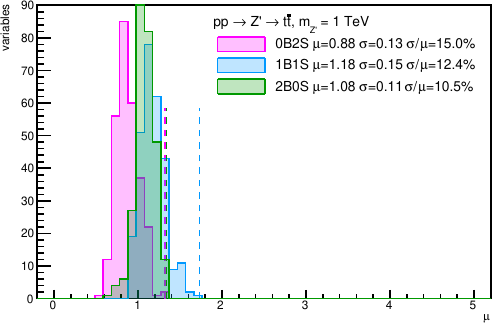} &
       \includegraphics[width=0.45\textwidth]{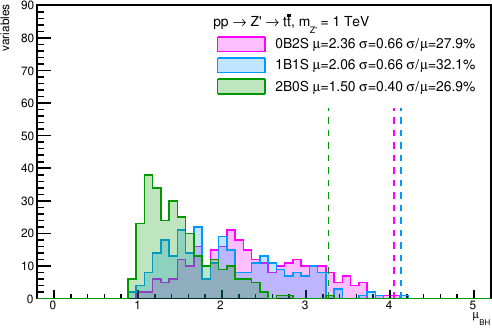}
       \end{tabular}
    \caption{The mean fitted signal strength over the full bin range (left) and over the best \BumpHunter{} area (right) over the variables of interest, averaged over replicas, for the vector $Z'$ model with $m_{Z'} = 1000\,$GeV. Green: 2B0S, blue: 1B1S and pink: 0B2S topologies. The vertical dashed lines indicate the largest mean fitted $\mu$.}
\label{fig_mucmp}
\end{center}
\end{figure*}

The distribution of the fitted signal strength $\mu$ over the variables of interest, averaged over the replicas, was observed to peak around~1 for the $y_0$ and $Z'$ signal models with strong injected signal but more flat and reaching higher values due to statistical fluctuations for the DM pair signal model with smaller injected signal. The signal strength fitted over the \BumpHunter{} area was found around~2 and more broadly distributed as different variables enhance the signal in the best BH area by a~different amount. See Figure~\ref{fig_mucmp} for an example of the $\mu$ and $\mu_\mathrm{BH}$ comparison over variables studied for the vector $Z'$ model.

\begin{figure*}
  \begin{center}
    \includegraphics[width=0.99\textwidth]{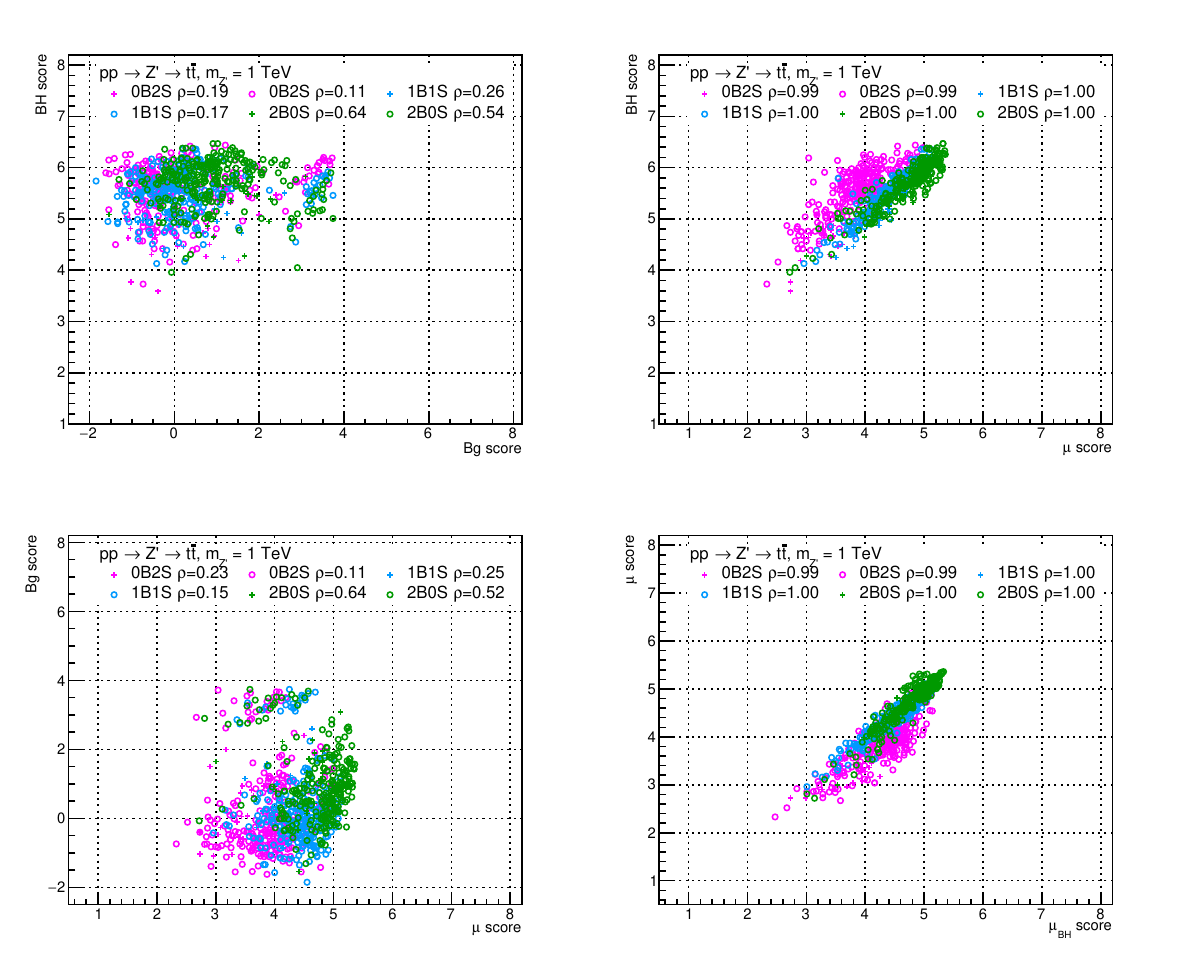} \\
    \caption{Correlations between the BH, Bg and $\mu$ scores over the variables for
      the vector $Z'$ model with $m_{Z'} = 1000\,$GeV
      The circle (cross) markers represent results using 2D (1D) variables. Green: 2B0S, blue: 1B1S and pink: 0B2S topologies.}
\label{fig_scores2dcmp}
\end{center}
\end{figure*}

The comparisons of the resulting BH, Bg, $\mu$ and $\mu_\mathrm{BH}$ scores for the $Z'$ vector signal model are shown in~Figure~\ref{fig_scores2dcmp}.
One can observe a~large correlation of the BH and $\mu$-based scores, proving the strong signal search power of the \BumpHunter{} algorithm with regard to a~likelihood fit over a~full range.
The $\mu_\mathrm{BH}$ score is even larger, reflecting its stronger power to reject the background only hypothesis over the area found by the \BumpHunter{} algorithm in both 1D and 2D versions.

The 2D version of the \BumpHunter{} algorithm can be regarded as a~sort of MV algorithm and the likelihood fit over the BH-selected area as a~fit to a~MV discriminant based on two (or more, in case of ratios) input variables.
In addition, the best performing BH areas can be visually checked and compared to expectations of a~particular model.

\clearpage
\section{Stability to systematics uncertainties}

In order to evaluate the stability of results and variables sensitivity to systematic uncertainties, the following study has been performed.
Jet momentum is first randomly smeared on the jet-by-jet basis by a Gaussian term with the resolution taken from the \Delphes{} hadronic calorimeter resolution formula in the default ATLAS card which reads
\begin{eqnarray}
|\eta| \leq 1.7                     &:& \sigma_E/E = \sqrt{0.0302^2   + 0.5205^2/ E + 1.59^2/E^2} \\
|\eta| > 1.7 \land |\eta| \leq 3.2  &:& \sigma_E/E =  \sqrt{0.0500^2   + 0.706^2/ E} \\
|\eta| > 3.2 \land |\eta| \leq 4.9  &:& \sigma_E/E =  \sqrt{0.09420^2   + 1/E}\,.
\end{eqnarray}

\begin{figure*}[!h]
\begin{center}
  \begin{tabular}{cc}
    {\includegraphics[width=0.47\textwidth]{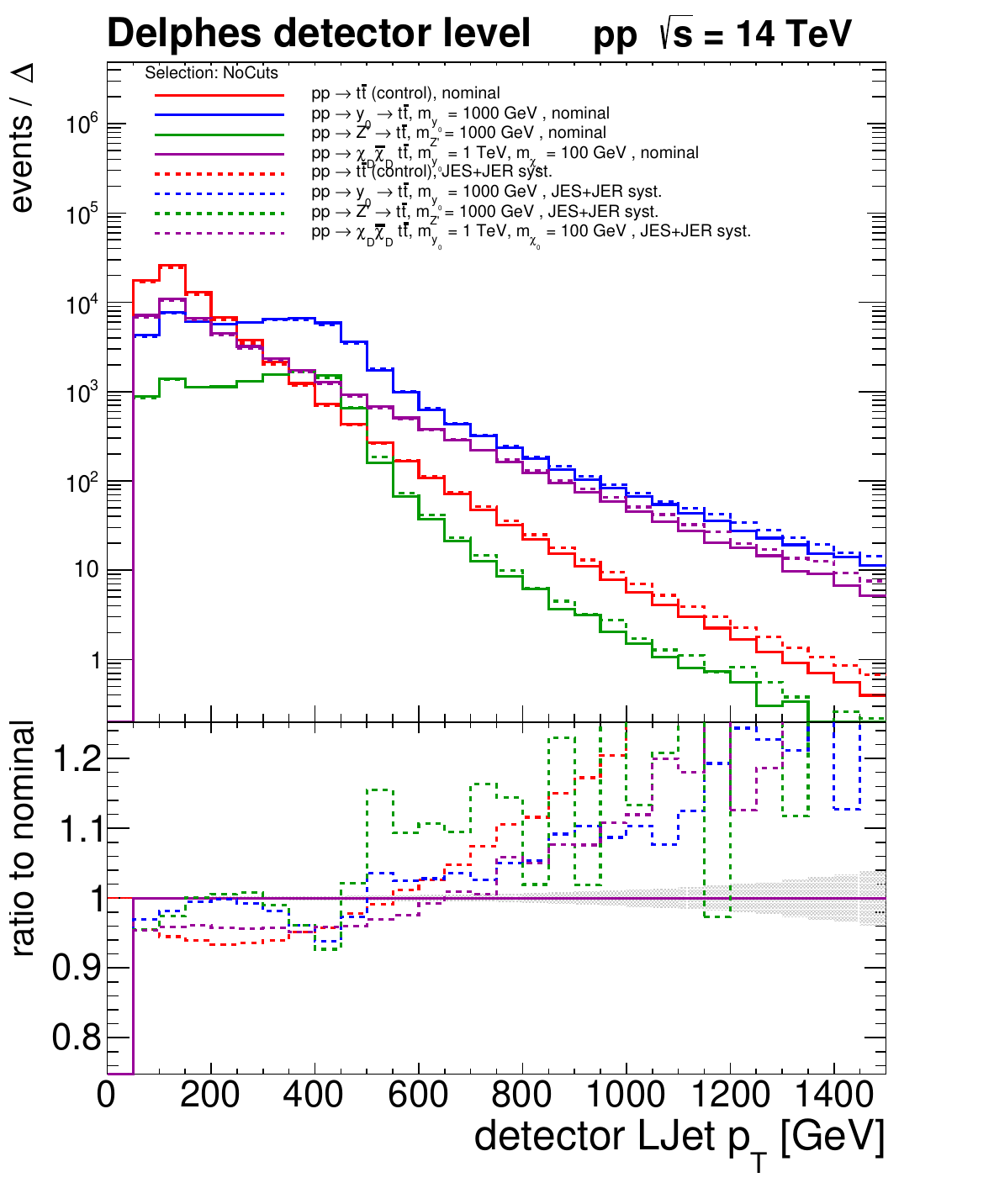}} &
    {\includegraphics[width=0.47\textwidth]{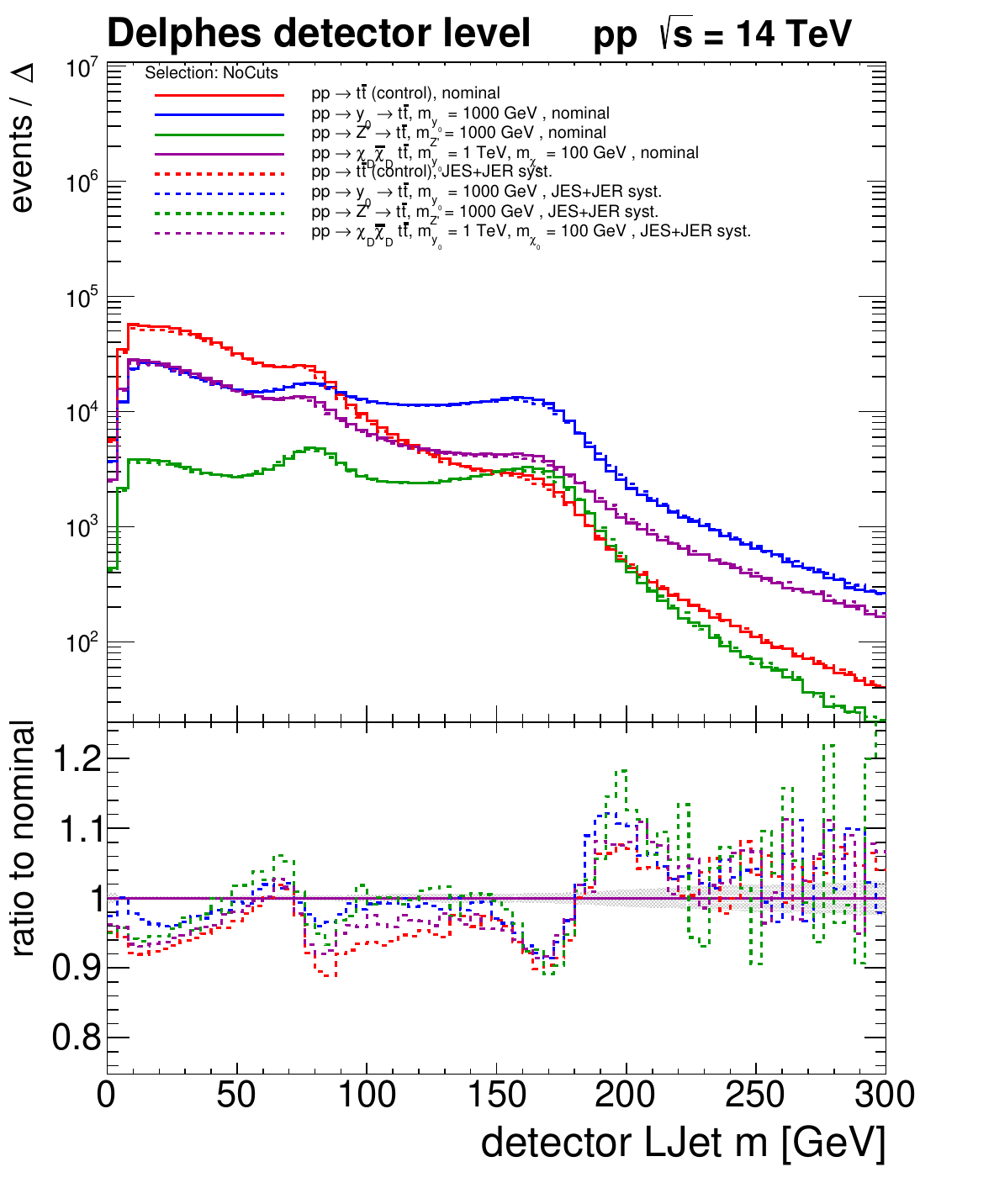}} \\
    {\includegraphics[width=0.47\textwidth]{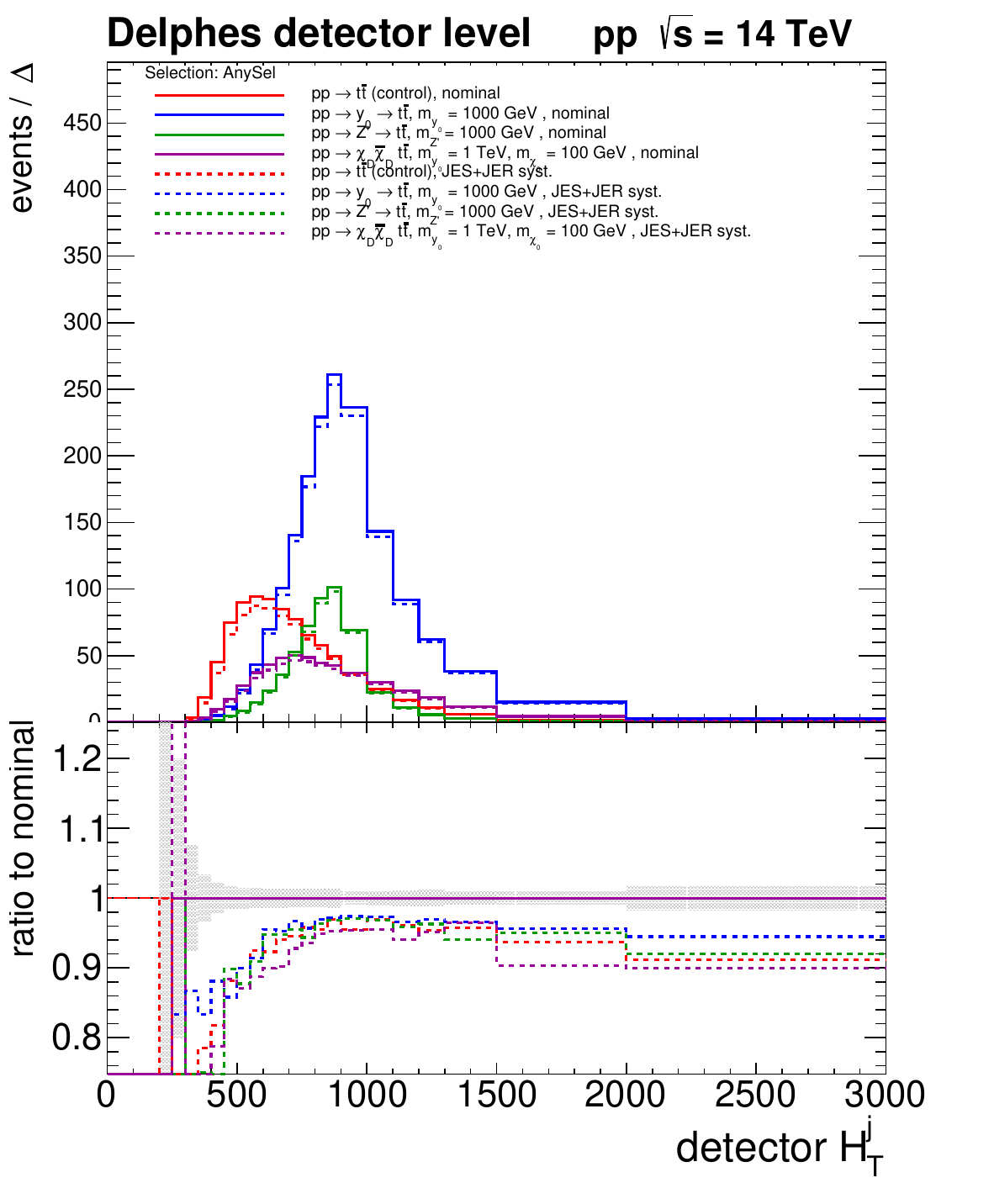}} &
    {\includegraphics[width=0.47\textwidth]{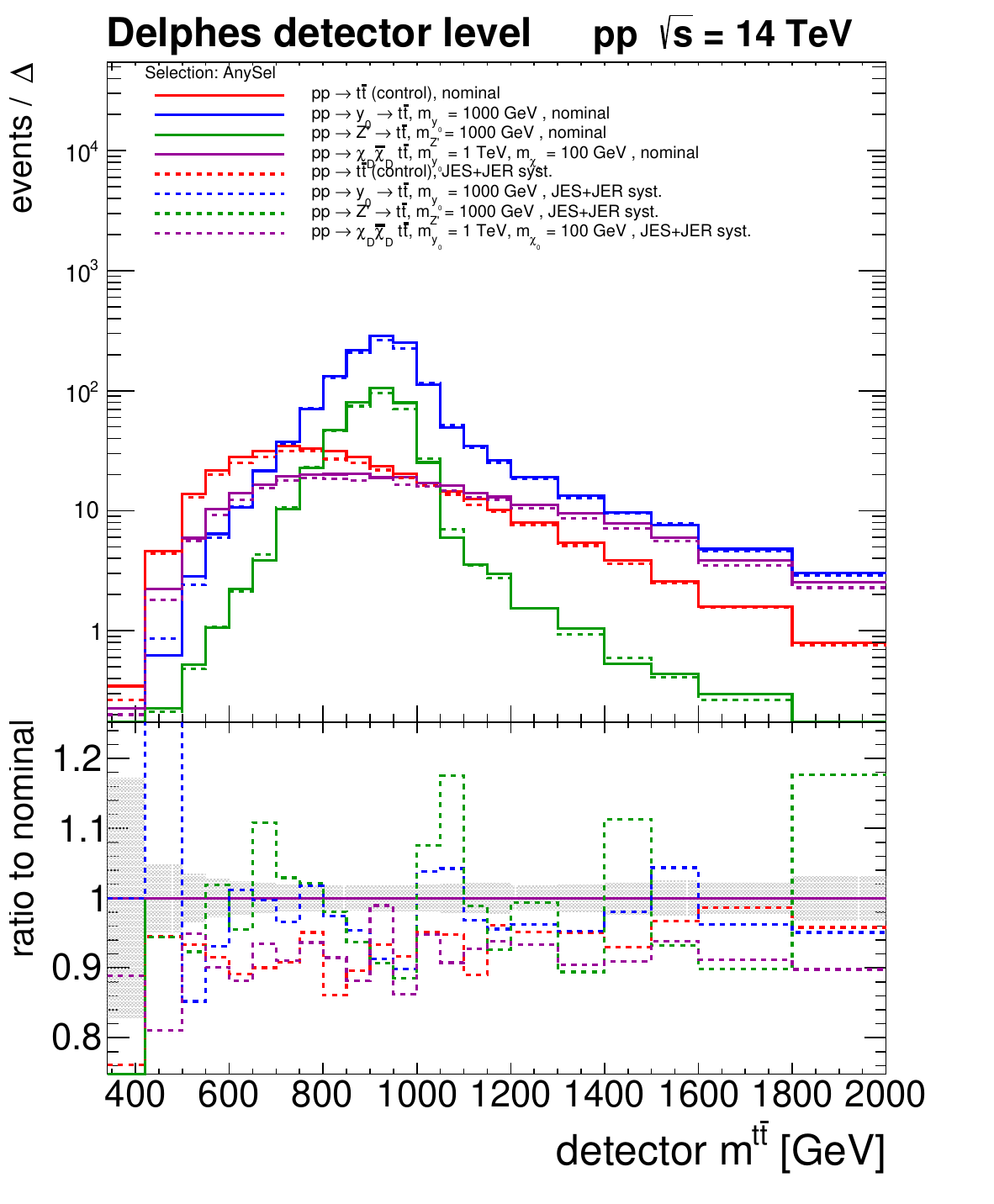}} \\
   \end{tabular}
  \caption{A comparison of selected nominal (solid) and systematics-varied (dashed) spectra for the three main BSM signal samples using in this study. Top left: the large-$R$ jet transverse momentum, top right: the large-$R$ jet mass, bottom left: the scalar sum of the small-$R$ jets transverse momenta, bottom right: invariant mass of the reconstructed \ttbar{} pair mass; all at the \Delphes{} detector level for events passing any of the 0B2S, 1B1S or 2B0S selections.}
   \label{fig_syst_shapes}
\end{center}
\end{figure*}

The jet momenta are then also modified by a shape using a multiplicative factor $1 + \gamma \, (\pt - p_\mathrm{T}^0) - \beta/\pt$, with $p_\mathrm{T}^0 = 400\,$GeV, $\gamma = 0.0001$ and $\beta = 1.2$, resulting in a positive-slope variation of the jet \pt{} by $-5$\% to $+5$\% over the range of 100--1000~GeV.
The overall effect is checked in~Figure~\ref{fig_syst_shapes} for the large-$R$ jet \pt{} (amounting from $-5$\% to $+15$\%) and other variables as a comparison of selected nominal and systematics-varied spectra for the three BSM signal models used in this study and for the $\ttbar{}$ sample. Similar trends are observed, depending on the shape of the original spectrum as the resulting shape is effectively a convolution of the nominal spectrum with the resolution which alters kinematics of large-$R$ jets on an event-by-event basis as well as by the \pt{} dependent multiplicative factor. The imprint on the $\HTj$ spectrum is due to selection effects on the large-$R$ jets.
Both event-by-event and jet-by-jet systematic variations are thus taken into account, as well as an overall slope change as function of the large-$R$ jet \pt{}, modelling the jet energy resolution and scale uncertainties, respectively. This then affects other distributions, too, as well as the selection efficiency, which decreases by about 5\%.

The following mixed exercise is then performed: the pseudo-data spectra are composed from the systematics-varied samples as described above, and are compared to the prediction composed from the nominal samples.
In the process, each varied sample is normalized to the integral of its corresponding nominal sample.
This is motivated by the fact that in real analyses, dedicated background-dominated control regions (CR) are used to constrain background normalizations and so even in the case of non-matching effects or treatment of data and simulation, backgrounds would be rescaled to match the data in CRs. Still, samples normalization is allowed to float in the pseudo-experiments just as in the case of the nominal exercise.

\begin{figure*}[h!]
\begin{center}
\begin{tabular}{ccc}
\includegraphics[width=0.33\textwidth]{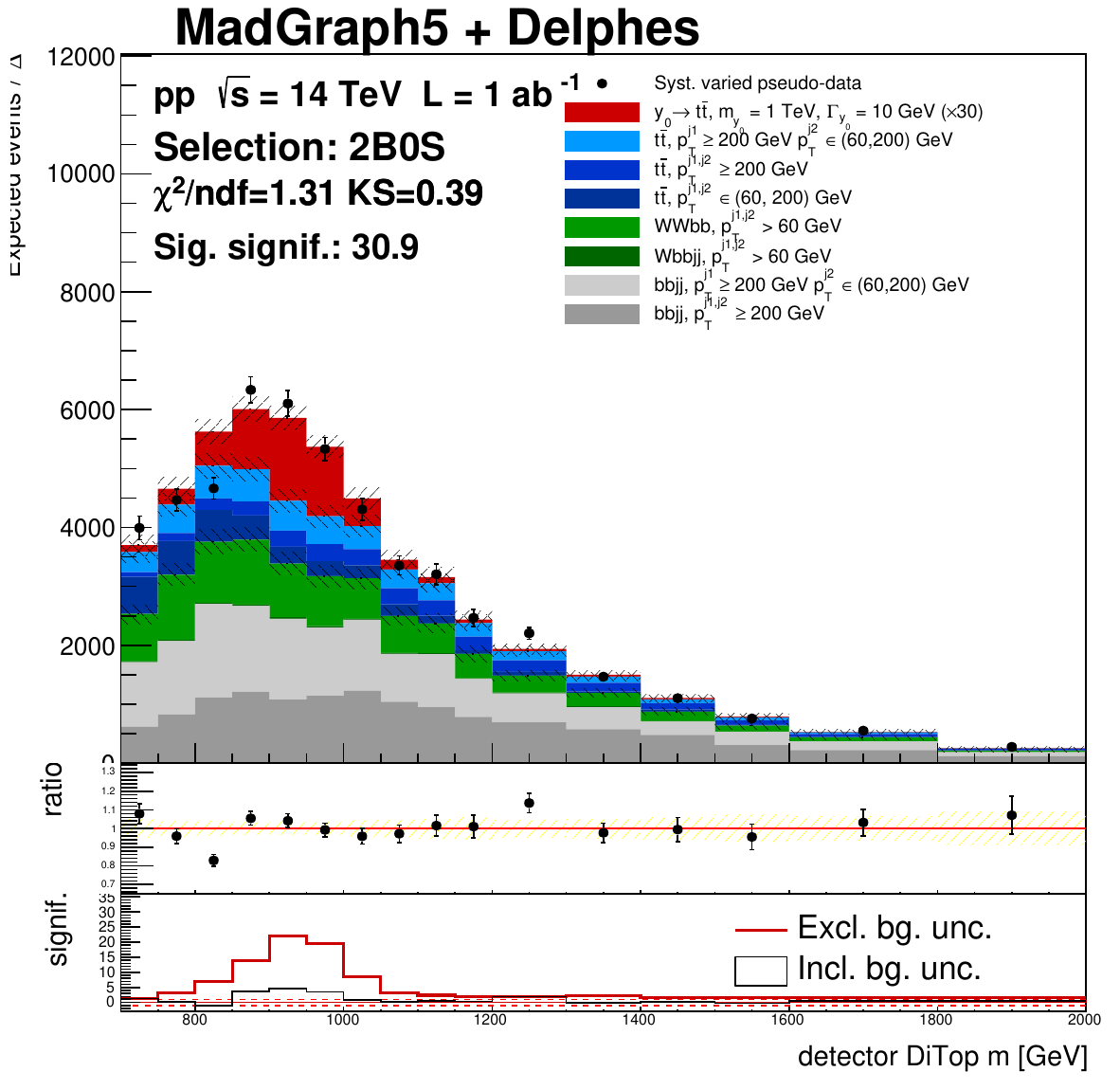} &
\includegraphics[width=0.33\textwidth]{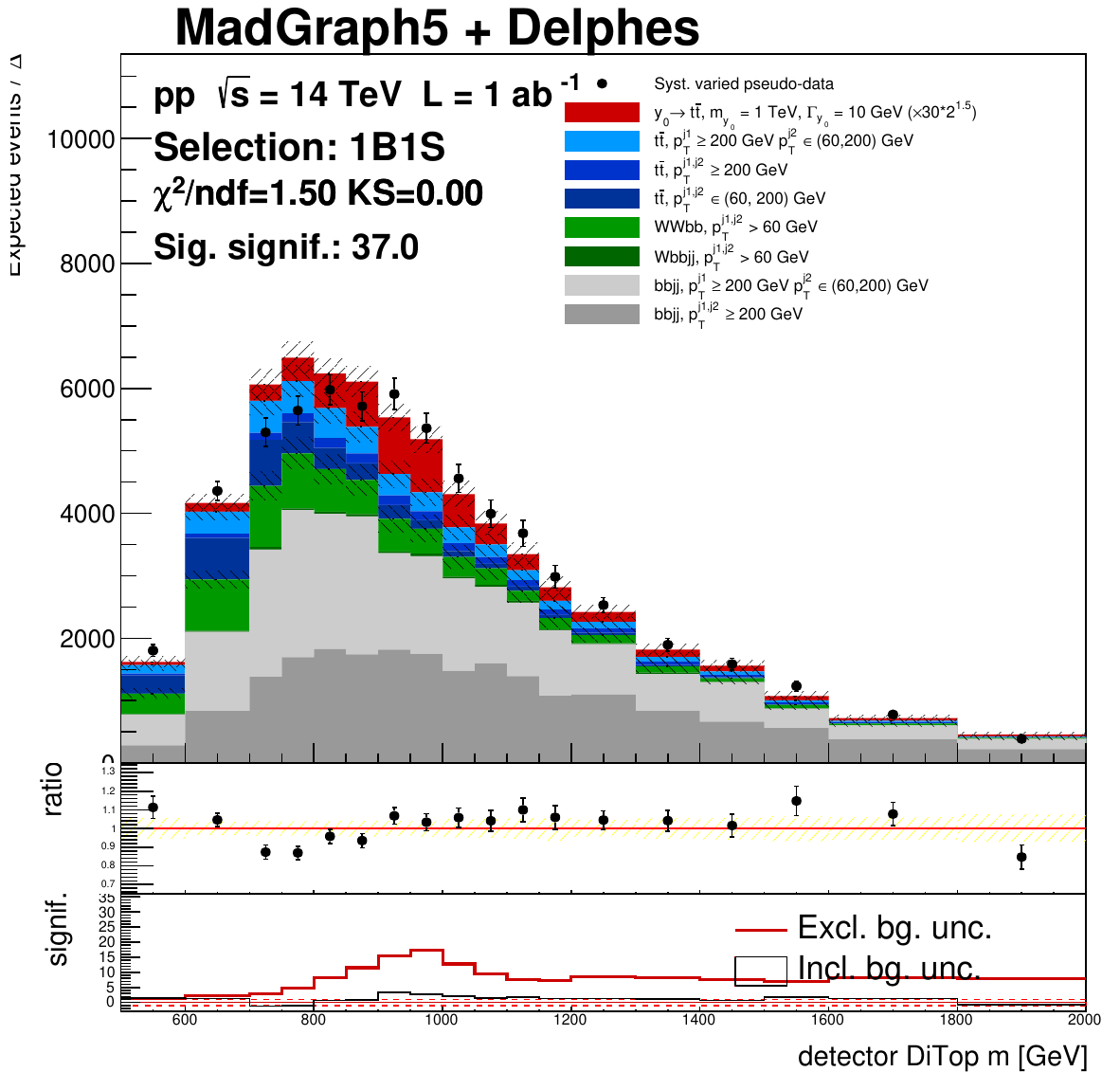} &
\includegraphics[width=0.33\textwidth]{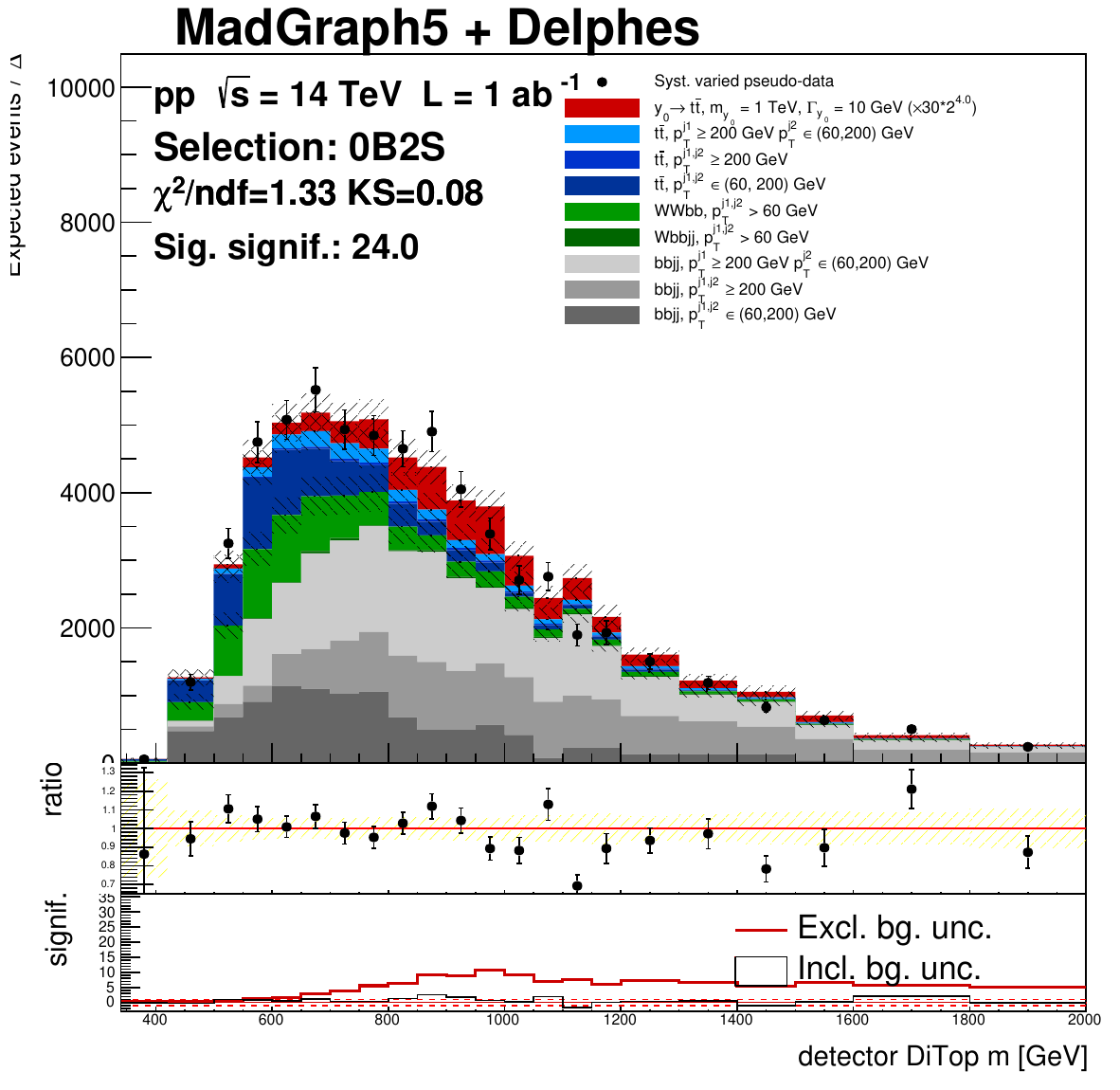} \\
\includegraphics[width=0.33\textwidth]{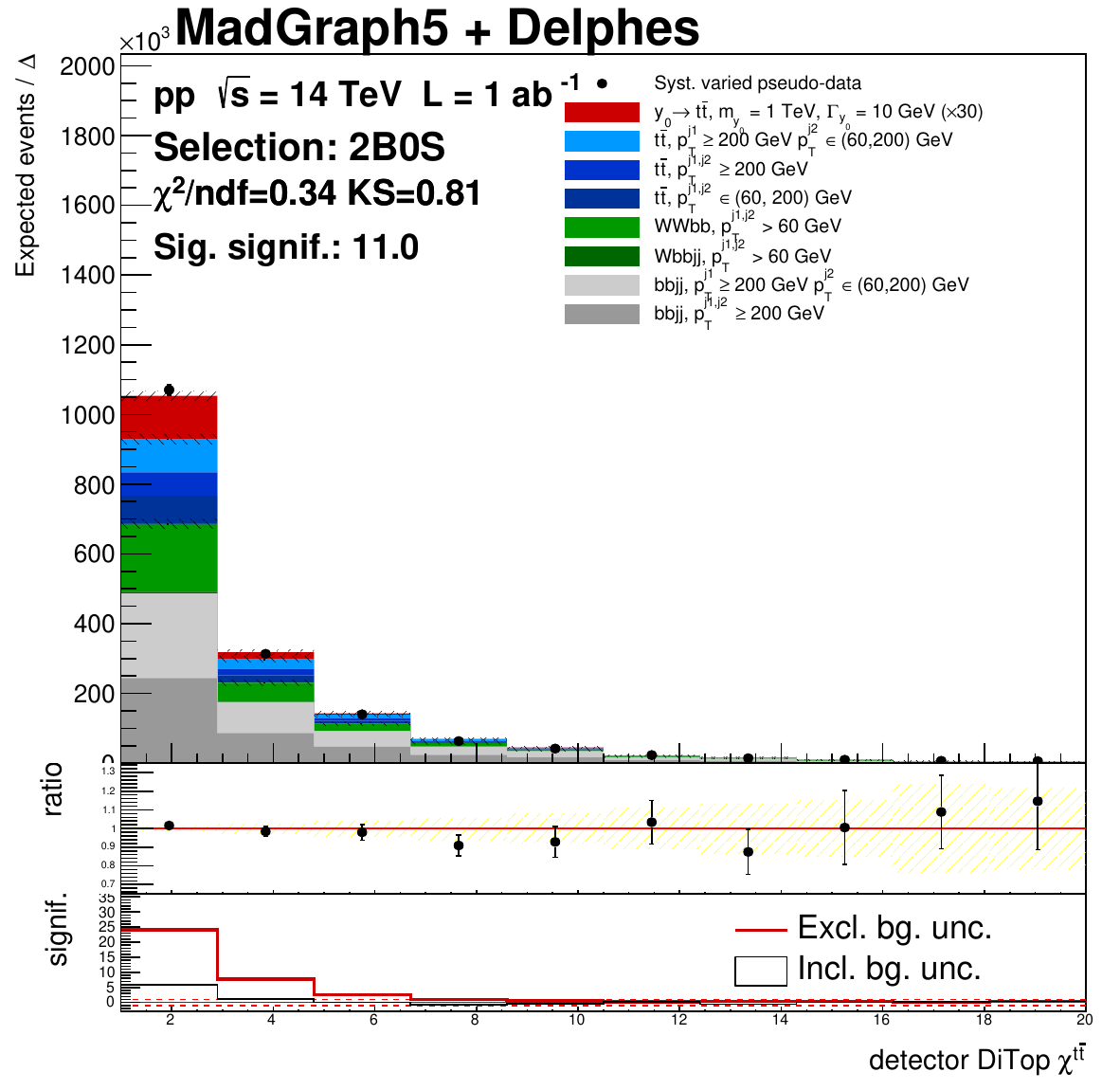} &
\includegraphics[width=0.33\textwidth]{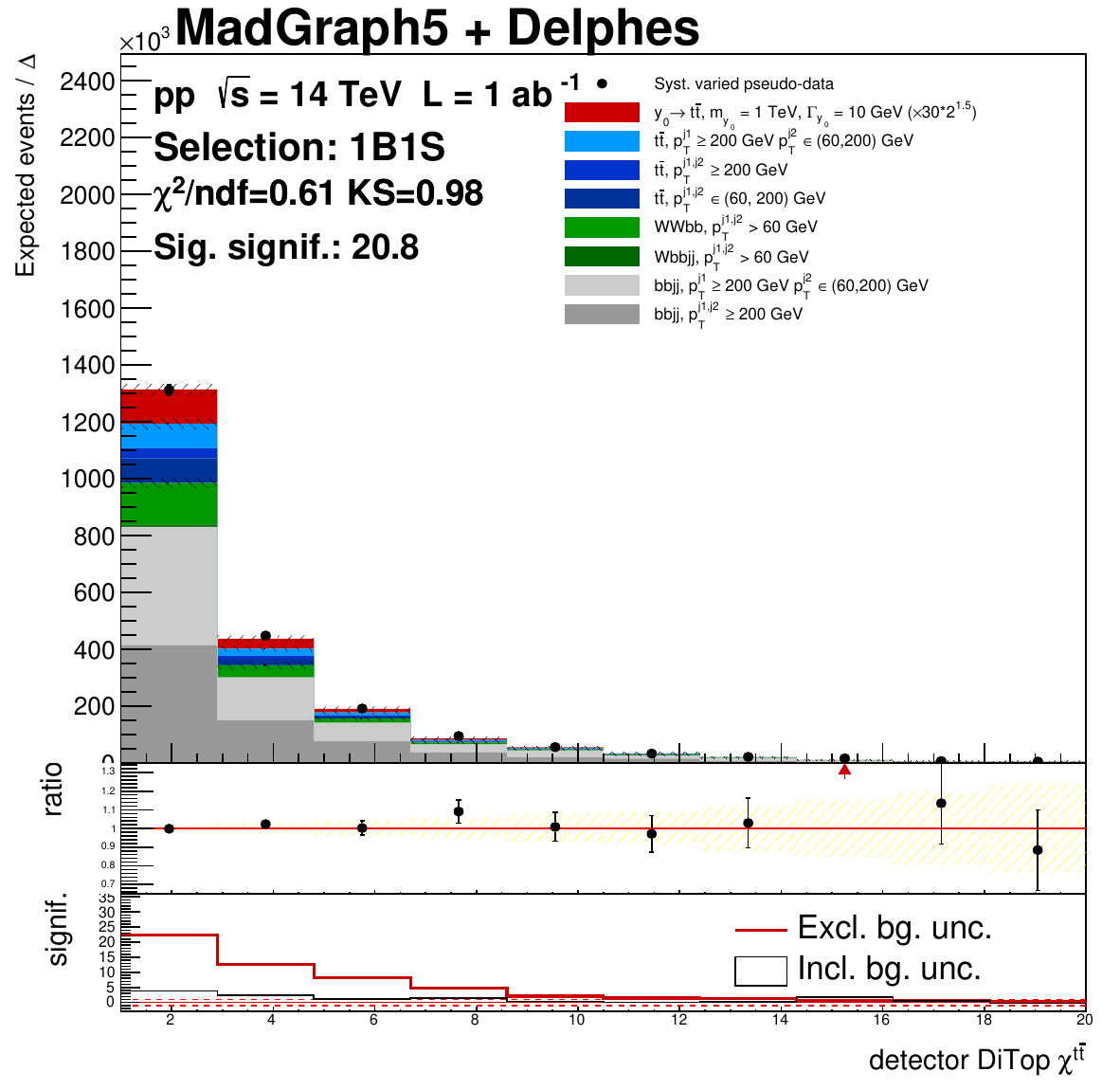} &
\includegraphics[width=0.33\textwidth]{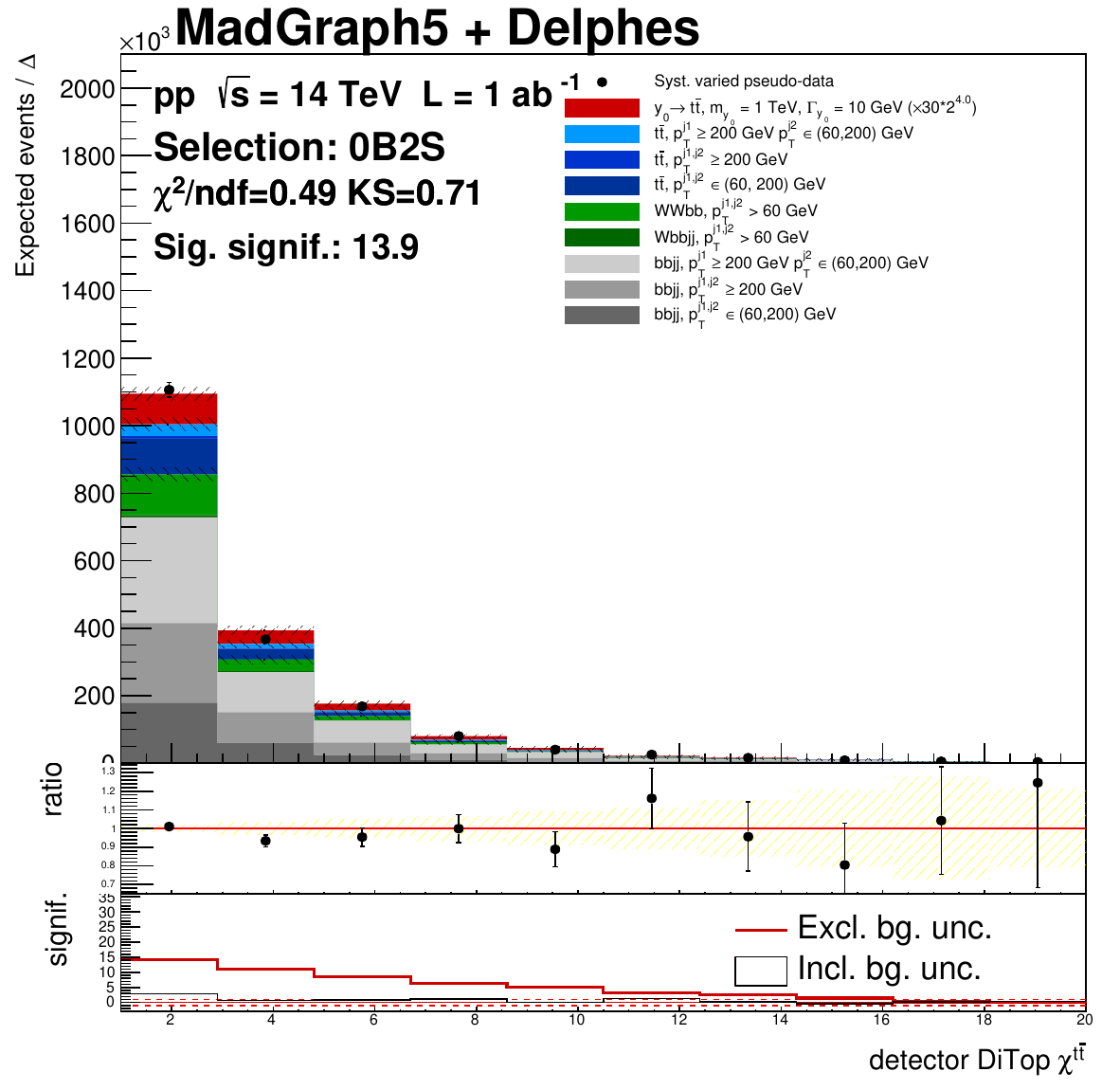} \\
2B0S  &
1B1S  &
0B2S  \\
\end{tabular}
\caption{Mixed comparison with systematics-varied pseudo-data build from sub-samples normalized to their nominal integrals: Stacked invariant mass of the $\ttbar$ pair (top) and the $\Chittbar$ variable (bottom) at the \Delphes{} detector level for the scaled $y_0$ model with $m_{y_0} = 1\,$TeV, with both the pseudo-data/prediction ratio and the signal significance shown in lower pads, for the boosted-boosted (2B0S, left), boosted-semiboosted (1B1S, middle) and semiboosted-semiboosted (0B2S, right) topologies.
}
\label{fig_stack_MIX_model_y0_compact_RS}
\end{center}
\end{figure*}

\begin{figure*}[h!]
\begin{center}
\begin{tabular}{ccc}
\includegraphics[width=0.33\textwidth]{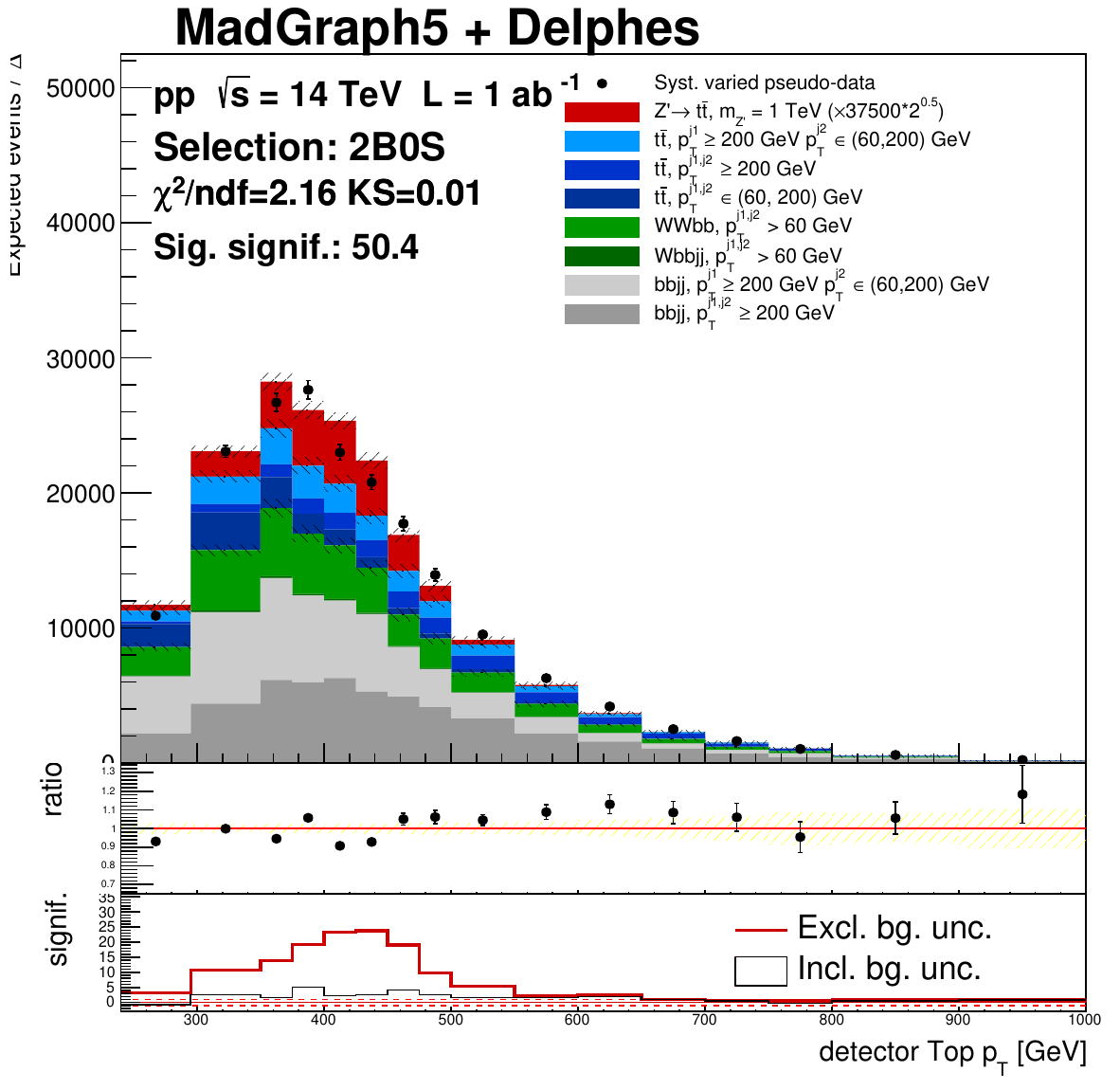} &
\includegraphics[width=0.33\textwidth]{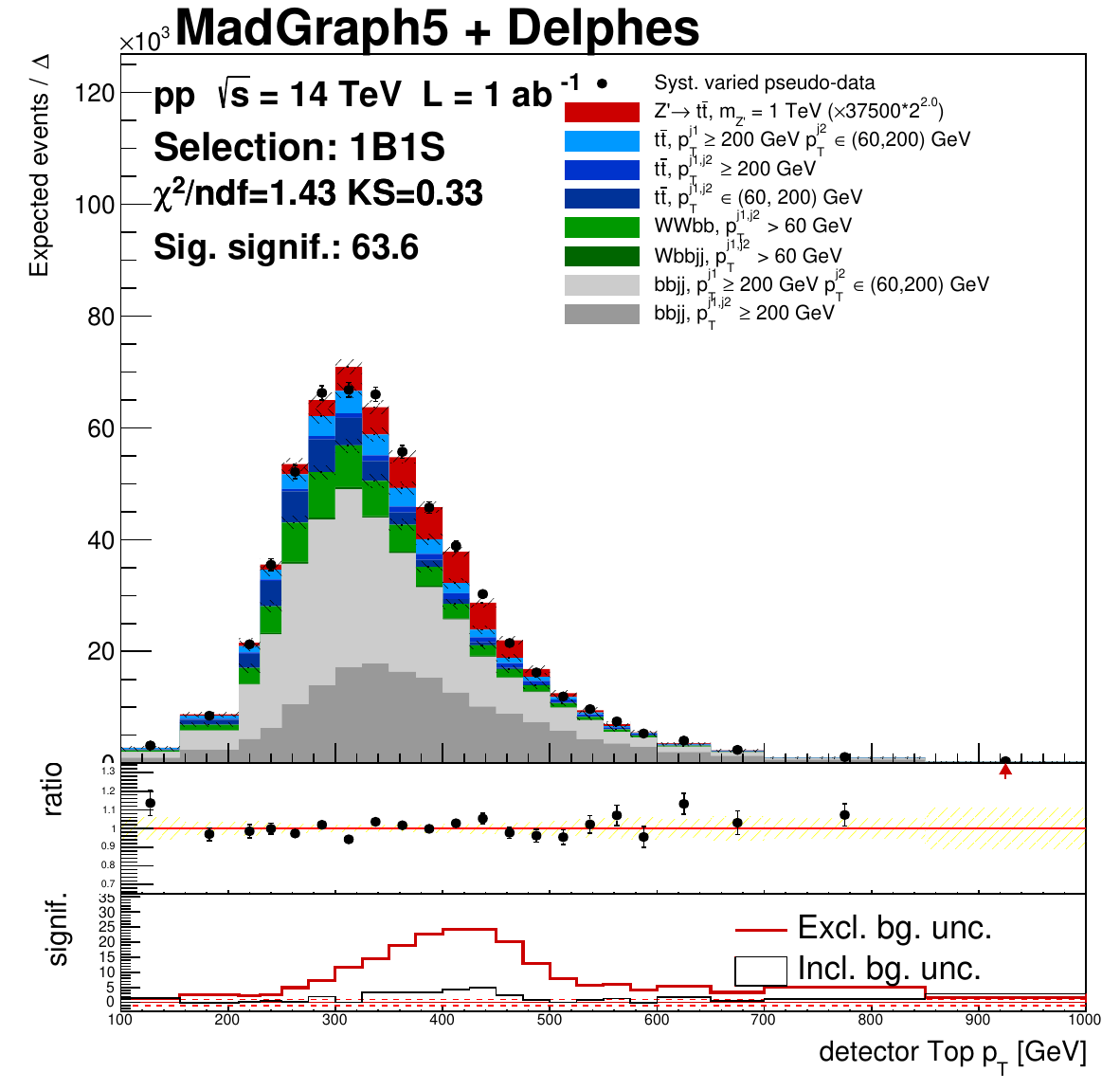} &
\includegraphics[width=0.33\textwidth]{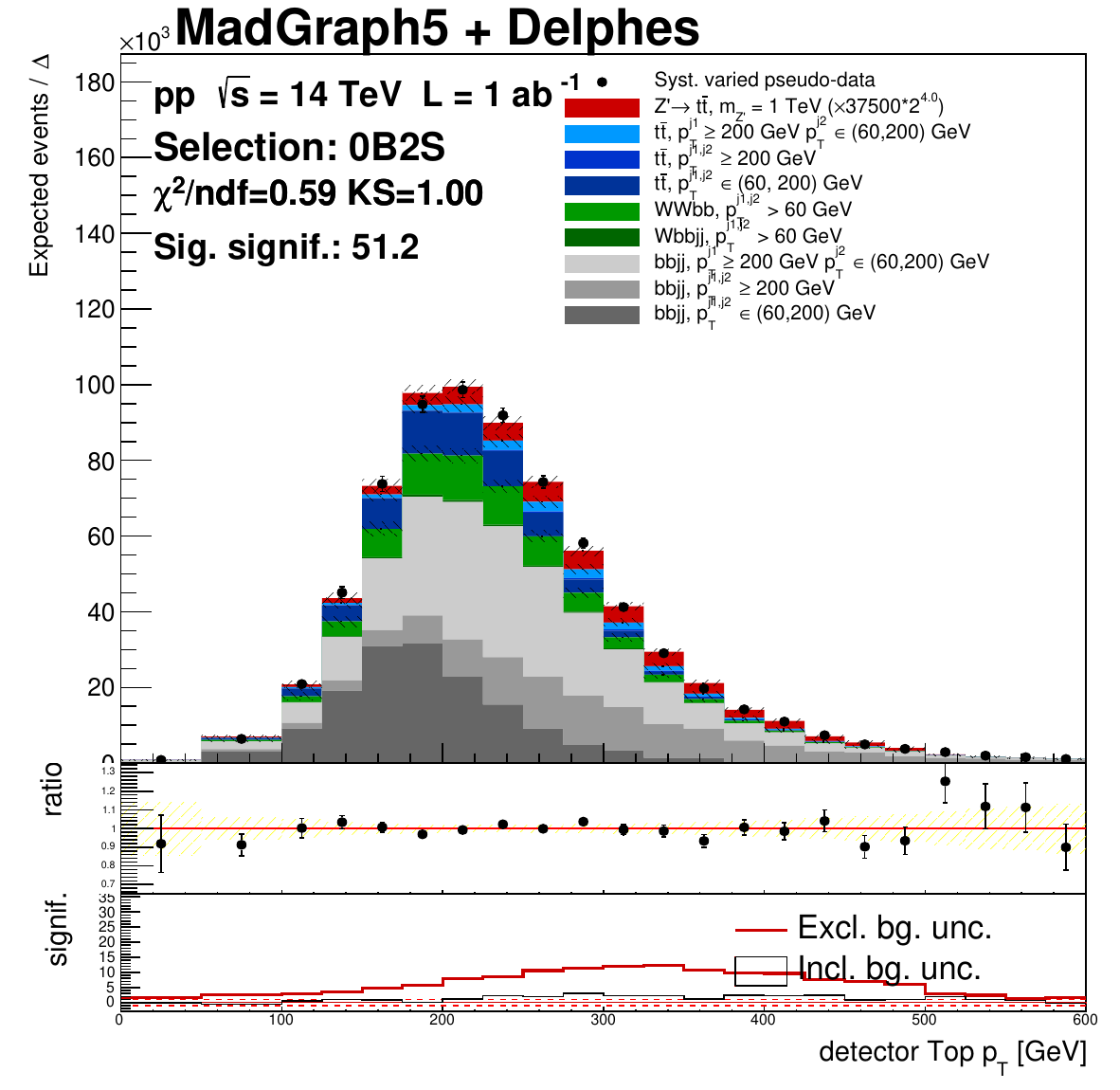} \\
\includegraphics[width=0.33\textwidth]{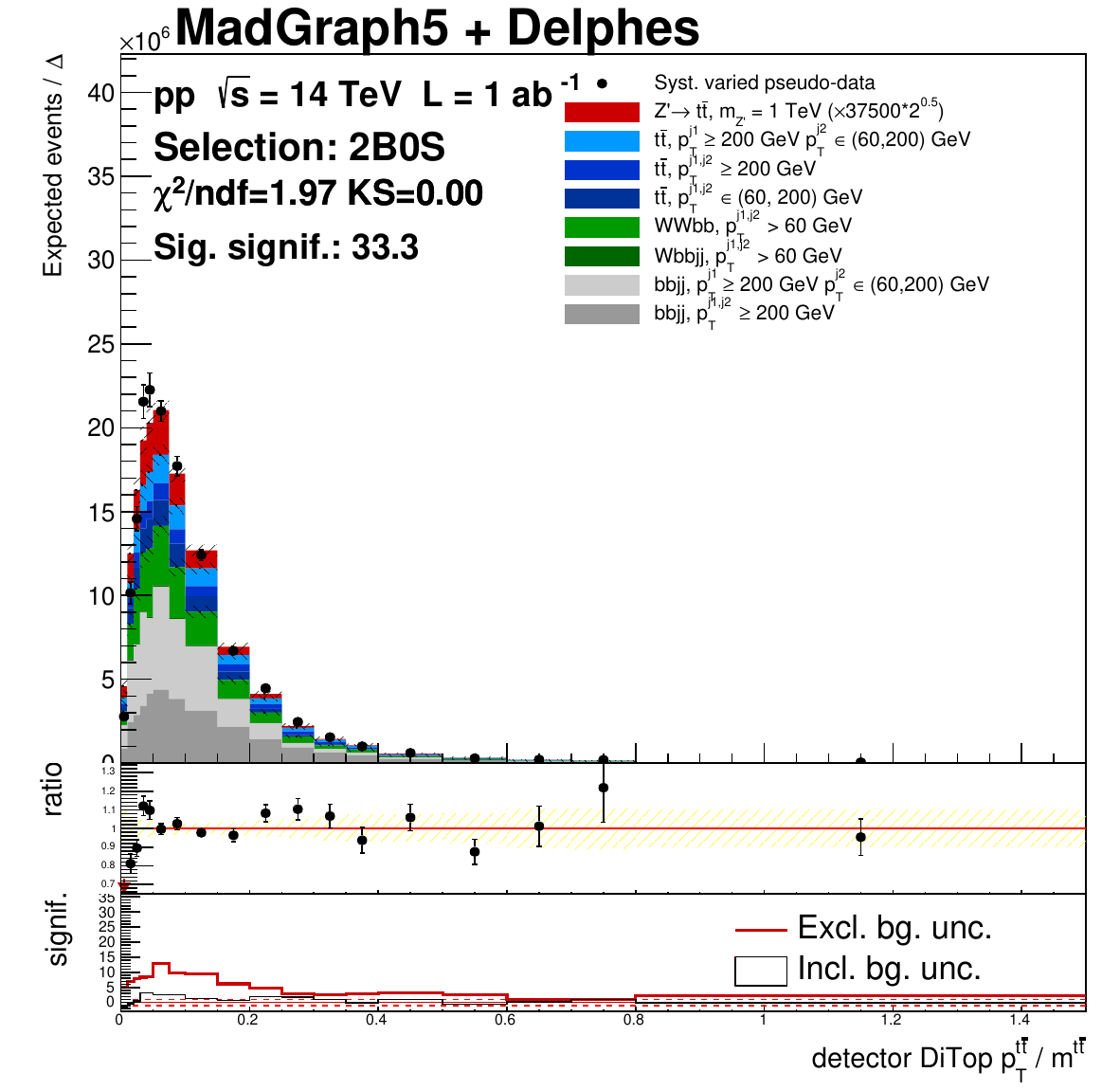} &
\includegraphics[width=0.33\textwidth]{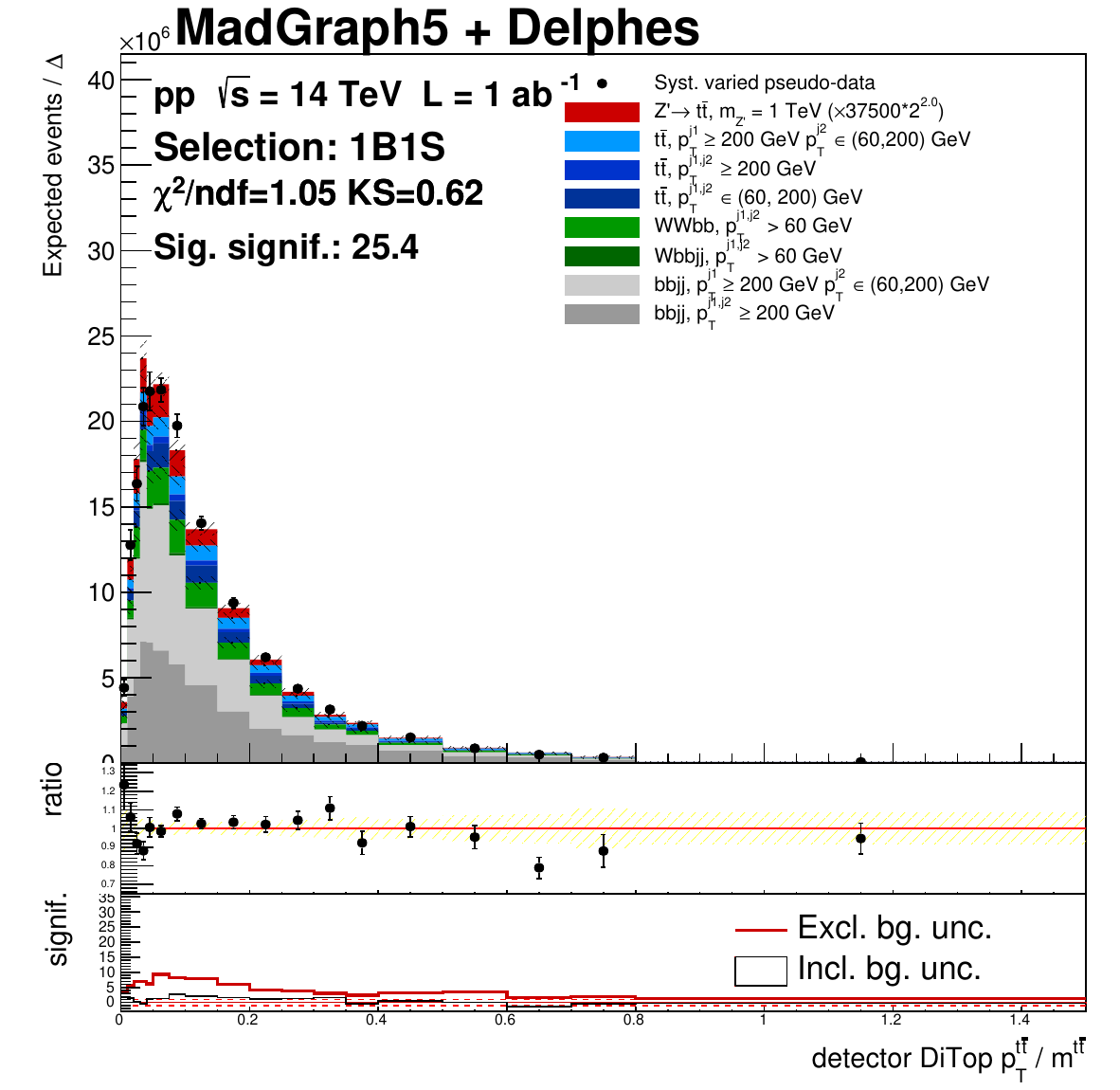} &
\includegraphics[width=0.33\textwidth]{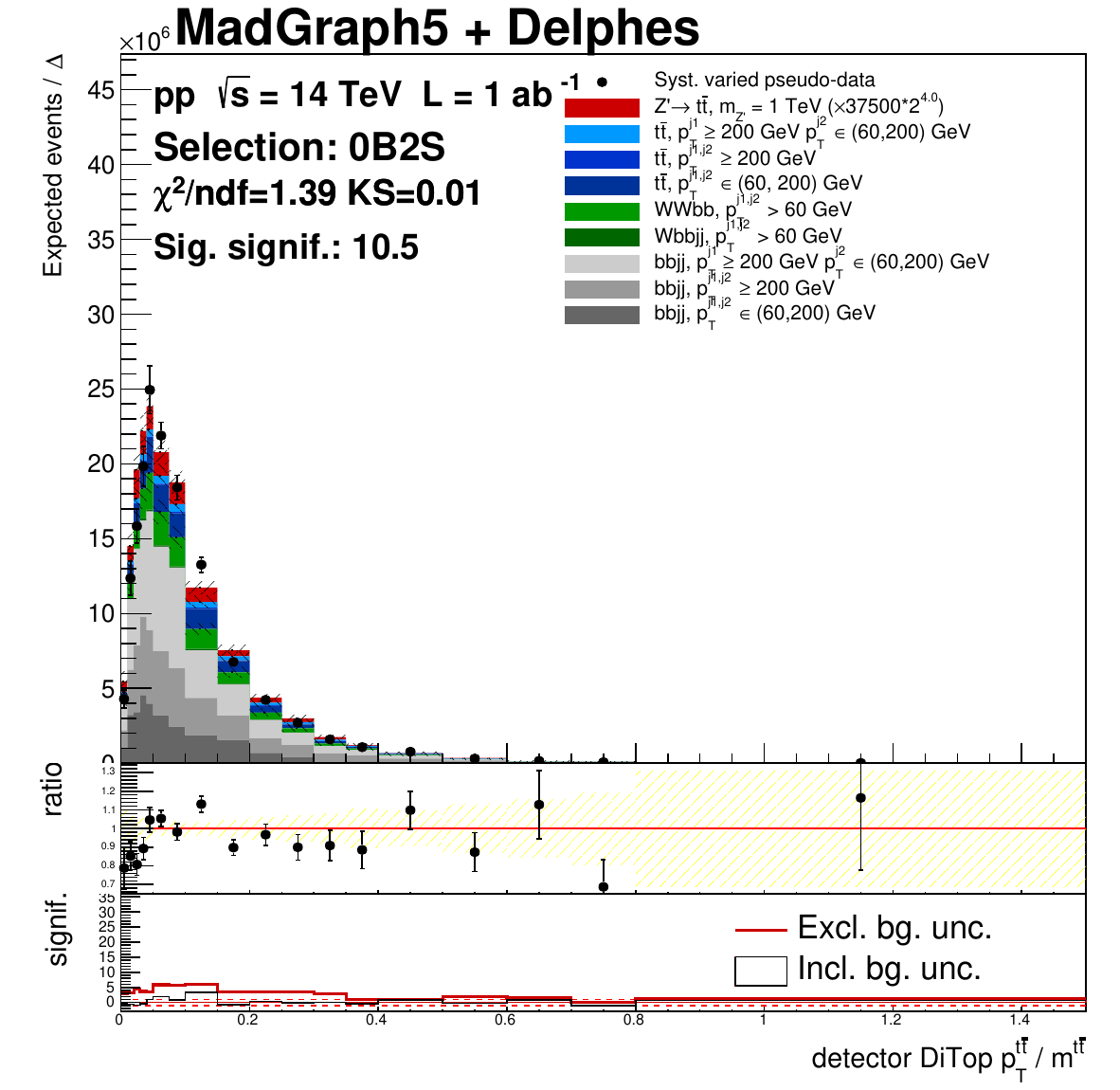} \\
2B0S  &
1B1S  &
0B2S  \\
\end{tabular}
\caption{Mixed comparison with systematics-varied pseudo-data build from sub-samples normalized to their nominal integrals: Stacked $\pt{}$ of the reconstructed top quarks (top) and the $\ttbar$ pair $\pt$ relative to the pair mass (bottom) at the \Delphes{} detector level for the scaled $Z'$ model with $m_{Z'}=1\,$TeV with both the pseudo-data/prediction ratio and the signal significance shown in lower pads, for the boosted-boosted (2B0S, left), boosted-semiboosted (1B1S, middle) and semiboosted-semiboosted (0B2S, right) topologies.
}
\label{fig_stack_MIX_model_zp_compact_RS}
\end{center}
\end{figure*}

\begin{figure*}[h!]
\begin{center}
\begin{tabular}{ccc}
\includegraphics[width=0.33\textwidth]{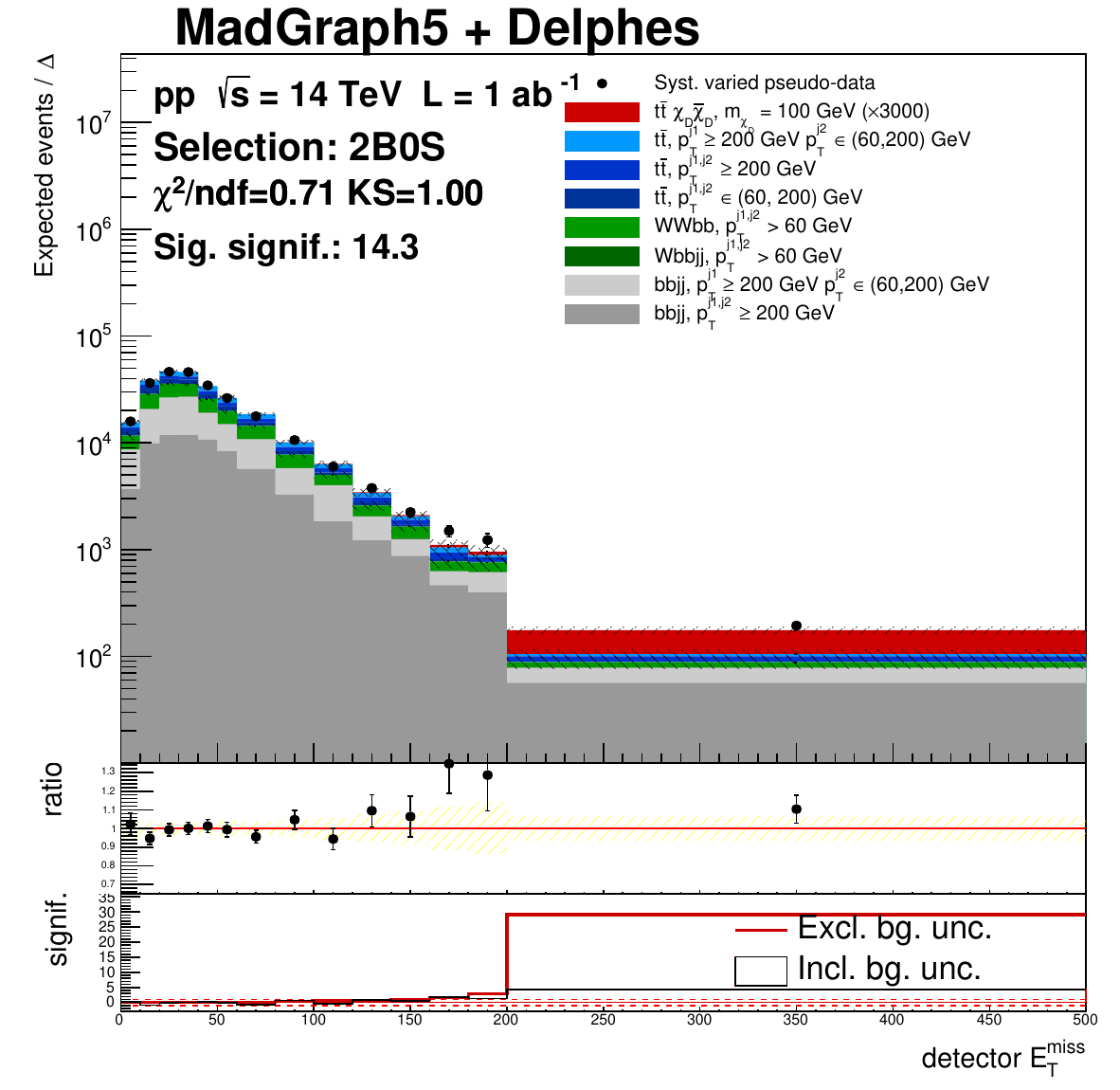} &
\includegraphics[width=0.33\textwidth]{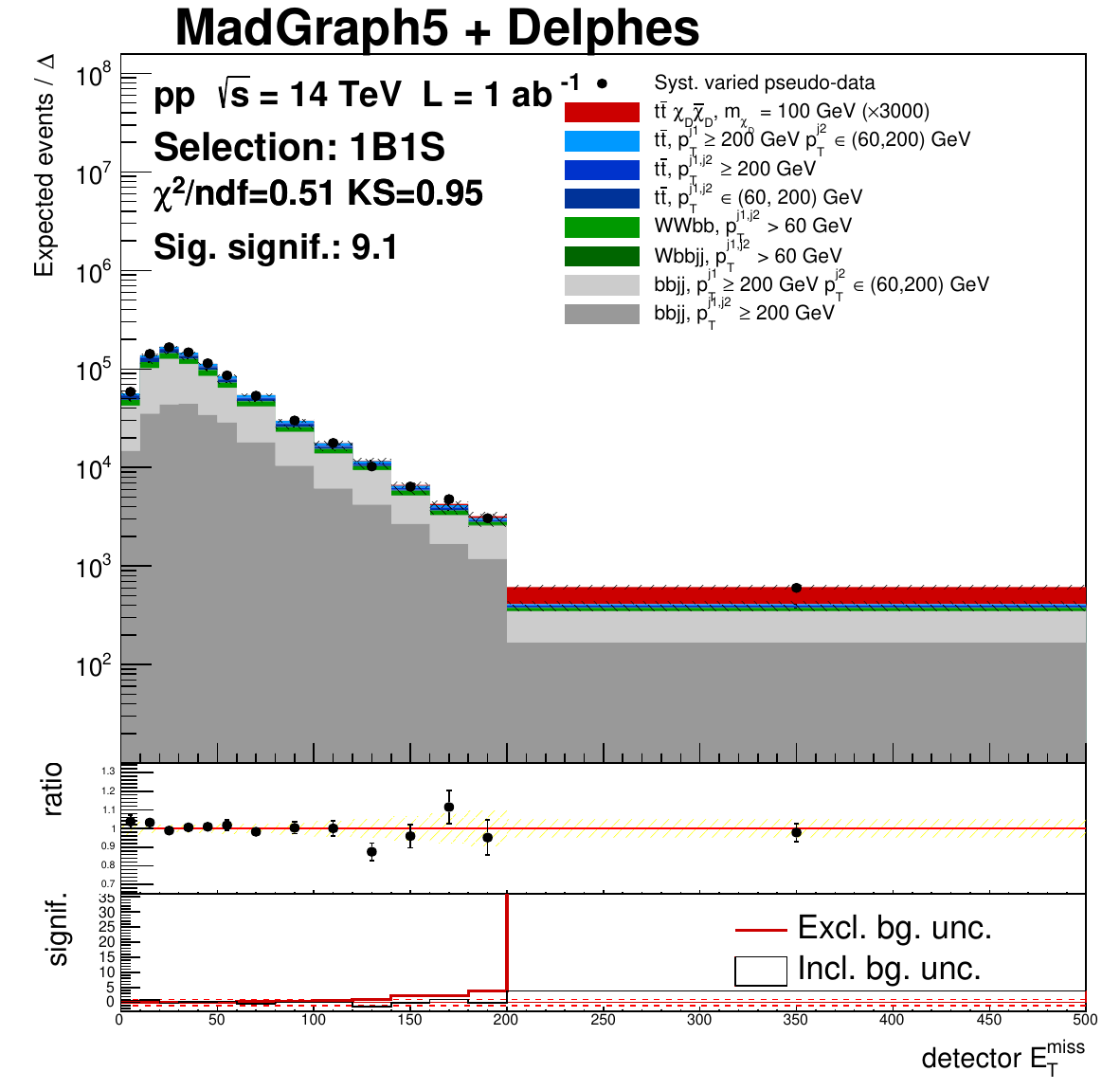} &
\includegraphics[width=0.33\textwidth]{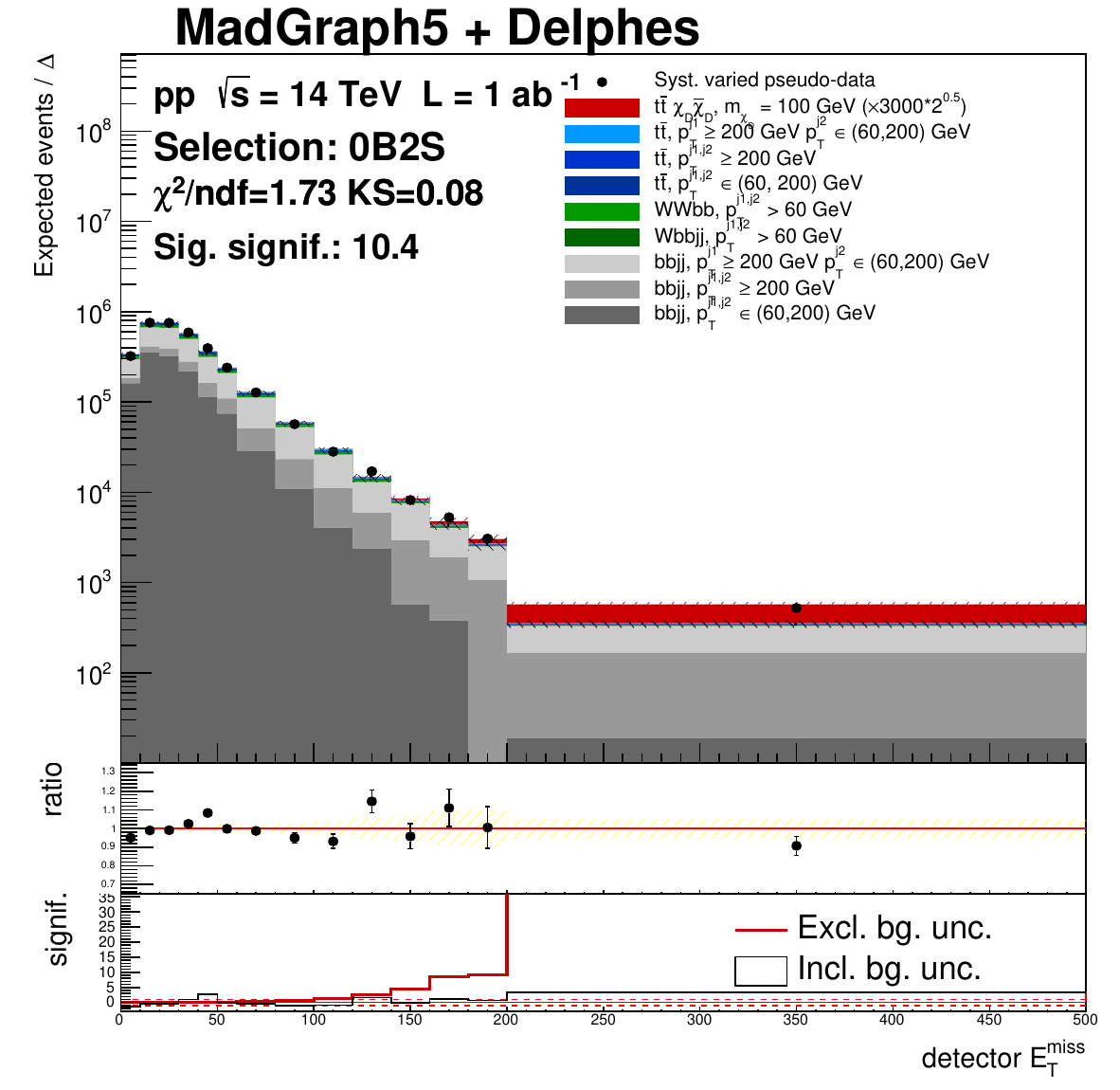} \\
2B0S  &
1B1S  &
0B2S  \\
\end{tabular}
\caption{Mixed comparison with systematics-varied pseudo-data build from sub-samples normalized to their nominal integrals: Stacked absolute value of the missing transverse energy at the \Delphes{} detector level for the scaled $y_0 \rightarrow \chi_D \bar{\chi}_{D},\, m_{\chi_D} = 100\,$ GeV model with both the pseudo-data/prediction ratio and the signal significance shown in lower pads, for the boosted-boosted (2B0S, left), boosted-semiboosted (1B1S, middle) and semiboosted-semiboosted (0B2S, right) topologies.
}
\label{fig_stack_MIX_models_Met_compact_RS}
\end{center}
\end{figure*}

For illustration, the 1D comparison for selected spectra between the systematics-varied pseudo-data and nominal prediction are shown in Figures~\ref{fig_stack_MIX_model_y0_compact_RS}--\ref{fig_stack_MIX_models_Met_compact_RS}.

Finally, the full BH procedure is repeated over pseudo-experiments drawn from the systematics-varied pseudo-data compared to the nominal background prediction.
The best scoring variables resulting from this mixed exercise are shown in Tables~\ref{tab:bestBHscores_y0_MIX}--\ref{tab:bestBHscores_xdxd_MIX}.
There are two competing effects concerning the number of scoring variables in the mixed exercise compared to the nominal one: larger disagreement between the data and prediction versus the smaller selection efficiency, leading to larger statistical uncertainties which are used to build pseudo-experiments. The former leads to less scoring variables while the latter to larger compatibility of more scoring variables.
The same variables are often found to be the best scoring ones, or are replaced by those that scored within $0.5$-$\sigma$ (and actually much less) to the best one in the nominal case.

The sensitivity of variables which contain a relative observable, \emph{i.e.} such that is a ratio to either the $\mtt{}$ or the geometric mean of the two top quark transverse momenta, is evaluated as follows.
The ratio of the number of relative variables to the number of absolute variables, in both cases scoring within 1-$\sigma$ to the best variable, is computed for both analysis paths. A double ratio is then computed between the mixed and nominal exercises ratios. As show in~Table~\ref{tab:ratios_varCounts}, most of the double ratios are above unity, meaning that there are more relative variables scoring within 1-$\sigma$ to the best variable in the mixed exercise path compared to the nominal case.
This proves that, within our model, the relative variables appear more frequently in the mixed exercise and are thus more stable to the systematics differences between the pseud-data and prediction.

Observables robust to the systematics variations of the jet energy scale and resolution appear to be the \ttbar{} system transverse momentum or the $\pout$ variable relative to the \ttbar{} system mass or to the geometric mean of the two top quark transverse momenta, in combination, as a 2D spectrum, with other variables like again the \ttbar{} mass, the mass of the large-$R$ jets four-vectors sum or the sum of the small-$R$ jets transverse momenta and the missing transverse energy.

\clearpage
\clearpage
\begin{table*}[p]
  \centering
\resizebox*{0.48\textheight}{!}{
     \begin{tabular}{l|lll}
\hline 
variable & 2B0S & 1B1S & 0B2S \\ \hline 
  $E^\mathrm{miss}_\mathrm{T}$ vs. $H_\mathrm{T}^{j} + E^\mathrm{miss}_\mathrm{T}$ &    &    &   $\mathbf{6.37} \pm 0.18$  \\ 
  $E^\mathrm{miss}_\mathrm{T}$ vs. $H_\mathrm{T}^{j}$                    &    &   ${6.11} \pm 0.2$  &   $\mathbf{6.43} \pm 0.19$  \\ 
  $E^\mathrm{miss}_\mathrm{T}$ vs. $m^{t\bar{t}}$                        &    &    &   ${6.25} \pm 0.19$  \\ 
  $H_\mathrm{T}^{j} + E^\mathrm{miss}_\mathrm{T}$ vs. $R^{t2,t1}$        &   ${5.99} \pm 0.21$  &    &    \\ 
  $H_\mathrm{T}^{j} + E^\mathrm{miss}_\mathrm{T}$ vs. $m^{t\bar{t}}$     &   ${5.97} \pm 0.19$  &   ${6.11} \pm 0.22$  &   ${6.34} \pm 0.19$  \\ 
  $H_\mathrm{T}^{j} + E^\mathrm{miss}_\mathrm{T}$ vs. $p_\mathrm{T}^{t\bar{t}}$ &    &   ${6.1} \pm 0.2$  &    \\ 
  $H_\mathrm{T}^{j} + E^\mathrm{miss}_\mathrm{T}$ vs. $y_\mathrm{boost}^{t\bar{t}}$ &    &   ${6.15} \pm 0.24$  &    \\ 
  $H_\mathrm{T}^{j} + E^\mathrm{miss}_\mathrm{T}$ vs. jets Sphericity    &   ${5.96} \pm 0.2$  &   ${6.1} \pm 0.22$  &   ${6.24} \pm 0.24$  \\ 
  $H_\mathrm{T}^{j}$ vs. $R^{t2,t1}$                                     &   ${6.01} \pm 0.2$  &    &    \\ 
  $H_\mathrm{T}^{j}$ vs. $m^{t\bar{t}}$                                  &   ${5.99} \pm 0.21$  &   ${6.13} \pm 0.22$  &   ${6.3} \pm 0.22$  \\ 
  $H_\mathrm{T}^{j}$ vs. $p_\mathrm{T}^{t\bar{t}}$                       &    &   ${6.11} \pm 0.22$  &    \\ 
  $H_\mathrm{T}^{j}$ vs. $y_\mathrm{boost}^{t\bar{t}}$                   &    &   ${6.18} \pm 0.24$  &    \\ 
  $H_\mathrm{T}^{j}$ vs. jets Sphericity                                 &    &   ${6.08} \pm 0.22$  &    \\ 
  $\sum m^{J}$ vs. $H_\mathrm{T}^{j} + E^\mathrm{miss}_\mathrm{T}$       &    &   ${6.16} \pm 0.21$  &   ${6.27} \pm 0.19$  \\ 
  $\sum m^{J}$ vs. $H_\mathrm{T}^{j}$                                    &    &   ${6.2} \pm 0.21$  &    \\ 
  $m_{\sum J}^\mathrm{vis}$ vs. $E^\mathrm{miss}_\mathrm{T}$             &    &    &   ${6.23} \pm 0.2$  \\ 
  $m_{\sum J}^\mathrm{vis}$ vs. $H_\mathrm{T}^{j} + E^\mathrm{miss}_\mathrm{T}$ &   $\textcolor{red}{\mathbf{6.19}} \pm 0.17$  &   $\mathbf{6.37} \pm 0.18$  &   $\mathbf{6.42} \pm 0.18$  \\ 
  $m_{\sum J}^\mathrm{vis}$ vs. $H_\mathrm{T}^{j}$                       &   $\mathbf{6.13} \pm 0.19$  &   $\textcolor{red}{\mathbf{6.37}} \pm 0.19$  &   $\mathbf{6.36} \pm 0.21$  \\ 
  $m_{\sum J}^\mathrm{vis}$ vs. $R^{t2,t1}$                              &   ${5.93} \pm 0.24$  &    &    \\ 
  $m_{\sum J}^\mathrm{vis}$ vs. $m^{t\bar{t}}$                           &    &   ${6.15} \pm 0.22$  &    \\ 
  $m_{\sum J}^\mathrm{vis}$ vs. $y_\mathrm{boost}^{t\bar{t}}$            &    &   ${6.07} \pm 0.24$  &    \\ 
  $m_{\sum J}^\mathrm{vis}$ vs. jets Aplanarity                          &   ${5.94} \pm 0.19$  &    &    \\ 
  $p_\mathrm{T}^{t\bar{t}} / \sqrt{p_\mathrm{T}^{t1} p_\mathrm{T}^{t2}}$ vs. $E^\mathrm{miss}_\mathrm{T}$ &    &    &   $\mathbf{6.37} \pm 0.18$  \\ 
  $p_\mathrm{T}^{t\bar{t}} / \sqrt{p_\mathrm{T}^{t1} p_\mathrm{T}^{t2}}$ vs. $H_\mathrm{T}^{j} + E^\mathrm{miss}_\mathrm{T}$ &   ${6.01} \pm 0.23$  &   ${6.19} \pm 0.24$  &   $\mathbf{6.45} \pm 0.16$  \\ 
  $p_\mathrm{T}^{t\bar{t}} / \sqrt{p_\mathrm{T}^{t1} p_\mathrm{T}^{t2}}$ vs. $H_\mathrm{T}^{j}$ &   ${6.02} \pm 0.21$  &   $\mathbf{6.27} \pm 0.22$  &   $\mathbf{6.43} \pm 0.19$  \\ 
  $p_\mathrm{T}^{t\bar{t}} / \sqrt{p_\mathrm{T}^{t1} p_\mathrm{T}^{t2}}$ vs. $\Delta\phi^{t\bar{t}}$ &    &   ${6.08} \pm 0.22$  &    \\ 
  $p_\mathrm{T}^{t\bar{t}} / \sqrt{p_\mathrm{T}^{t1} p_\mathrm{T}^{t2}}$ vs. $\sum m^{J}$ &    &    &   ${6.28} \pm 0.21$  \\ 
  $p_\mathrm{T}^{t\bar{t}} / \sqrt{p_\mathrm{T}^{t1} p_\mathrm{T}^{t2}}$ vs. $m^{t\bar{t}}$ &    &   ${6.12} \pm 0.21$  &   ${6.28} \pm 0.22$  \\ 
  $p_\mathrm{T}^{t\bar{t}} / \sqrt{p_\mathrm{T}^{t1} p_\mathrm{T}^{t2}}$ vs. $m_{\sum J}^\mathrm{vis}$ &   ${6.04} \pm 0.2$  &   ${6.15} \pm 0.24$  &   ${6.28} \pm 0.24$  \\ 
  $p_\mathrm{T}^{t\bar{t}} / \sqrt{p_\mathrm{T}^{t1} p_\mathrm{T}^{t2}}$ vs. jets Sphericity &    &    &   $\mathbf{6.38} \pm 0.21$  \\ 
  $p_\mathrm{T}^{t\bar{t}} / m^{t\bar{t}}$ vs. $E^\mathrm{miss}_\mathrm{T}$ &    &    &   $\mathbf{6.39} \pm 0.19$  \\ 
  $p_\mathrm{T}^{t\bar{t}} / m^{t\bar{t}}$ vs. $H_\mathrm{T}^{j} + E^\mathrm{miss}_\mathrm{T}$ &   $\mathbf{6.17} \pm 0.21$  &   $\mathbf{6.35} \pm 0.2$  &   $\mathbf{6.45} \pm 0.16$  \\ 
  $p_\mathrm{T}^{t\bar{t}} / m^{t\bar{t}}$ vs. $H_\mathrm{T}^{j}$        &   $\mathbf{6.18} \pm 0.2$  &   $\mathbf{6.34} \pm 0.22$  &   $\textcolor{red}{\mathbf{6.48}} \pm 0.15$  \\ 
  $p_\mathrm{T}^{t\bar{t}} / m^{t\bar{t}}$ vs. $m^{t\bar{t}}$            &    &   ${6.21} \pm 0.22$  &   ${6.3} \pm 0.21$  \\ 
  $p_\mathrm{T}^{t\bar{t}} / m^{t\bar{t}}$ vs. $m_{\sum J}^\mathrm{vis}$ &   $\mathbf{6.07} \pm 0.2$  &   $\mathbf{6.27} \pm 0.22$  &   ${6.25} \pm 0.23$  \\ 
  $p_\mathrm{T}^{t\bar{t}} / m^{t\bar{t}}$ vs. $y_\mathrm{boost}^{t\bar{t}}$ &    &   ${6.13} \pm 0.22$  &    \\ 
  $p_\mathrm{T}^{t\bar{t}} / m^{t\bar{t}}$ vs. jets Sphericity           &    &    &   $\mathbf{6.37} \pm 0.2$  \\ 
  $y_\mathrm{boost}^{t\bar{t}}$ vs. $p_\mathrm{T}^{t\bar{t}}$            &    &   ${6.09} \pm 0.22$  &    \\ 
  $|p_\mathrm{out}| / m^{t\bar{t}}$ vs. $E^\mathrm{miss}_\mathrm{T}$     &    &    &   ${6.22} \pm 0.22$  \\ 
  $|p_\mathrm{out}| / m^{t\bar{t}}$ vs. $H_\mathrm{T}^{j} + E^\mathrm{miss}_\mathrm{T}$ &    &    &   ${6.28} \pm 0.23$  \\ 
  $|p_\mathrm{out}| / m^{t\bar{t}}$ vs. $H_\mathrm{T}^{j}$               &    &    &   ${6.28} \pm 0.22$  \\ 
  $|p_\mathrm{out}| / m^{t\bar{t}}$ vs. $p_\mathrm{T}^{t\bar{t}} / \sqrt{p_\mathrm{T}^{t1} p_\mathrm{T}^{t2}}$ &    &    &   ${6.33} \pm 0.22$  \\ 
  $|p_\mathrm{out}| / m^{t\bar{t}}$ vs. $p_\mathrm{T}^{t\bar{t}} / m^{t\bar{t}}$ &    &    &   ${6.24} \pm 0.26$  \\ 
  $|p_\mathrm{out}| / m^{t\bar{t}}$ vs. jets Sphericity                  &    &    &   ${6.24} \pm 0.21$  \\ 
     \hline
\end{tabular}

  }
  \caption{The mixed excercise: systematics-varied pseudodata and nominal backgrounds; the best 1D and/or 2D \BumpHunter{} variables scores ($\log t$) and the statistical uncertainty over 100 pseudo-experiments for stacked backgrounds (including the $\ttbar{}$ samples) and the $y_0$ signal model with $m_{y_0} = 1\,$TeV in the pseudo-data over the topologies. Empty fields indicate that the variable did not score within $1\sigma$ to the best scoring variable in given topology (red) while those scoring within $0.5\sigma$ are marked in bold.}
  \label{tab:bestBHscores_y0_MIX}
\end{table*}

\clearpage

\begin{table*}[p]
    \centering
\resizebox*{0.48\textheight}{!}{
     \begin{tabular}{l|lll}
\hline 
variable & 2B0S & 1B1S & 0B2S \\ \hline 
  $H_\mathrm{T}^{j} + E^\mathrm{miss}_\mathrm{T}$ vs. $m^{t\bar{t}}$     &   ${6.29} \pm 0.17$  &   ${6.25} \pm 0.2$  &   ${6.27} \pm 0.25$  \\ 
  $H_\mathrm{T}^{j}$ vs. $m^{t\bar{t}}$                                  &    &   ${6.28} \pm 0.19$  &   $\mathbf{6.33} \pm 0.23$  \\ 
  $\delta^{t\bar{t}}$ vs. $m^{t\bar{t}}$                                 &    &    &   ${6.15} \pm 0.27$  \\ 
  $m^{t\bar{t}}$ vs. $\Delta\phi^{t\bar{t}}$                             &    &    &   ${6.15} \pm 0.25$  \\ 
  $m_{\sum J}^\mathrm{vis}$ vs. $H_\mathrm{T}^{j} + E^\mathrm{miss}_\mathrm{T}$ &   $\textcolor{red}{\mathbf{6.48}} \pm 0.12$  &   $\textcolor{red}{\mathbf{6.46}} \pm 0.15$  &   $\mathbf{6.3} \pm 0.25$  \\ 
  $m_{\sum J}^\mathrm{vis}$ vs. $H_\mathrm{T}^{j}$                       &   $\mathbf{6.45} \pm 0.12$  &   $\mathbf{6.44} \pm 0.16$  &   $\mathbf{6.39} \pm 0.22$  \\ 
  $m_{\sum J}^\mathrm{vis}$ vs. $\Delta\phi^{t\bar{t}}$                  &    &   ${6.22} \pm 0.21$  &    \\ 
  $m_{\sum J}^\mathrm{vis}$ vs. $\chi^{t\bar{t}}$                        &    &    &   ${6.14} \pm 0.25$  \\ 
  $m_{\sum J}^\mathrm{vis}$ vs. $m^{t\bar{t}}$                           &    &   ${6.34} \pm 0.18$  &   $\mathbf{6.33} \pm 0.2$  \\ 
  $m_{\sum J}^\mathrm{vis}$ vs. jets Aplanarity                          &   $\mathbf{6.4} \pm 0.15$  &    &    \\ 
  $m_{\sum J}^\mathrm{vis}$ vs. jets Sphericity                          &    &   ${6.23} \pm 0.22$  &    \\ 
  $p_\mathrm{T}^{t\bar{t}} / \sqrt{p_\mathrm{T}^{t1} p_\mathrm{T}^{t2}}$ vs. $H_\mathrm{T}^{j} + E^\mathrm{miss}_\mathrm{T}$ &   ${6.26} \pm 0.21$  &   ${6.19} \pm 0.27$  &   ${6.24} \pm 0.25$  \\ 
  $p_\mathrm{T}^{t\bar{t}} / \sqrt{p_\mathrm{T}^{t1} p_\mathrm{T}^{t2}}$ vs. $H_\mathrm{T}^{j}$ &   ${6.31} \pm 0.19$  &   ${6.25} \pm 0.22$  &   ${6.28} \pm 0.24$  \\ 
  $p_\mathrm{T}^{t\bar{t}} / \sqrt{p_\mathrm{T}^{t1} p_\mathrm{T}^{t2}}$ vs. $m^{t\bar{t}}$ &    &   ${6.22} \pm 0.21$  &   $\mathbf{6.37} \pm 0.23$  \\ 
  $p_\mathrm{T}^{t\bar{t}} / \sqrt{p_\mathrm{T}^{t1} p_\mathrm{T}^{t2}}$ vs. $m_{\sum J}^\mathrm{vis}$ &   $\mathbf{6.45} \pm 0.15$  &   ${6.33} \pm 0.21$  &   $\mathbf{6.42} \pm 0.21$  \\ 
  $p_\mathrm{T}^{t\bar{t}} / \sqrt{p_\mathrm{T}^{t1} p_\mathrm{T}^{t2}}$ vs. jets Sphericity &    &    &   $\mathbf{6.32} \pm 0.26$  \\ 
  $p_\mathrm{T}^{t\bar{t}} / m^{t\bar{t}}$ vs. $H_\mathrm{T}^{j} + E^\mathrm{miss}_\mathrm{T}$ &   $\mathbf{6.4} \pm 0.17$  &   ${6.25} \pm 0.23$  &   ${6.29} \pm 0.24$  \\ 
  $p_\mathrm{T}^{t\bar{t}} / m^{t\bar{t}}$ vs. $H_\mathrm{T}^{j}$        &   $\mathbf{6.39} \pm 0.17$  &   ${6.29} \pm 0.22$  &   $\mathbf{6.33} \pm 0.24$  \\ 
  $p_\mathrm{T}^{t\bar{t}} / m^{t\bar{t}}$ vs. $\delta^{t\bar{t}}$       &    &    &   ${6.14} \pm 0.25$  \\ 
  $p_\mathrm{T}^{t\bar{t}} / m^{t\bar{t}}$ vs. $m^{t\bar{t}}$            &    &   ${6.21} \pm 0.21$  &   $\textcolor{red}{\mathbf{6.44}} \pm 0.17$  \\ 
  $p_\mathrm{T}^{t\bar{t}} / m^{t\bar{t}}$ vs. $m_{\sum J}^\mathrm{vis}$ &   $\mathbf{6.47} \pm 0.13$  &   $\mathbf{6.35} \pm 0.19$  &   $\mathbf{6.37} \pm 0.21$  \\ 
  $p_\mathrm{T}^{t\bar{t}} / m^{t\bar{t}}$ vs. jets Sphericity           &    &    &   $\mathbf{6.34} \pm 0.24$  \\ 
  $p_\mathrm{T}^{t\bar{t}}$ vs. $m^{t\bar{t}}$                           &    &    &   ${6.28} \pm 0.23$  \\ 
  $y_\mathrm{boost}^{t\bar{t}}$ vs. $m^{t\bar{t}}$                       &    &    &   ${6.18} \pm 0.23$  \\ 
  $|p_\mathrm{out}| / m^{t\bar{t}}$ vs. $H_\mathrm{T}^{j} + E^\mathrm{miss}_\mathrm{T}$ &    &    &   ${6.21} \pm 0.27$  \\ 
  $|p_\mathrm{out}| / m^{t\bar{t}}$ vs. $H_\mathrm{T}^{j}$               &    &    &   ${6.18} \pm 0.23$  \\ 
  $|p_\mathrm{out}| / m^{t\bar{t}}$ vs. $m^{t\bar{t}}$                   &    &    &   ${6.13} \pm 0.26$  \\ 
  jets Sphericity vs. $m^{t\bar{t}}$                                     &    &    &   ${6.17} \pm 0.24$  \\ 
     \hline
\end{tabular}

}
  \caption{The mixed excercise: systematics-varied pseudodata and nominal backgrounds; the best 1D and/or 2D \BumpHunter{} variables scores ($\log t$) and the statistical uncertainty over 100 pseudo-experiments for stacked backgrounds (including the $\ttbar{}$ samples) and for the $Z'$ signal model with $m_{Z'} = 1\,$TeV in the pseudo-data over the topologies. Empty fields indicate that the variable did not score within $1\sigma$ to the best scoring variable in given topology (red) while those scoring within $0.5\sigma$ are marked in bold.}
  \label{tab:bestBHscores_zp_MIX}
\end{table*}

\clearpage
\begin{table*}[p]
    \centering
\resizebox*{0.415\textheight}{!}{
     \begin{tabular}{l|lll}
\hline 
variable & 2B0S & 1B1S & 0B2S \\ \hline 
  $E^\mathrm{miss}_\mathrm{T}$                                           &   ${4.76} \pm 0.42$  &    &    \\ 
  $E^\mathrm{miss}_\mathrm{T}$ vs. $H_\mathrm{T}^{j} + E^\mathrm{miss}_\mathrm{T}$ &   $\mathbf{5.66} \pm 0.29$  &   ${5.62} \pm 0.27$  &   ${5.79} \pm 0.31$  \\ 
  $E^\mathrm{miss}_\mathrm{T}$ vs. $H_\mathrm{T}^{j}$                    &   $\mathbf{5.69} \pm 0.28$  &   ${5.7} \pm 0.24$  &   $\mathbf{5.99} \pm 0.32$  \\ 
  $E^\mathrm{miss}_\mathrm{T}$ vs. $R^{t2,t1}$                           &   ${5.33} \pm 0.28$  &    &    \\ 
  $E^\mathrm{miss}_\mathrm{T}$ vs. $\chi^{t\bar{t}}$                     &   $\mathbf{5.42} \pm 0.27$  &    &    \\ 
  $E^\mathrm{miss}_\mathrm{T}$ vs. $\delta^{t\bar{t}}$                   &   ${5.35} \pm 0.29$  &   ${5.61} \pm 0.26$  &    \\ 
  $E^\mathrm{miss}_\mathrm{T}$ vs. $\sum m^{J}$                          &   ${5.37} \pm 0.28$  &   ${5.46} \pm 0.29$  &    \\ 
  $E^\mathrm{miss}_\mathrm{T}$ vs. $p_\mathrm{T}^{t\bar{t}}$             &   $\mathbf{5.5} \pm 0.27$  &    &    \\ 
  $E^\mathrm{miss}_\mathrm{T}$ vs. $y_\mathrm{boost}^{t\bar{t}}$         &   ${5.39} \pm 0.26$  &    &    \\ 
  $E^\mathrm{miss}_\mathrm{T}$ vs. jets Aplanarity                       &    &   ${5.65} \pm 0.25$  &    \\ 
  $E^\mathrm{miss}_\mathrm{T}$ vs. jets Sphericity                       &   ${5.37} \pm 0.24$  &    &    \\ 
  $H_\mathrm{T}^{j} + E^\mathrm{miss}_\mathrm{T}$ vs. $H_\mathrm{T}^{j}$ &   $\textcolor{red}{\mathbf{6.48}} \pm 0.14$  &   $\textcolor{red}{\mathbf{6.6}} \pm 0.0$  &   $\textcolor{red}{\mathbf{6.46}} \pm 0.16$  \\ 
  $H_\mathrm{T}^{j} + E^\mathrm{miss}_\mathrm{T}$ vs. $R^{t2,t1}$        &    &   $\mathbf{5.66} \pm 0.29$  &    \\ 
  $H_\mathrm{T}^{j} + E^\mathrm{miss}_\mathrm{T}$ vs. $\chi^{t\bar{t}}$  &   ${5.06} \pm 0.33$  &   $\mathbf{5.56} \pm 0.3$  &    \\ 
  $H_\mathrm{T}^{j} + E^\mathrm{miss}_\mathrm{T}$ vs. $m^{t\bar{t}}$     &    &   ${5.53} \pm 0.27$  &    \\ 
  $H_\mathrm{T}^{j} + E^\mathrm{miss}_\mathrm{T}$ vs. $p_\mathrm{T}^{t\bar{t}}$ &    &   $\mathbf{5.59} \pm 0.29$  &    \\ 
  $H_\mathrm{T}^{j}$ vs. $R^{t2,t1}$                                     &   ${5.36} \pm 0.28$  &   $\mathbf{5.65} \pm 0.28$  &    \\ 
  $H_\mathrm{T}^{j}$ vs. $\chi^{t\bar{t}}$                               &    &   ${5.6} \pm 0.25$  &    \\ 
  $H_\mathrm{T}^{j}$ vs. $\delta^{t\bar{t}}$                             &    &    &   ${5.72} \pm 0.34$  \\ 
  $H_\mathrm{T}^{j}$ vs. $m^{t\bar{t}}$                                  &    &   ${5.58} \pm 0.29$  &    \\ 
  $H_\mathrm{T}^{j}$ vs. $p_\mathrm{T}^{t\bar{t}}$                       &    &   $\mathbf{5.8} \pm 0.26$  &    \\ 
  $H_\mathrm{T}^{j}$ vs. $y_\mathrm{boost}^{t\bar{t}}$                   &   ${5.26} \pm 0.28$  &    &    \\ 
  $m^{t\bar{t}} / \sqrt{p_\mathrm{T}^{t1} p_\mathrm{T}^{t2}}$ vs. $E^\mathrm{miss}_\mathrm{T}$ &   ${5.32} \pm 0.27$  &    &    \\ 
  $m^{t\bar{t}} / \sqrt{p_\mathrm{T}^{t1} p_\mathrm{T}^{t2}}$ vs. $H_\mathrm{T}^{j} + E^\mathrm{miss}_\mathrm{T}$ &    &   ${5.56} \pm 0.28$  &    \\ 
  $m^{t\bar{t}} / \sqrt{p_\mathrm{T}^{t1} p_\mathrm{T}^{t2}}$ vs. $H_\mathrm{T}^{j}$ &    &   ${5.59} \pm 0.26$  &    \\ 
  $m^{t\bar{t}} / \sqrt{p_\mathrm{T}^{t1} p_\mathrm{T}^{t2}}$ vs. $\Delta\phi^{t\bar{t}}$ &    &   ${5.36} \pm 0.33$  &    \\ 
  $m^{t\bar{t}} / \sqrt{p_\mathrm{T}^{t1} p_\mathrm{T}^{t2}}$ vs. $m^{t\bar{t}}$ &    &    &   ${5.52} \pm 0.46$  \\ 
  $m_{\sum J}^\mathrm{vis}$ vs. $E^\mathrm{miss}_\mathrm{T}$             &   ${5.38} \pm 0.27$  &   ${5.68} \pm 0.25$  &    \\ 
  $m_{\sum J}^\mathrm{vis}$ vs. $H_\mathrm{T}^{j} + E^\mathrm{miss}_\mathrm{T}$ &   ${5.47} \pm 0.24$  &   ${5.72} \pm 0.25$  &    \\ 
  $m_{\sum J}^\mathrm{vis}$ vs. $H_\mathrm{T}^{j}$                       &   ${5.28} \pm 0.27$  &   $\mathbf{5.72} \pm 0.26$  &   ${5.75} \pm 0.33$  \\ 
  $m_{\sum J}^\mathrm{vis}$ vs. $R^{t2,t1}$                              &   ${5.22} \pm 0.31$  &   ${5.53} \pm 0.27$  &    \\ 
  $m_{\sum J}^\mathrm{vis}$ vs. $\delta^{t\bar{t}}$                      &    &   ${5.19} \pm 0.36$  &    \\ 
  $m_{\sum J}^\mathrm{vis}$ vs. $m^{t\bar{t}}$                           &    &   ${5.52} \pm 0.27$  &    \\ 
  $m_{\sum J}^\mathrm{vis}$ vs. jets Sphericity                          &    &   ${5.45} \pm 0.32$  &    \\ 
  $p_\mathrm{T}^{t\bar{t}} / \sqrt{p_\mathrm{T}^{t1} p_\mathrm{T}^{t2}}$ vs. $E^\mathrm{miss}_\mathrm{T}$ &   $\mathbf{5.69} \pm 0.23$  &   ${5.67} \pm 0.24$  &    \\ 
  $p_\mathrm{T}^{t\bar{t}} / \sqrt{p_\mathrm{T}^{t1} p_\mathrm{T}^{t2}}$ vs. $H_\mathrm{T}^{j} + E^\mathrm{miss}_\mathrm{T}$ &   ${5.26} \pm 0.28$  &   ${5.57} \pm 0.27$  &   $\mathbf{5.97} \pm 0.32$  \\ 
  $p_\mathrm{T}^{t\bar{t}} / \sqrt{p_\mathrm{T}^{t1} p_\mathrm{T}^{t2}}$ vs. $H_\mathrm{T}^{j}$ &    &   $\mathbf{5.74} \pm 0.26$  &   $\mathbf{5.96} \pm 0.3$  \\ 
  $p_\mathrm{T}^{t\bar{t}} / \sqrt{p_\mathrm{T}^{t1} p_\mathrm{T}^{t2}}$ vs. $\Delta\phi^{t\bar{t}}$ &    &   ${5.44} \pm 0.29$  &    \\ 
  $p_\mathrm{T}^{t\bar{t}} / \sqrt{p_\mathrm{T}^{t1} p_\mathrm{T}^{t2}}$ vs. $m^{t\bar{t}}$ &    &   ${5.48} \pm 0.28$  &    \\ 
  $p_\mathrm{T}^{t\bar{t}} / \sqrt{p_\mathrm{T}^{t1} p_\mathrm{T}^{t2}}$ vs. $m_{\sum J}^\mathrm{vis}$ &   ${5.4} \pm 0.24$  &   $\mathbf{5.62} \pm 0.28$  &   $\mathbf{5.86} \pm 0.33$  \\ 
  $p_\mathrm{T}^{t\bar{t}} / \sqrt{p_\mathrm{T}^{t1} p_\mathrm{T}^{t2}}$ vs. jets Sphericity &    &    &   $\mathbf{5.97} \pm 0.31$  \\ 
  $p_\mathrm{T}^{t\bar{t}} / m^{t\bar{t}}$ vs. $E^\mathrm{miss}_\mathrm{T}$ &   ${5.43} \pm 0.25$  &    &    \\ 
  $p_\mathrm{T}^{t\bar{t}} / m^{t\bar{t}}$ vs. $H_\mathrm{T}^{j} + E^\mathrm{miss}_\mathrm{T}$ &   ${5.39} \pm 0.25$  &   ${5.55} \pm 0.27$  &   ${5.81} \pm 0.31$  \\ 
  $p_\mathrm{T}^{t\bar{t}} / m^{t\bar{t}}$ vs. $H_\mathrm{T}^{j}$        &    &   $\mathbf{5.7} \pm 0.28$  &   ${5.78} \pm 0.31$  \\ 
  $p_\mathrm{T}^{t\bar{t}} / m^{t\bar{t}}$ vs. $m^{t\bar{t}}$            &    &   $\mathbf{5.59} \pm 0.3$  &    \\ 
  $p_\mathrm{T}^{t\bar{t}} / m^{t\bar{t}}$ vs. $m_{\sum J}^\mathrm{vis}$ &    &    &   ${5.79} \pm 0.34$  \\ 
  $p_\mathrm{T}^{t\bar{t}} / m^{t\bar{t}}$ vs. jets Sphericity           &   ${5.34} \pm 0.25$  &    &   $\mathbf{5.86} \pm 0.31$  \\ 
  $p_\mathrm{T}^{t\bar{t}}$ vs. $m^{t\bar{t}}$                           &    &   ${5.38} \pm 0.32$  &    \\ 
  $y_\mathrm{boost}^{t\bar{t}}$ vs. $p_\mathrm{T}^{t\bar{t}}$            &    &   ${5.46} \pm 0.3$  &    \\ 
  $|p_\mathrm{out}| / m^{t\bar{t}}$ vs. $H_\mathrm{T}^{j} + E^\mathrm{miss}_\mathrm{T}$ &    &   ${5.48} \pm 0.29$  &   $\mathbf{5.98} \pm 0.31$  \\ 
  $|p_\mathrm{out}| / m^{t\bar{t}}$ vs. $H_\mathrm{T}^{j}$               &    &    &   ${5.91} \pm 0.25$  \\ 
  $|p_\mathrm{out}| / m^{t\bar{t}}$ vs. $m^{t\bar{t}} / \sqrt{p_\mathrm{T}^{t1} p_\mathrm{T}^{t2}}$ &    &   $\mathbf{5.66} \pm 0.31$  &    \\ 
  $|p_\mathrm{out}| / m^{t\bar{t}}$ vs. $m_{\sum J}^\mathrm{vis}$        &    &    &   ${5.82} \pm 0.3$  \\ 
  $|p_\mathrm{out}| / m^{t\bar{t}}$ vs. $p_\mathrm{T}^{t\bar{t}} / \sqrt{p_\mathrm{T}^{t1} p_\mathrm{T}^{t2}}$ &    &   $\mathbf{5.61} \pm 0.31$  &   $\mathbf{5.89} \pm 0.3$  \\ 
  $|p_\mathrm{out}| / m^{t\bar{t}}$ vs. $p_\mathrm{T}^{t\bar{t}} / m^{t\bar{t}}$ &    &    &   $\mathbf{5.75} \pm 0.38$  \\ 
  $|p_\mathrm{out}| / m^{t\bar{t}}$ vs. $p_\mathrm{T}^{t\bar{t}}$        &    &   ${5.36} \pm 0.32$  &    \\ 
  $|p_\mathrm{out}| / m^{t\bar{t}}$ vs. jets Sphericity                  &    &   ${5.46} \pm 0.31$  &    \\ 
  jets Sphericity vs. $R^{t2,t1}$                                        &   ${5.24} \pm 0.3$  &    &    \\ 
     \hline
\end{tabular}

}
\caption{The mixed excercise: systematics-varied pseudodata and nominal backgrounds; the best 1D and/or 2D \BumpHunter{} variable scores ($\log t$) and the statistical uncertainty over 100 pseudo-experiments for stacked backgrounds (including the $\ttbar{}$ samples) and for the DM \ttbar{}-associated pair production $\ttbar{} + \chi_D \bar{\chi}_D$ signal model with $m_{\chi_D} = 100\,$GeV in the pseudo-data over the topologies. Empty fields indicate that the variable did not score within $2.5\sigma$ ($4\sigma$) in the 0B2S (1B1S and 2B0S) topologies to the best scoring variable in given topology (red) while those scoring within $2.2\sigma$ ($3.5\sigma$) are marked in bold.}
  \label{tab:bestBHscores_xdxd_MIX}
\end{table*}

\begin{table}[!ht]
\begin{tabular}{cc|r}
model & topology & variables counts double ratio \\ \hline 
$y_0$, $m_{y_0} = 1000\,$GeV &  2B0S   & 0.667/0.562 = 6/9 / 9/16 = \textbf{1.185} \\
$y_0$, $m_{y_0} = 1000\,$GeV &  1B1S   & 0.625/0.645 = 10/16 / 20/31 = \textbf{0.969} \\
$y_0$, $m_{y_0} = 1000\,$GeV &  0B2S   & 1.900/1.474 = 19/10 / 28/19 = \textbf{1.289} \\ \hline
$Z'$, $m_{Z'} = 1000\,$GeV  &  2B0S   & 1.500/1.000 = 6/4 / 8/8 = \textbf{1.500} \\
$Z'$, $m_{Z'} = 1000\,$GeV  &  1B1S   & 1.143/0.571 = 8/7 / 8/14 = \textbf{2.000} \\
$Z'$, $m_{Z'} = 1000\,$GeV  &  0B2S   & 1.273/1.500 = 14/11 / 12/8 = \textbf{0.848} \\ \hline
$\chi_D \bar{\chi}_D + \ttbar{}$, $m_{\chi_D} = 100\,$GeV &  2B0S  & 0.368/0.333 = 7/19 / 6/18 = \textbf{1.105} \\
$\chi_D \bar{\chi}_D + \ttbar{}$, $m_{\chi_D} = 100\,$GeV &  1B1S  & 0.739/0.667 = 17/23 / 8/12 = \textbf{1.109} \\
$\chi_D \bar{\chi}_D + \ttbar{}$, $m_{\chi_D} = 100\,$GeV &  0B2S  & 2.800/1.667 = 14/5 / 5/3 = \textbf{1.680} \\
  \end{tabular}
\caption{Double ratios between the mixed and the nominal exercise cases of the counts of relative over non-relative variables scoring within 1-$\sigma$ to the best scoring variable fro each signal model in each topology.}
\label{tab:ratios_varCounts}
\end{table}

\section{Conclusions}

We have presented a~realistic analysis of $\ttbar{}$ final states using SM and BSM-enhanced signal samples and non-$\ttbar{}$ backgrounds illustrated on 1D pseudo-data stacked spectra corresponding to the integrated luminosity of $1\,\mathrm{ab}^{-1}$ (corresponding to the early High Luminosity LHC programme which should lead to about $3\,\mathrm{ab}^{-1}$ in 2029--2032). Focusing in next on variables with the potential to distinguish the signal of a~resonant $\ttbar$ production or $\ttbar$-associated DM pair production at the future LHC energy of 14~TeV, a detailed comparison is performed on a smaller luminosity of $0.01\,\mathrm{ab}^{-1}$ which matches the generated samples luminosities, with the same signal strengths.

The models with \ttbar{} production mediated by a~vector ($Z'$) or scalar ($y_0$) predict the expected peak in the \ttbar{} invariant mass in all studied 2B0S, 1B1S and 0B2S topologies. The model with \ttbar{}-associated production of a~pair of invisible DM particles provides a~broad excess in the \ttbar{} mass in all considered topologies. All models predict a~broad excess in top quark transverse momentum but also in other spectra. The addition of the semiboosted topologies into the selection can add 10--15\% events in the $Z'$ mass range of 1--1.5~TeV.

The 2D variable $\HT+\Etmiss$ versus the invariant mass of the 4-vector sum of large-$R$ jets $m^\mathrm{vis}_{\sum J}$ performs very well in terms of the signal significance. 
Other powerful variables are the $\ttbar$ rest frame angular variables like $\Chittbar$ or $\Yboost$ 
and relative dimensionless $\pout / \mtt$, $\pttt / \mtt$ and $\pttt / \sqrt{\ptto{1}\ptto{2}}$ which exhibit a~large signal significance potential while they are also less sensitive to experimental uncertainties related to the jet energy scale and resolution, thus being possibly robust also in real measurements. The signal separation for the $\ttbar$-associated production of a~pair of invisible DM particles in terms of the 2D distribution of $\Etmiss$ vs. $\HT + \Etmiss$ seems especially simple yet powerful. We note that the discriminating power of a~2D variable is not a~straightforward function of the statistical correlation factor between the two 1D variables. 

The sensitivity study of the results and variables to effects of experimental systematic uncertainties shows a good stability and concludes that observables containing a relative quantity, \emph{i.e.} an observable ratio to the $\mtt{}$ or the top quark transverse momenta geometric mean, are more frequently scoring close to the best-scoring variable in the \BumpHunter{} test, proving that such variables are indeed more systematics-robust.

The proposed new 1D or 2D variables may prove useful for signal bump hunting in LHC searches or serve as input for more complex classification approaches.

In conclusions, we have identified statistically and systematics-robust variables with the potential to enhance signal significance in sub-regions of these variables. 
The 2D variables offer a~clear physics insight into regions which can be avoided to enhance the signal significance. Such variables can be also used as input for more powerful classifiers.
The application and extension of the \BumpHunter{} algorithm presented in two dimensions are well-performing and robust, highly correlated to results of a~likelihood fit of the signal strength and can be regarded as an intermediate approach which bridges the gaps between classical searches using a~single sensitive variable and multidimensional techniques followed by a~likelihood fit.

\section*{Acknowledgments}

The author would like to thank the Czech Science Foundation project GA\v{C}R 19-21484S for the support of this work.

\bigskip
\noindent 
\textbf{Data Availability Statement}
\\ The code, scripts, configuration and simulation card files supporting this study can be found in~\cite{github}.


\bibliography{main}{}
\bibliographystyle{unsrt}
\appendix
\clearpage
\section{Grooming methods comparison}
\label{app:groom_cmp}

\begin{figure}[!hp]
\begin{center}
\begin{tabular}{c}
\includegraphics[width=0.95\textwidth]{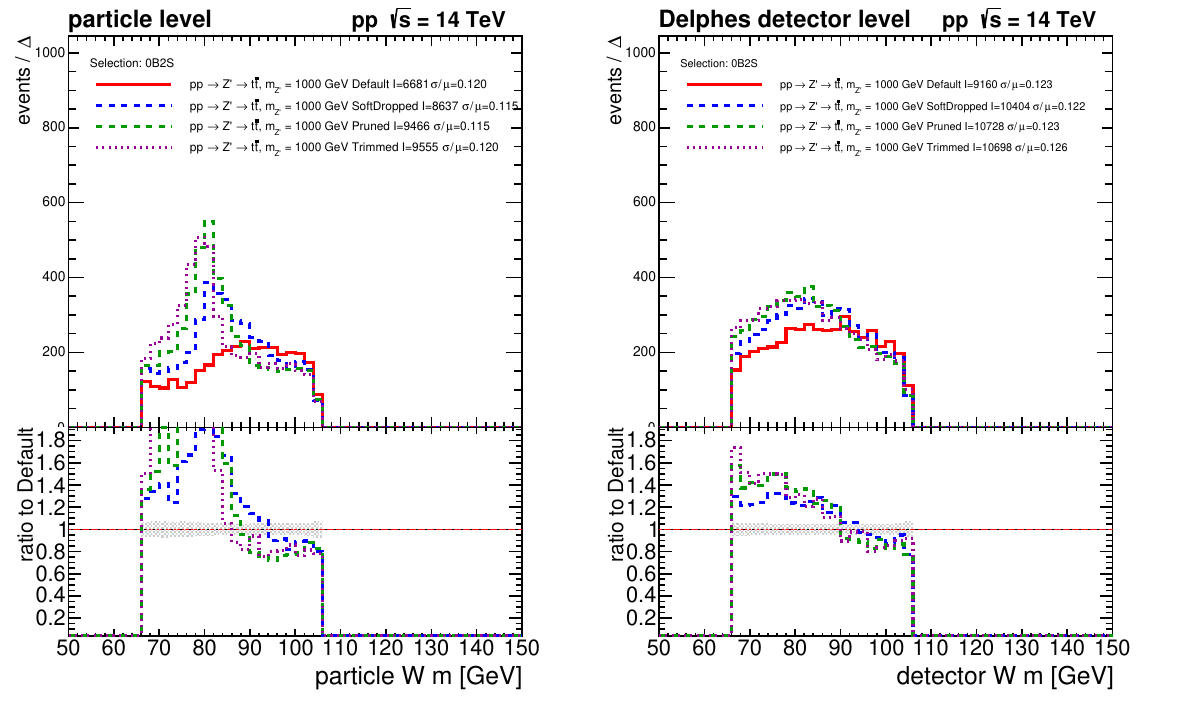}\\
\end{tabular}
\caption{Comparison between variously groomed jets for the mass of the reconstructed $W$ bosons in the semiboosted-semiboosted (0B2S) topology. Left: particle level, right: Delphes ATLAS detector level.}
\label{fig_shapes_WMass_0B2S}
\end{center}
\end{figure}

\begin{figure}[!hp]
\begin{center}
\begin{tabular}{c}
\includegraphics[width=0.95\textwidth]{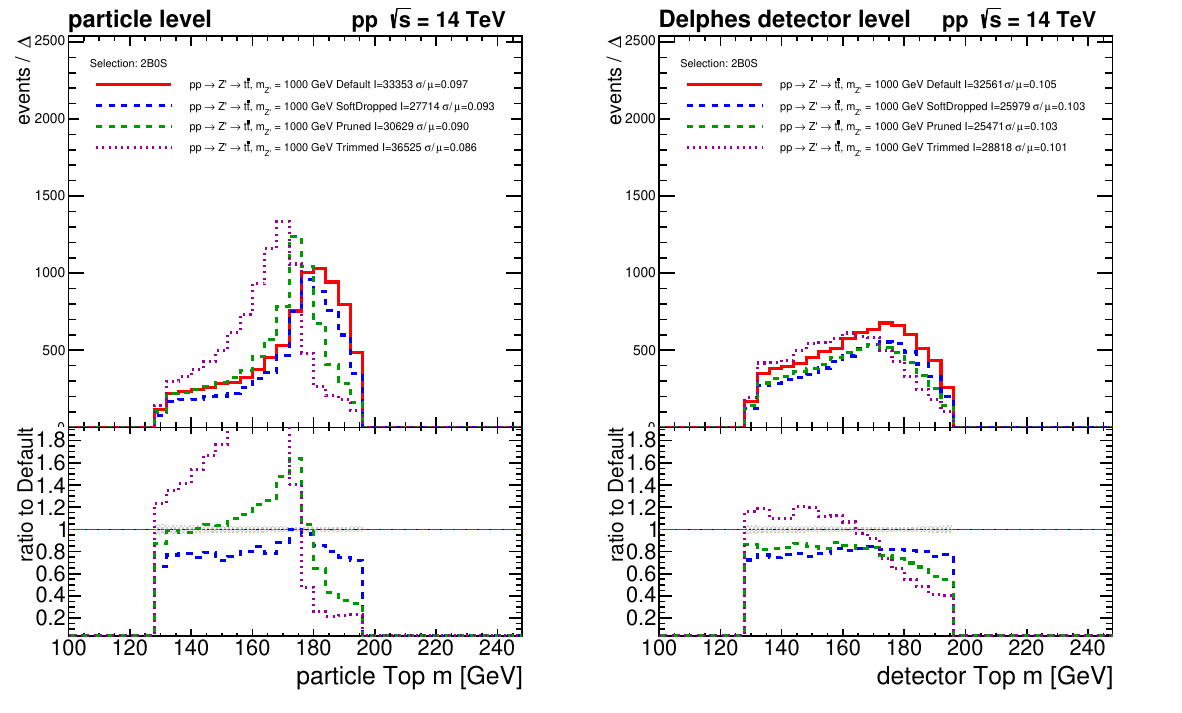}\\
\end{tabular}
\caption{Comparison between variously groomed jets for the mass of the reconstructed top quarks in the boosted-boosted (2B0S) topology. Left: particle level, right: Delphes ATLAS detector level.}
\label{fig_shapes_TopMass_2B0S}
\end{center}
\end{figure}

\begin{figure}[!hp]
\begin{center}
\begin{tabular}{c}
\includegraphics[width=0.95\textwidth]{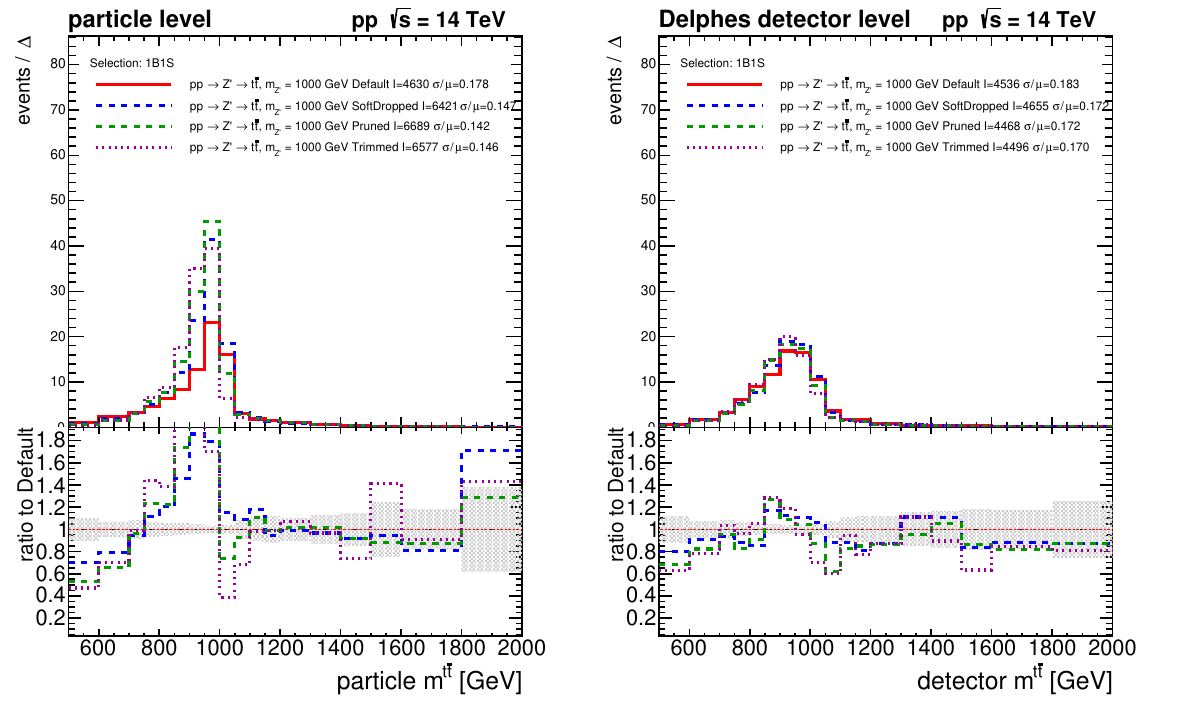}\\
\end{tabular}
\caption{Comparison between variously groomed jets for the invariant mass of the $\ttbar$ pair in the boosted-semiboosted (1B1S) topology. Left: particle level, right: Delphes ATLAS detector level.}
\label{fig_shapes_DiTopMass_denser_1B1S}
\end{center}
\end{figure}

The performance of several grooming methods is evaluated in this Appendix in the case of the line shape of the $W$-jet candidates in Figure~\ref{fig_shapes_WMass_0B2S} in the 0B2S topology, top-jets in Figure~\ref{fig_shapes_TopMass_2B0S} in the 2B0S topology, and also in terms of the reconstructed mass of the $Z' \rightarrow \ttbar{}$ in Figure~\ref{fig_shapes_DiTopMass_denser_1B1S} in the 1B1S topology.
From these figures, the Trimmed jets case was selected as the default one, optimizing the mass width for the top jets but also the event yield.


\clearpage
\section{Jet energy scale}
\label{app:JES}

Although small-$R$ jets should be provided also as a~calibrated collection by \Delphes{}, we have found some additional calibration is needed, and the jet energy scale for both the precalibrated small and uncalibrated large-$R$ jets was derived in the same way.

The detector-level jets calibration is performed w.r.t. to the matching particle level jets based on the $\pt$ ratio
$$R_{\pt} \equiv \frac{\pt^\mathrm{det}}{\pt^\mathrm{ptcl}}$$
for jets matched angularly within $\Delta R < 0.2$.
The ratio is evaluated as a~function of the detector jet energy and pseudorapidity and is fitted using 2D formulas with exponential energy dependence for small-$R$ jets, as only a~small additional correction was needed, and with logarithmic energy dependence in case of large-$R$ jets, i.e.
$$ f(E,\eta) \equiv \left[ \mathcal{C}_E + \mathcal{A} \, \log\beta {E} \right] \cdot \left[ \mathcal{C}_\eta + \mathcal{D}\,\exp\frac{-(|\eta|-\eta_0)^2}{2\sigma_\eta^2} \right] \,,$$
with $\mathcal{A}$ being a~linear function of the jets $\eta$ while $\mathcal{D}$ of jet energies $E$. The $\eta$ dependence of the $\beta$ parameter was tried but the slope parameter was consistent with zero.
The fitted parameters for the small-$R$ and large-$R$ jets are presented in~Table~\ref{tab:JESfit_pars}.
While the $\chi^2/$ndf are of the order of 20, the 2D fit residuals were checked to be centered at 1 with standard deviation of 0.01. The inverse of the ratio is the jet energy scale (JES) correction defined as a~function of the non-calibrated jets energy and $\eta$.

\begin{figure*}
\begin{center}
\begin{tabular}{c}
  \includegraphics[width=0.42\textwidth]{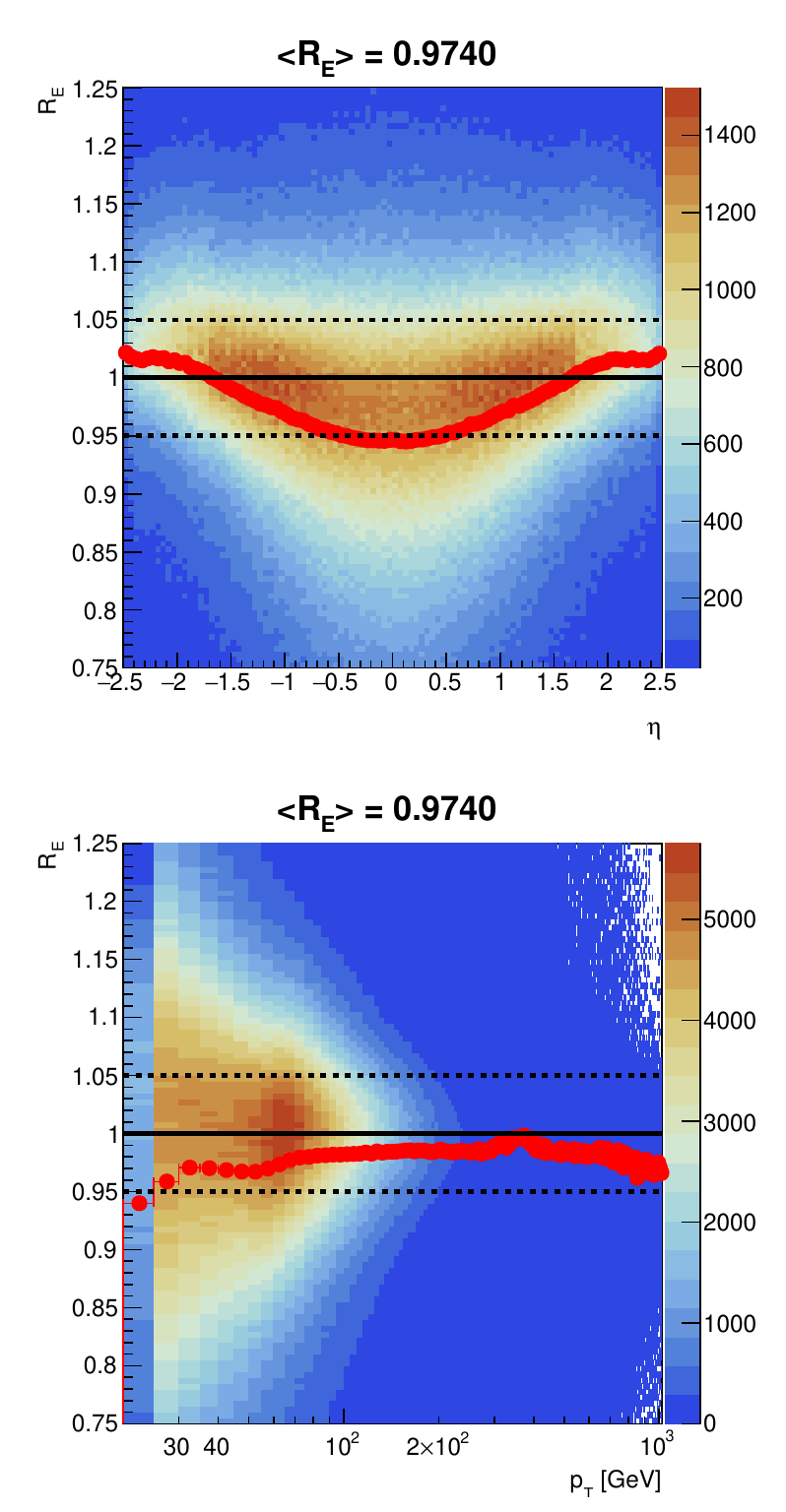}
  \includegraphics[width=0.42\textwidth]{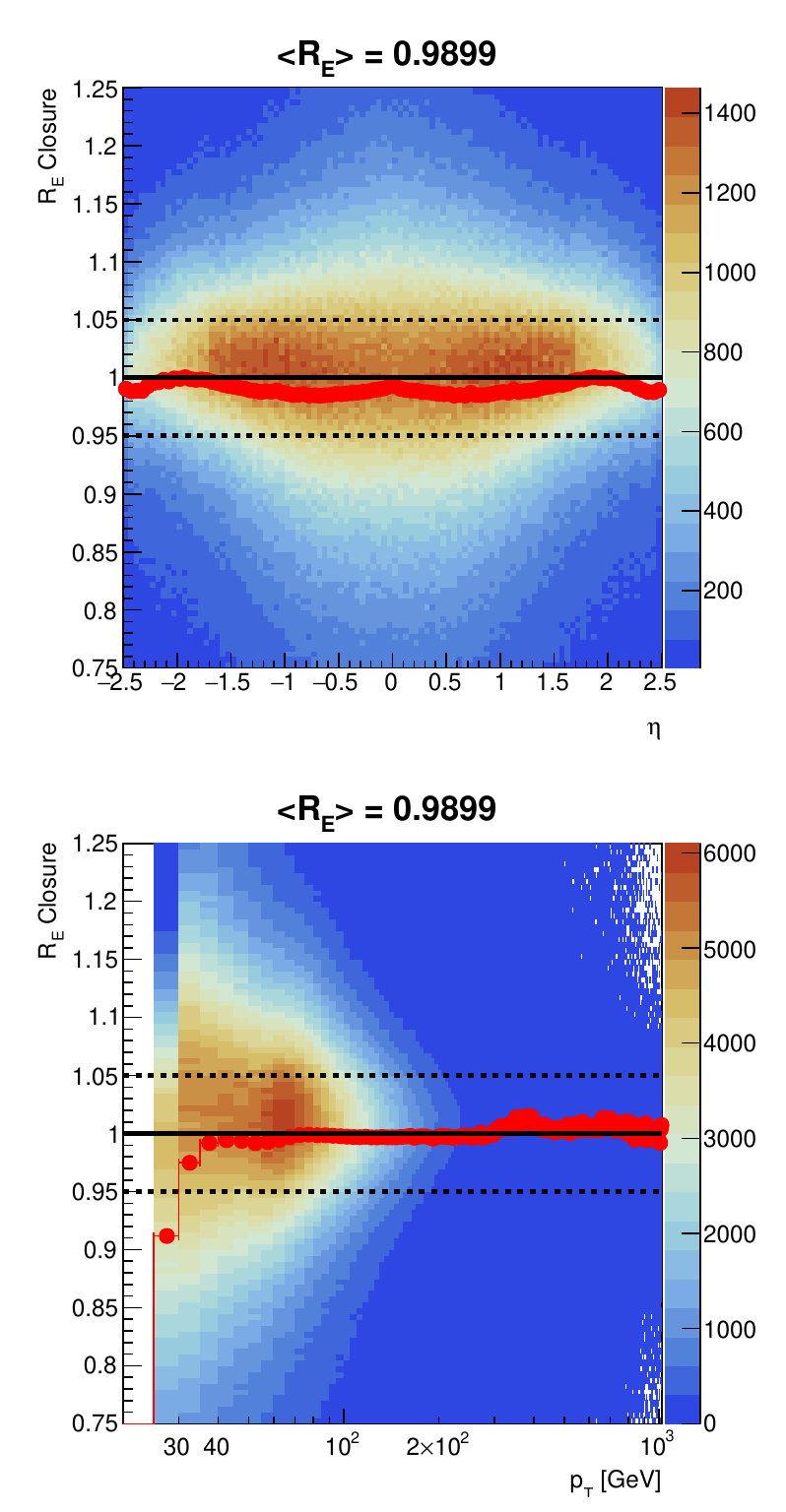}
\end{tabular}
\caption{Left: before JES, right: after the additional (on top of the \Delphes{} JES) small-$R$ jets jet energy scale correction: energy ratios between the detector and particle-level matched jets as a~function of the jet pseudorapidity $\eta$ (top) and transverse momentum $\pt$ (bottom). The black dashed lines indicate the $1 \pm 0.05$ $y$-axis region and the red marks show the profiled distribution.}
\label{fig_JES_jets_imbalances}
\end{center}
\end{figure*}

\begin{figure*}
\begin{center}
\begin{tabular}{c}
  \includegraphics[width=0.42\textwidth]{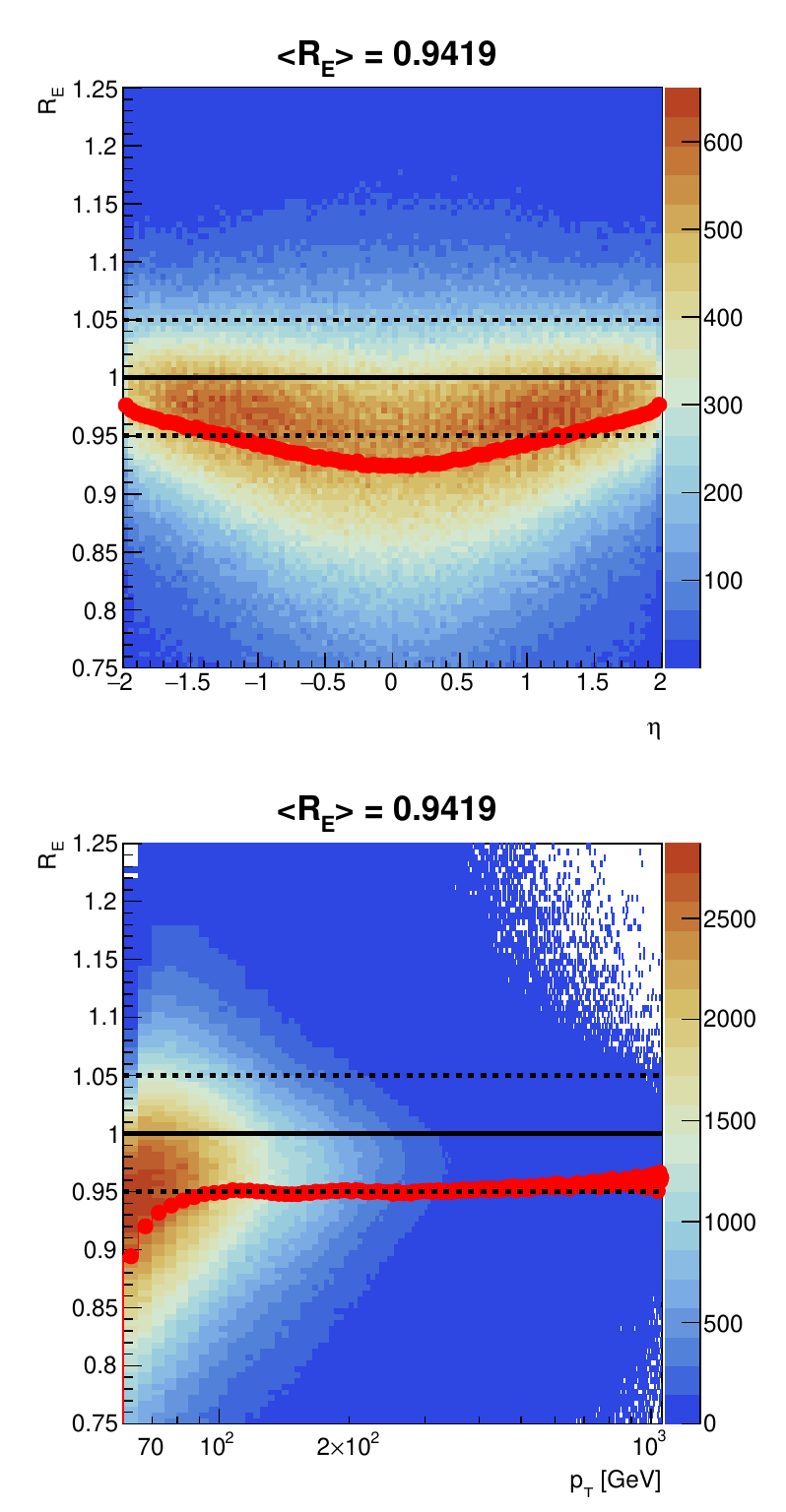}
  \includegraphics[width=0.42\textwidth]{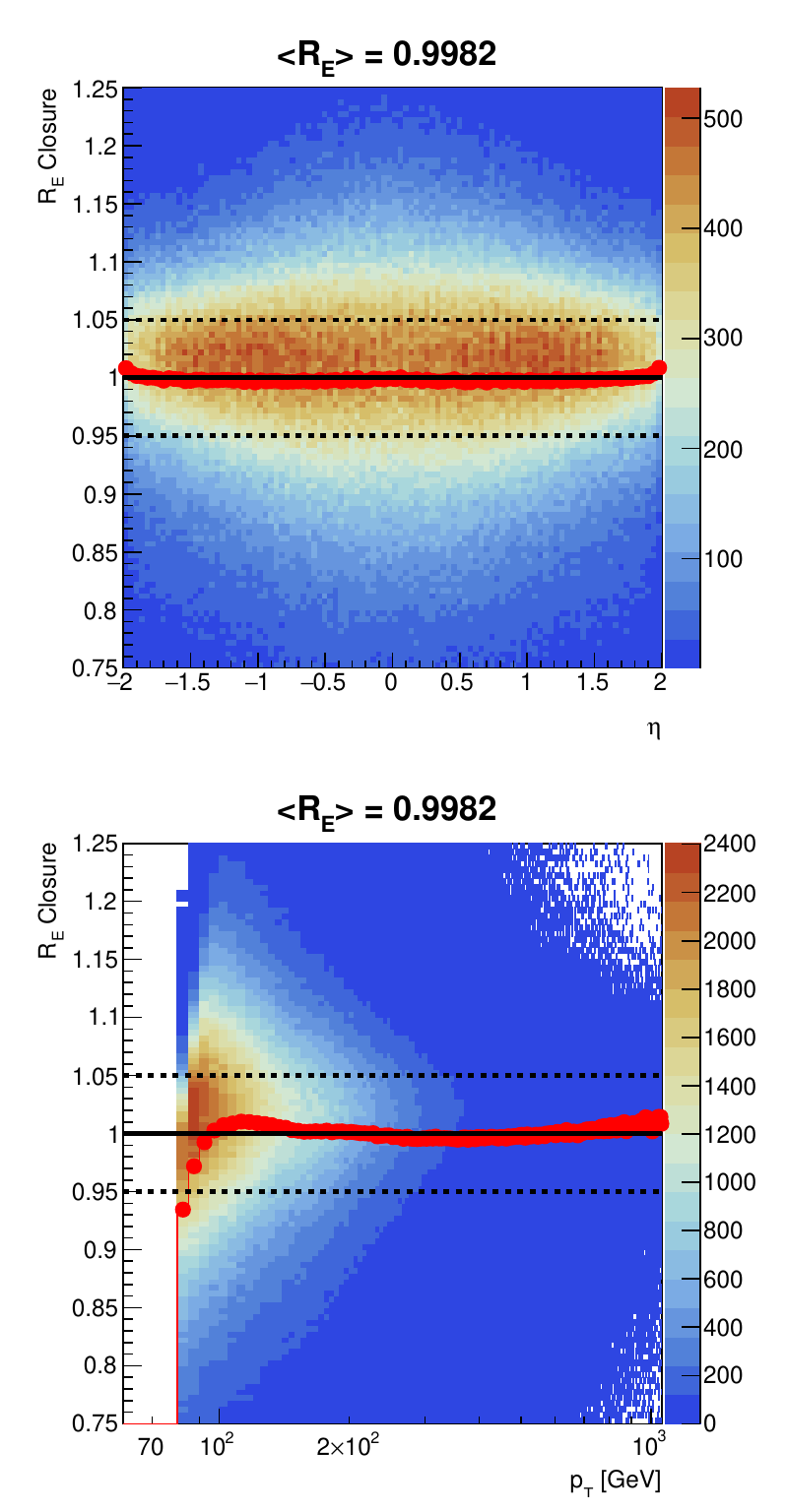}
\end{tabular}
\caption{Left: before JES, right: after the large-$R$ jets jet energy scale correction: energy ratios between the detector and particle-level matched jets as a~function of the jet pseudorapidity $\eta$ (top) and transverse momentum $\pt$ (bottom). The black dashed lines indicate the $1 \pm 0.05$ $y$-axis region, while the red marks show the profiled distribution.}
\label{fig_JES_ljets_imbalances}
\end{center}
\end{figure*}

For the JES event-by-event and jet-by-jet closure test, the
$$R_E \equiv \frac{E^\mathrm{det}}{E^\mathrm{ptcl}}$$
variable was studied and the ratio was evaluated as a~function of jets $\pt$ and $\eta$, see Figures~\ref{fig_JES_jets_imbalances}--\ref{fig_JES_ljets_imbalances}.

\begin{table*}
  \centering
\begin{tabular}{|l|ll|} \hline
  parameter & value & stat. unc. \\ \hline
  $\mathcal{A}^{(0)}$ & -0.19397 & 0.00107 \\
$\mathcal{A}^{(1)}$ & 0.11116 & 0.00085 \\
$\beta^{(0)}$ & -0.01812 & 0.00016 \\
$\beta^{{1}}$ & 0. (fixed) & --- \\
$\mathcal{D}^{(0)}$ & -0.78014 & 0.00004 \\
$\mathcal{D}^{(1)}$ & -0.00001 & 0.00000 \\
$\eta_0$ & 0.62113 & 0.00814 \\
$\sigma_\eta$ & 6.49777 & 0.02842 \\
$\mathcal{C}_E$ & 1.52223 & 0.00004 \\
  $\mathcal{C}_\eta$ & 1.33235 & 0.00007 \\
   \hline
\end{tabular}
\begin{tabular}{|l|ll|} \hline
 parameter & value & stat. unc. \\ \hline
$\mathcal{A}^{(0)}$ & 0.02724 & 0.00015 \\
$\mathcal{A}^{(1)}$ & -0.00087 & 0.00011 \\
$\beta^{(0)}$ & 0.03594 & 0.00240 \\
$\beta^{(1)}$ & 0. (fixed) & --- \\
$\mathcal{D}^{(0)}$ & 3.99237 & 0.05791 \\
$\mathcal{D}^{(1)}$ & -0.00049 & 0.00001 \\
$\eta_0$ & 115.48512 & 0.17823 \\
$\sigma_\eta$ & 49.70973 & 0.09824 \\
$\mathcal{C}_E$ & 0.69936 & 0.00629 \\
$\mathcal{C}_\eta$ & 0.91786 & 0.00409 \\
 \hline
\end{tabular}
\caption{Fit parameters and their statistical uncertainties for the additional calibration of small-$R$ jets (left, $\chi^{2}/$ndf $= 29759.1/1840 = 16.2$), and full large-$R$ jets calibration (right, $\chi^{2}/$ndf $= 43460.8/1822 = 23.9$). The $\mathcal{A}$ dependence on the absolute value of the jet pseudorapidity $|\eta|$ is parameterized as  $\mathcal{A} = \mathcal{A}^{(0)} + |\eta| \, \mathcal{A}^{(1)} $ while the $\mathcal{D}$ dependence on the jet energy is parameterized as $\mathcal{D} = \mathcal{D}^{(0)} + E\,\mathcal{D}^{(1)}$.}
\label{tab:JESfit_pars}
\end{table*}

\clearpage

\section{Boosted objects tagging efficiencies}
\label{app:Tag}

\begin{figure*}
\begin{center}
\begin{tabular}{cc}
  \includegraphics[width=0.47\textwidth]{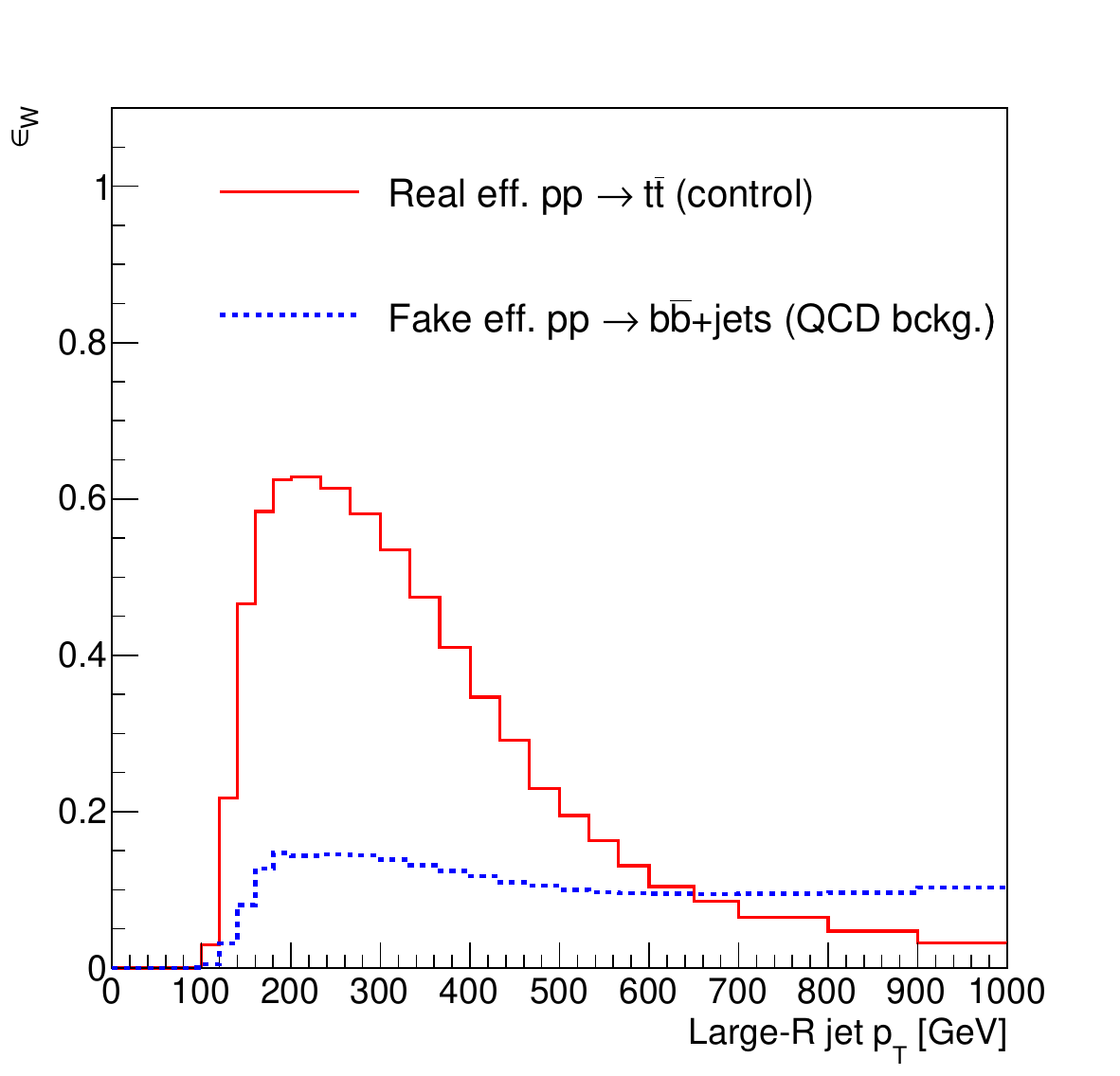} &
  \includegraphics[width=0.47\textwidth]{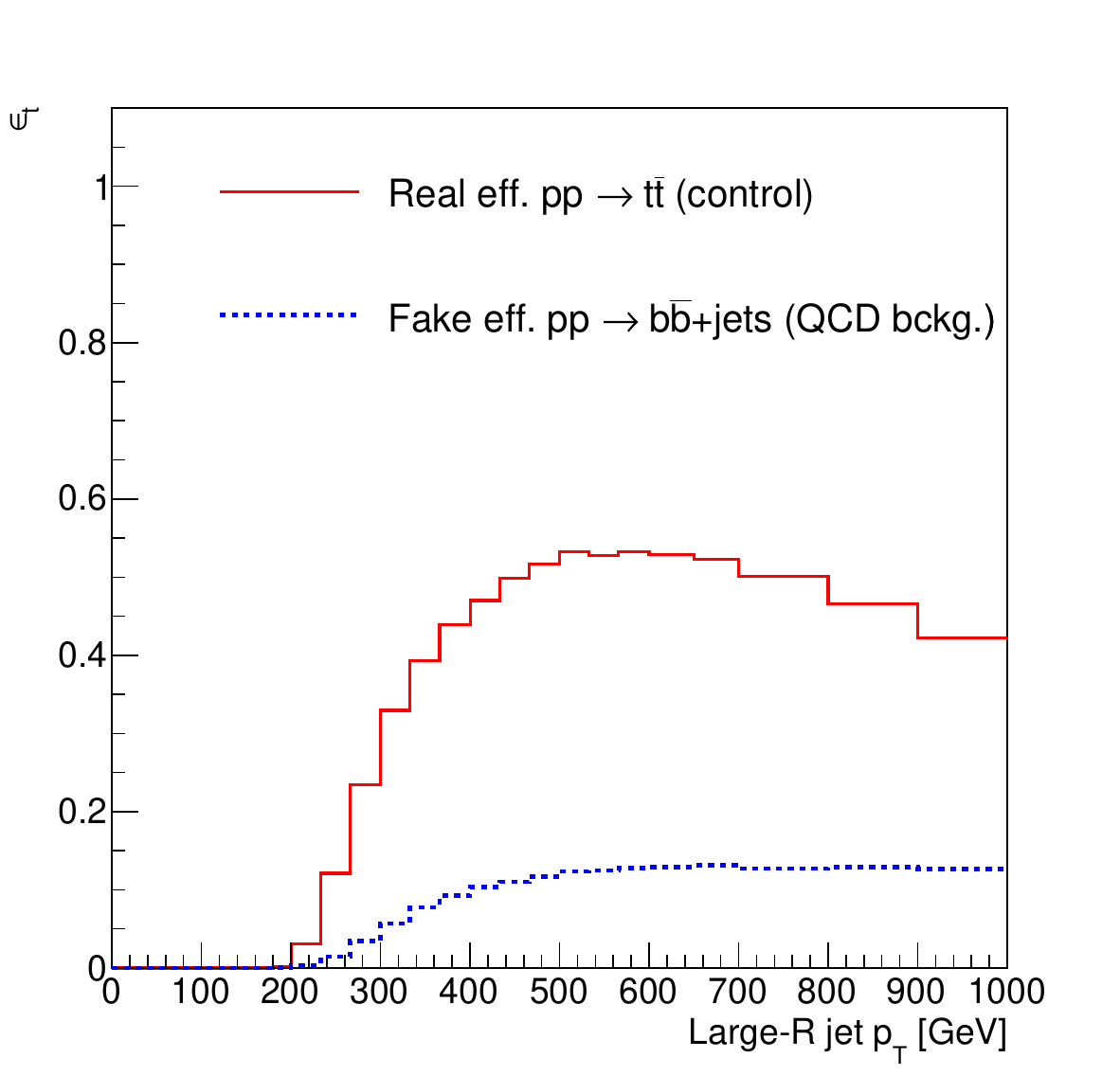} \\
  \end{tabular}
 \caption{The real (red) and fake (blue) $W$ (left) and $t$ (right) tagging efficiencies at the \Delphes{} detector level as function of the large-$R$ jet transverse momentum evaluated using the baseline $\ttbar{}$ sample and the $b\bar{b}+$jets QCD background sample.}
\label{fig_tagging_vs_pt}
\end{center}
\end{figure*}

\begin{figure*}
\begin{center}
\begin{tabular}{c}
  \includegraphics[width=0.65\textwidth]{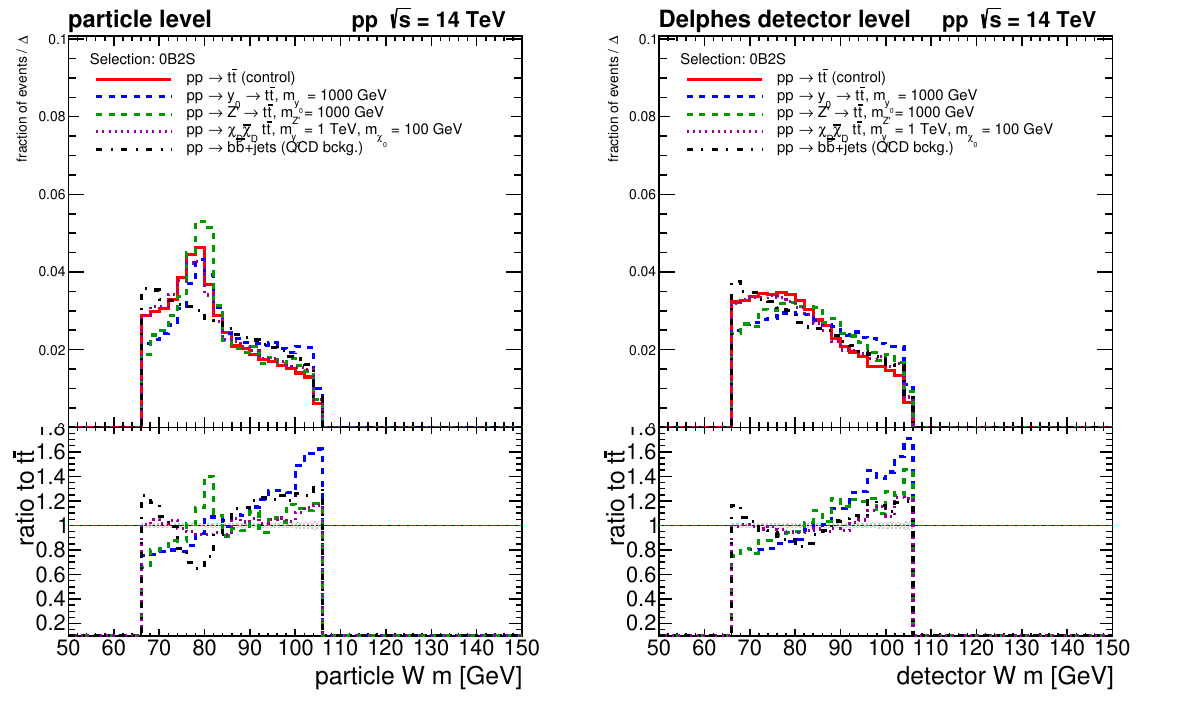} \\
  \includegraphics[width=0.65\textwidth]{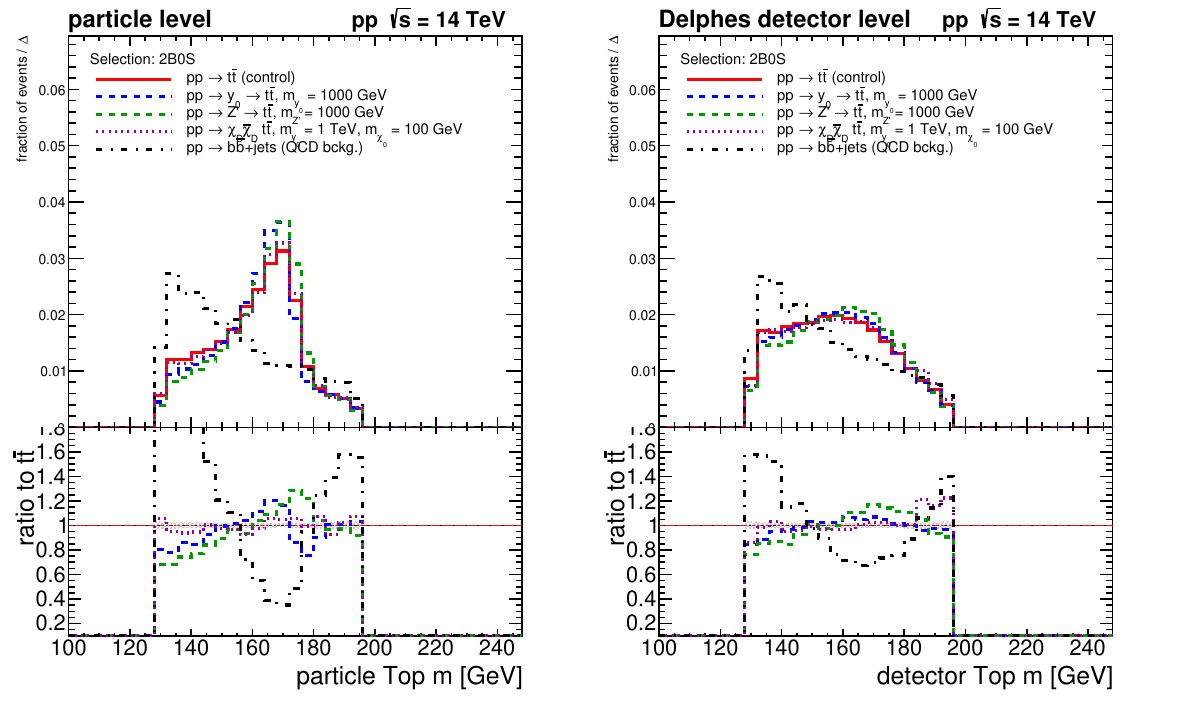} \\
  \end{tabular}
 \caption{The mass shapes of the tagged $W$ (top) and $t$ (bottom) jets in the 0B2S and 2B0S topologies, respectively, at the particle (left) and \Delphes{} detector (right) levels for various samples used in this study.}
\label{fig_jet_masses}
\end{center}
\end{figure*}

\begin{figure*}
\begin{center}
\begin{tabular}{c}
  \includegraphics[width=0.65\textwidth]{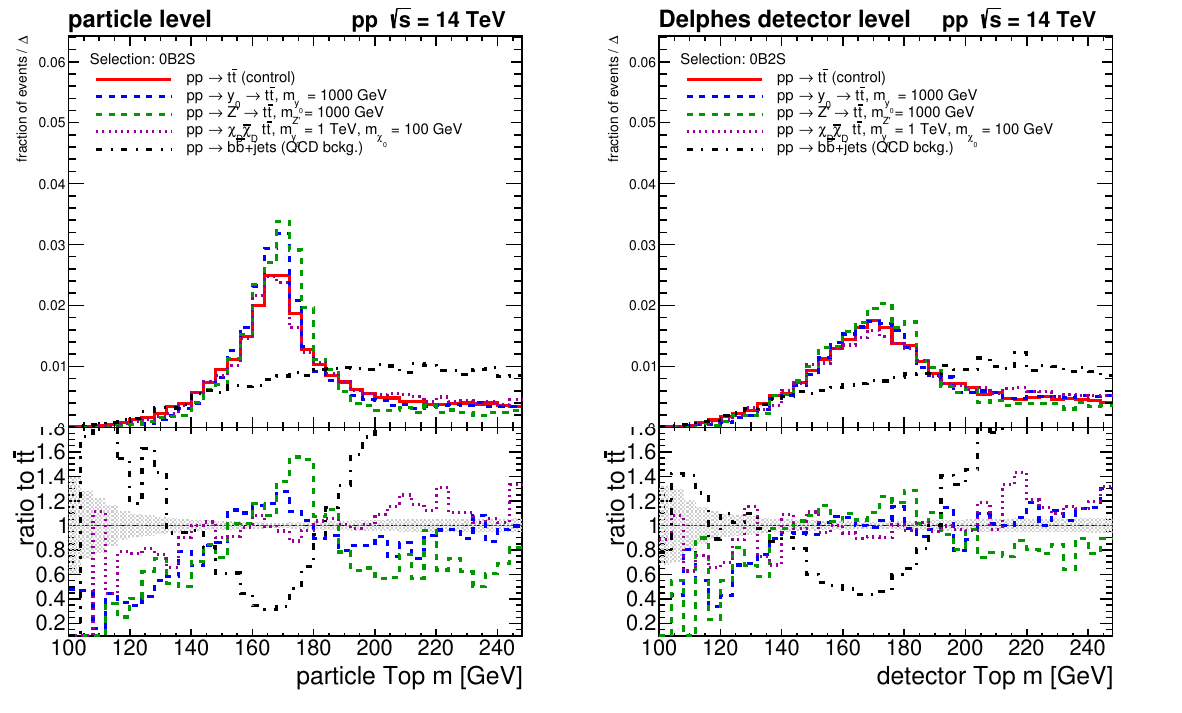} \\
  \end{tabular}
 \caption{The mass shapes of the top quark candidates in the 0B2S topology, where the four-vector of a~nearest $b$-jet is added to that of a~$W$-tagged jet at the particle (left) and \Delphes{} detector (right) levels for various samples used in this study.}
\label{fig_jet_masses_0b2s}
\end{center}
\end{figure*}

The matching of large-$R$ jets was done to final state $t$ and $W$ particles with a~status code of~52 in \Pythia{}\footnote{According to the \Pythia{}8 manual: "52 \ldots outgoing copy of recoiler, with changed momentum''.}. The matching cut on the $\Delta R$ between the large-$R$ jet and the $W$ or $t$ parton of 0.15 and 0.12, respectively, was determined using the minima of the $\Delta R$ between the jet and the partons.
The corresponding true and fake tagging efficiencies for $W$ and $t$ tagging are depicted in~Figure~\ref{fig_tagging_vs_pt}, and the resulting shapes of the tagged large-$R$ jet masses are shown in~Figure~\ref{fig_jet_masses}, including the mass of the top quark candidates in the 0B2S topology where its four-vector is formed as a~combination of four-vectors of the nearest $b$-jet to the $W$-tagged jet (Figure~\ref{fig_jet_masses_0b2s}).

\end{document}